\documentclass{article}

\usepackage{arxiv}

\ifpdf \usepackage[pdftex]{graphicx} \pdfcompresslevel=9
\else \usepackage[dvips]{graphicx} \fi

\graphicspath{ {./img/} }
\DeclareGraphicsExtensions{.jpg,.png,.pdf}

\usepackage{amsfonts}
\usepackage{amsmath}
\usepackage{amssymb}
\usepackage{array}
\usepackage{bm}
\usepackage{booktabs}
\usepackage[font=small,labelfont=bf]{caption}
\usepackage{color}
\usepackage{contour}
\usepackage{doi}
\usepackage{enumitem}
\usepackage{float}
\usepackage[T1]{fontenc}    
\usepackage{gensymb}
\usepackage{graphicx}
\usepackage{hyperref}
\usepackage{cleveref}
\usepackage[utf8]{inputenc} 
\usepackage{makecell} 
\usepackage{mathrsfs}  
\usepackage{microtype}      
\usepackage{multirow}
\usepackage{nicefrac}       
\usepackage{overpic}
\usepackage{stackengine}
\usepackage[font=small]{subcaption}
\usepackage{url}            
\usepackage[table]{xcolor}
\usepackage{xfrac}

\newcommand{\bp}{\mathbf{p}}

\newcommand{\bnr}[1]{\mathcal{N}_r(#1)}
\newcommand{\bn}[2]{\mathcal{N}_{#1}(#2)}
\newcommand{\bni}[1]{\mathscr{N}(#1)}
\newcommand{\bnp}{\mathbf{n}_\bp}
\newcommand{\bdist}[2]{d(#1,#2)}

\crefname{chapter}{chapter}{chapters}
\crefname{section}{section}{sections}
\crefname{subsection}{subsection}{subsections}
\crefname{figure}{figure}{figures}
\crefname{equation}{equation}{equations}

\usepackage{ifthen}
\newboolean{DRAFT}
\setboolean{DRAFT}{false}  
\newcommand{\marc}[1]{{\ifthenelse{\boolean{DRAFT}}{\color{red}\textsuperscript{Marc }}{}#1}}
\newcommand{\toni}[1]{{\ifthenelse{\boolean{DRAFT}}{\color{green}\textsuperscript{Toni }}{}#1}}
\newcommand{\carlos}[1]{{\ifthenelse{\boolean{DRAFT}}{\color{blue}\textsuperscript{Carlos }}{}#1}}
\usepackage{dcolumn}

\title{Revisiting Poisson-disk Subsampling for Massive Point Cloud Decimation}

\author{\href{https://orcid.org/0000-0001-5621-7565}{
        \includegraphics[scale=0.06]{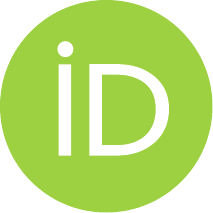}\hspace{1mm}
        Marc Comino-Trinidad} \\
	Unversidad Rey Juan Carlos, Spain \\
	\texttt{marc.comino@urjc.es} 
	\And
	\href{https://orcid.org/0000-0003-0270-2332}{
        \includegraphics[scale=0.06]{orcid.pdf}\hspace{1mm} 
        Antonio Chica} \\
	Universitat Polit\`ecnica de Catalunya, Spain \\
	\texttt{achica@cs.upc.edu} 
	\And
	\href{https://orcid.org/0000-0002-8480-4713}{
        \includegraphics[scale=0.06]{orcid.pdf}\hspace{1mm} 
        Carlos Andújar} \\
	Universitat Polit\`ecnica de Catalunya, Spain \\
	\texttt{andujar@cs.upc.edu} 
}

\begin{document}

\maketitle
\begin{abstract}
Scanning devices often produce point clouds exhibiting highly uneven distributions of point samples across the surfaces being captured. Different point cloud subsampling techniques have been proposed to generate more evenly distributed samples. Poisson-disk sampling approaches assign each sample a cost value so that subsampling reduces to sorting the samples by cost and then removing the desired ratio of samples with the highest cost. Unfortunately, these approaches compute the sample cost using pairwise distances of the points within a constant search radius, which is very costly for massive point clouds with uneven densities.  
In this paper, we revisit Poisson-disk sampling for point clouds. Instead of optimizing for equal densities, we propose to maximize the distance to the closest point, which is equivalent to estimating the local point density as a value inversely proportional to this distance. This algorithm can be efficiently implemented using k nearest-neighbors searches. Besides a kd-tree, our algorithm also uses a voxelization to speed up the searches required to compute per-sample costs. We propose a new strategy to minimize cost updates that is amenable for out-of-core operation. We demonstrate the benefits of our approach in terms of performance, scalability, and output quality. We also discuss extensions based on adding orientation-based and color-based terms to the cost function. 

\end{abstract}  
\section{Introduction}

Most 3D scanning technologies used nowadays involve a light sensor that captures scene data from a specific location. The data acquired might include just color (depth can be reconstructed later through photogrammetry techniques) or color and depth (for example, in triangulation-based and time-of-flight scanners). Although some scanning devices can be mounted on moving systems (e.g., robotic arms, vehicles, drones, airplanes, and satellites), for many scenarios (e.g., digitization of buildings), the highest accuracy is achieved by using stationary equipment such as LiDAR stations.

Considering a single scan, surfaces visible from the sensor location result in unevenly distributed samples due to multiple reasons. Some factors are intrinsic to the surface being captured. For example, surface patches with challenging materials (e.g., mirror-reflective) are poorly captured. Nevertheless, the most significant sources of uneven sampling are extrinsic factors, namely surface orientation and distance to the sensor~\cite{hackel2016}. Even if we could assume the sensor to capture sample data at regular solid angles, sampling spacing on a surface would be inversely proportional to sensor distance and the incident angle's cosine. 

\begin{figure}[!htb]
\centering
   \begin{subfigure}[t]{0.20\linewidth}
    \includegraphics[width=1\linewidth]{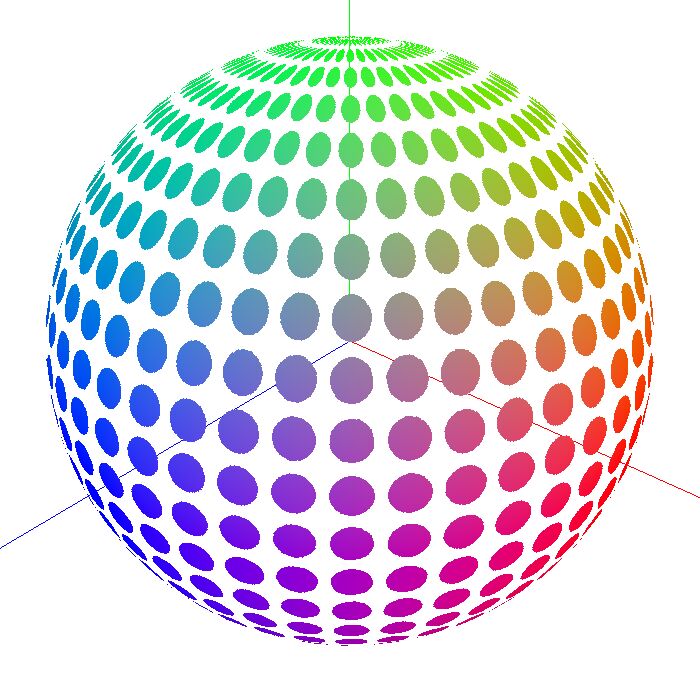}
    \caption{}
    \label{fig:c4f1i1}
  \end{subfigure}
  \hspace{5mm}
  \begin{subfigure}[t]{0.20\linewidth}
    \includegraphics[width=1\linewidth]{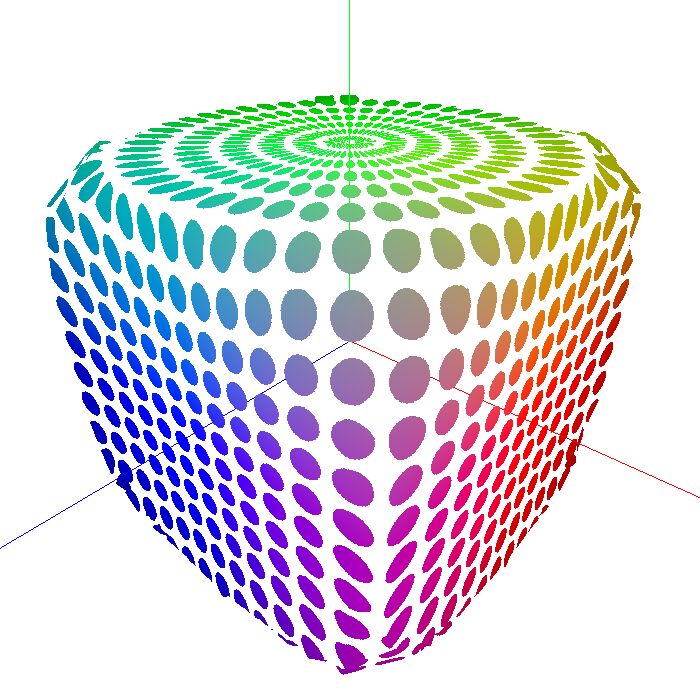}
    \caption{}
   \label{fig:c4f1i2}
  \end{subfigure}
  \hspace{5mm}
  \begin{subfigure}[t]{0.20\linewidth}
    \includegraphics[width=1\linewidth]{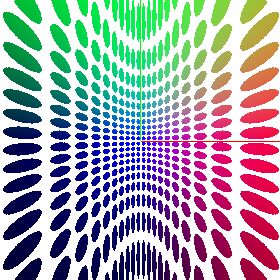}
    \caption{}
   \label{fig:c4f1i3}
  \end{subfigure}
\caption{Distribution of samples captured by a terrestrial LiDAR scanner on a sphere \textit{(a)}, a roughly cubical volume \textit{(b)} and a vertical wall \textit{(c)}. Samples are generated at regular intervals of the polar angles $(\theta, \psi)$, producing very irregular point densities even on a surrounding sphere \textit{(a)}. The volume in \textit{(b)} better shows the high density of samples around the poles in detriment to regions close to the Equator. Another scenario is the vertical wall in \textit{(c)} whose center is orthogonal to the beam direction. As we move towards the borders, the incident angle has a substantial influence on point density.} 
\label{fig:c4f1}

\vspace{-4mm}
\end{figure}

Surprisingly, high-end terrestrial LiDAR scanners often sample the surfaces at non-uniform solid angles, thus introducing an additional source of uneven sampling. Most of these LiDAR devices use a rotating head and a rotating mirror to capture $360^{\circ}$ data. The rotating mirror deflects an infrared laser beam at regular intervals in the vertical direction, whereas the rotating base allows for horizontal sweeping. The device measures the time for the beam to hit a surface and return to the sensor, determining its depth. In most LiDAR equipment, the number of acquired samples per horizontal line does not depend on the vertical angle. Therefore, in terms of angular resolution, raw scans contain much more samples around the poles than across the Equator (\Cref{fig:c4f1}).

Some scenarios allow for taking multiple scans by placing the scanner at different locations. The resulting point cloud improves surface coverage and alleviates the effect of occlusions but does not reduce the uneven sampling problem. Unevenly distributed samples are undesirable for several reasons. First, sample density in raw point clouds does not correlate to valuable surface attributes such as subjective relevance, color detail, or feature richness. Second, many algorithms (e.g., normal estimation, segmentation, reconstruction) are less robust when operating on highly uneven samples.  

Given a point cloud, we wish to get another point cloud roughly representing the same underlying surface (in terms of coverage) but having a more even distribution of samples across the surface. 

Increasing sampling in undersampled regions can be addressed, e.g., through simple interpolation or deep learning approaches that attempt to reconstruct missing data in a plausible manner considering the examples used to train the network. However, the most popular approach is to use a Moving Least Squares algorithm~\cite{lancaster1981,alexa2001,fleishman2005,kolluri2008}. These upsampling approaches are out of the scope of this paper since we instead focus on how to deal with oversampled regions in point clouds containing millions or billions of points. 

The problem we address is thus a particular instance of point cloud simplification, which can be approached through re-sampling (i.e., creating new points not included in the original point cloud) and sub-sampling (i.e., selecting a subset of the original samples). We adopt a sub-sampling approach because this avoids attribute (e.g., color, position) interpolation and thus leads to more accurate results. 

Poisson-disk sampling approaches~\cite{yuksel2015} seem to offer a convenient solution for selecting subsets with more evenly distributed samples in multiple domains. This is accomplished by defining a ranking criterion based on the distance of a point to its neighbors and then create Poisson disk samples by removing the desired ratio of samples with the highest rank. Intuitively, by removing points with many nearby neighbors, these approaches minimize the maximum point density. 
As we shall see, the running times of these approaches depend on the choice of a search radius. For highly unevenly sampled data, finding a suitable search radius is not possible. If too large, heavily oversampled areas will kill performance; if too small, the neighborhood will be empty for most parts of the cloud. Therefore, these methods do not scale well in real-world scanned datasets. 

This paper presents a scalable, out-of-core algorithm for generating evenly distributed point clouds through sub-sampling.  Instead of optimizing for equal densities, we propose a sample ranking criterion to maximize the distance between closest points, replacing radius searches with k-nearest-neighbors searches with k = 1. Namely, estimating the local point density as a value inversely proportional to the distance to the closest point. This setting allows us to speed up nearest neighbor searches through a simple voxelization approach.

We demonstrate the benefits of our approach in terms of performance and output quality. We also show that our rank criterion can be trivially extended by adding orientation-based and color-based terms, improving sampling density preservation in those regions where the surface exhibits high-frequency variation in shape or appearance.  

\section{Previous Work}

Strategies for point cloud simplification in the literature can be grouped into two major classes. The first one is \textit{re-sampling}~\cite{pauly2002, wu2004, du2007, miao2009}, that is, computing a set of different points that represent the same underlying surface as the original one but with a smaller number of points. Methods in the second group perform \textit{sub-sampling}~\cite{linsen2001, alexa2001, song2009, corsini2012, yuksel2015, qi2019}, namely finding a criterion to select some points from the original set.

\textbf{Re-sampling:}  Wu et al.~\cite{wu2004} introduced a method that explicitly optimizes the output cloud for splat rendering. Points, represented as splats, are greedily selected to ensure maximum surface coverage and, then, a global optimization step refines their position. Instead, Miao et al.~\cite{miao2009} studied how to produce non-uniform samplings so that more samples are placed near high curvature regions. This is achieved through adaptive hierarchical mean-shift clustering.

Pauly et al.~\cite{pauly2002} study how to transfer several mesh-based simplification methods to point clouds. Specifically, they propose using clustering, iterative decimation through quadric error metrics, and particle simulation. Du et al.~\cite{du2007} also proposed a method based on quadric error metrics. They first divide the cloud into cells and, for each one, they compute the average representative and PCA normal. After this, they use surface variation~\cite{pauly2002} to find the boundaries of the model while using point-pair contraction with quadric error to simplify planar regions further.

Re-sampling methods usually produce results smoother than the original clouds. These may be useful in order to reduce the noise level but may also remove geometric features. Notice that most methods rely on normals which have been computed on the noisy raw data. Because of this, sub-sampling methods are often better to preserve finer geometric detail. 

A related problem consists in sampling a given surface to generate randomized distributions. Most techniques are based on Blue-noise sampling to generate randomly placed points that are approximately evenly spaced. Yan et al.~\cite{yan2015} survey Blue-noise sampling methods, including Poisson-disk sampling and relaxation-based approaches. Farthest-point optimization techniques~\cite{schlomer2011,yan2014} also attempt to maximize the minimal distance in a point cloud as we do, but in the context of surface re-sampling rather than cloud sub-sampling. Yan et al.~\cite{yan2014} report that farthest-point optimization generates point sets with excellent blue-noise properties for surface sampling.

\toni{
When the final application of point cloud reduction is rendering it may make sense to be able to produce different levels of detail~\cite{dachsbacher2003sequential, wimmer2006instant, elseberg2013one, goswami2013efficient}. In these cases an underlying data structure is used to compute these resamplings efficiently, but it also conditions the shape and quality of the produced point clouds.
}

\textbf{Sub-sampling:} The stratified simplification approach~\cite{Nehab:2004:SPS} shares the use of a voxelization with our method, but chooses the final representatives points using an exponential distribution function centered at the center of each voxel. A final pass removes any samples that are still too close. In order to tackle large datasets data structures based on BSP~\cite{gobbetti2004layered}, octrees~\cite{WAND2008204}, cluster trees~\cite{yu2010asm}, and others have been proposed.

Linsen~\cite{linsen2001} proposes ranking points using multiple descriptors that account for non-planarity, non-uniformity of the surface, and normal variation. Points are inserted into a priority queue and iteratively removed until the target number of points is reached. A contemporary work by Alexa et al.~\cite{alexa2001} ranks points based on the distance from a point to its moving least squares (MLS) projection. Then, those with the largest distances are iteratively removed. These methods aim at removing noisy and redundant points from high-density areas. However, they rely on the assumption that the underlying surface is smooth. 

Song et al.~\cite{song2009} present a feature-preserving simplification algorithm that first determines whether a point belongs to an edge or not. Then, it removes non-edge points based on a sorting criterion similar to Linsen's~\cite{linsen2001} non-planarity score.

A drawback of the previous methods is that they may produce uneven sampling distributions. Qi et al.~\cite{qi2019} present a method that leverages feature preservation and density uniformity on the simplified clouds. They approach the problem by applying graph signal processing to point clouds represented as graphs. For this, they encode the clouds into adjacency matrices using k-nearest neighborhood adjacency.

Instead of decimation, Moenning et al.~\cite{moenning2003} presented a point cloud simplification algorithm which, starting from an initial random small set of points, iteratively adds new points until a target density is reached (refinement). The algorithm selects the point which is "farthest" from the rest using intrinsic geodesic Voronoi diagrams.

Finally, Yuksel~\cite{yuksel2015} proposed an algorithm for generating Poisson disk samples by iteratively removing points. Similar to other sub-sampling methods, they define ranking criteria based on the distance of a point to its neighbors. Intuitively, by removing points with many close neighbors, they end up minimizing the maximum point density.

\toni{
An alternative way to perform subsampling on a point cloud is to consider it as a graph and apply a graph filter~\cite{chen2017fast}. This filter may be high-pass to enhance the contours or low-pass to capture the rough shape of the model. The resulting sampling is guaranteed to be shift, rotation and scale invariant, but cannot deal with gigantic point clouds efficiently. Graphs are also used by Bletterer et al~\cite{bletterer2018towards, bletterer2020local} as a representation for the incremental application of Poisson-disk sampling.

The generation of blue noise samplings may be parallelized to provide better performance. Wei~\cite{wei2008parallel} proposed a sampling on a multiresolution grid that avoided any expected biases. Performing the sampling process in a random non-conflicting order allows for a GPU implementation. Bowers et al~\cite{bowers2010parallel} extend this to sampling on arbitrary surfaces, while also introducing a new method to analyse the quality of the result. Kalantari and Sen~\cite{kalantari2012fast} proposed an alternative method. They produced an initial small patch of Poisson-disk samples. The resulting blue noise sampling is obtained by subsequently copying randomly rotated and translated versions of this initial sampling. 

In some applications, the generated sampling may benefit from considering the domain anisotropy. Li et al~\cite{li2010anisotropic} extended standard dart throwing and relaxation to support anisotropic sampling.
}

A common limitation of most of the methods above is that they are not designed to work out-of-core (except Du et al.~\cite{du2007}). While one could easily find how to implement clustering in an out-of-core fashion, it is less obvious how to do the same for those requiring a consistent global ordering (priority queue) or large data structures (adjacency matrices). Hence, to the best of the authors' knowledge, there is no method suitable for decimating clouds with billions of points while preserving some of the discussed properties. 

A related approach by Corsini et al.~\cite{corsini2012} generates Blue-noise samplings of meshes and can be easily adapted to a point cloud sub-sampling algorithm. The method randomly selects samples and removes those within a given radius, which produces good uniform distributions. 

Potree~\cite{schutz2016} uses an adapted modifiable nested octree (stored out-of-core) to allow the interactive inspection of point clouds. Points are distributed across the octree nodes while guaranteeing a specific spacing between the points assigned to each node. A user-defined spacing is assigned to the root node and halved at each subsequent level. When the number of points at a leaf node reaches a certain threshold, this node is subdivided. It keeps an approximate Poisson disk sample while distributing the remaining points among its children. This method builds multiple Poisson disk samples at different resolutions simultaneously but does not guarantee the spacing between points in different nodes, and sometimes point stripes and holes may appear.

While the previous two methods can work out-of-core, they do not allow for direct control of the output amount of samples and do not implement any feature-preserving strategy.

\section{Problem formulation}

Given a point cloud $C$ and some user-provided decimation factor $\lambda \in (0,1)$, most simplification algorithms compute a new point cloud $C'$ such that $|C'| \leq \lambda|C|$ and the surfaces represented by $C$ and $C'$ are similar. \textit{Re-sampling} methods compute points in $C'$ as representative points of some neighborhood, and thus optimize their placement and compute their attributes (e.g., color) through some interpolation or averaging scheme. We want to avoid creating new samples which can potentially be placed away from the underlying surface, decreasing the model accuracy. Hence, we adopt a sub-sampling approach by constraining $C'$ to be a subset of $C$. Notice that we are assuming low-noise samples (e.g., high-end LiDAR equipment report 0.4 mm RMS range noise at 10\,m) since high-noise data would rather benefit from re-sampling. 

We want the sub-sampling process to remove points in oversampled regions while preserving them in undersampled areas. Namely, we want local point densities in $C'$ to be as uniform as possible. More formally, given $C$ and $\lambda$, we could compute $C'$ as:

\begin{equation}
\label{eq:c4e1}
\begin{split}
\underset{C'}{\text{argmin}} ~\max 
        & \{\rho_\bp | \bp \in C'\} \\
\text{subject to }
        & C' \subset C \\
        & |C'| = \lambda|C| 
\end{split}
\end{equation}

where $\rho_\bp$ is some local density estimate at point $\bp \in C'$. 
The equality constraint above assumes the user-provided decimation ratio $\lambda$ has the form $m/|C|$ for some integer $m$. 

Let us start by considering local density estimates $\rho_\bp$ based on points within some fixed distance $r$ from $\bp \in C'$. If $\bnr{\bp}$ denotes the neighbors around $\bp$ within radius $r$, then $\rho_\bp$ should be directly proportional to $|\bnr{\bp}|$ and inversely proportional to $r$, $r^2$ or $r^3$ (depending on whether density is measured along a line, on the underlying surface or inside a volume). 

We can rewrite the problem in Equation~\ref{eq:c4e1} by replacing density estimates by distances to the closest samples. Let $d_\bp$ be the distance from $\bp \in C'$ to its closest point in $C'$, i.e. $$d_\bp = \min \{\bdist{\bp}{\bp'} | \bp' \in C' \},$$ where $\bdist{\bp}{\bp'}$ is just the Euclidean distance between points $\bp, \bp'$. Then the problem can be formulated as:

\begin{equation}
\label{eq:c4e2}
\begin{split}
\underset{C'}{\text{argmax}}~\min 
        & \{ d_\bp | \bp \in C' \} \\
\text{subject to }
        & C' \subset C \\
        & |C'| = \lambda|C|     
\end{split}
\end{equation}

In summary, we assume that minimizing the maximum density is the same as maximizing the minimum distance between two closest points. This problem is also known as finding the Poisson disk sample set with maximal radius.

\section{Computing per-sample costs}

Yuksel~\cite{yuksel2015} proposes an iterative greedy algorithm that approaches the problems formulated in~\Cref{eq:c4e1,eq:c4e2}. It consists in sorting all the points according to a cost function and iteratively removing the samples with the highest cost. At the same time, the costs associated with the neighbors of the removed samples are updated to account for the new local densities. In particular, Yuksel defines the cost $w_\bp$ of a point $\bp$ as:

\begin{equation}
\label{eq:c4e3}
    w_\bp = \sum_{p' \in \bnr{\bp}}  \left (1 - \frac{\hat{d}(\bp, \bp')}{r} \right )^\alpha
\end{equation}
\begin{equation}
\label{eq:c4e5}
\begin{split}
    \hat{d}(\bp, \bp')=
    \left\{\begin{array}{@{}cl}
            \bdist{\bp}{\bp'}, & \text{if } \bdist{\bp}{\bp'} > r_{min}\\
            r_{min}   & \text{if } \bdist{\bp}{\bp'} \leq r_{min}
    \end{array}\right .
\end{split}
\end{equation}
\begin{equation}
\label{eq:c4e6}
    r = 2\sqrt[3]{\frac{V}{4\sqrt{2}|C'|}}
\end{equation}
\begin{equation}
\label{eq:c4e7}
    r_{min} = r\left(1- \lambda^\gamma\right)\beta
\end{equation}

where $V$ is the volume of the sampling domain, $\alpha=8$, $\gamma=1.5$ and $\beta=0.65$. 

\subsection{Scaling to massive point clouds with uneven local densities}

Although Yuksel's method vastly improves local point densities, one of its shortcomings is that it does not scale to massive datasets with highly uneven point distributions. The performance bottleneck is radius-based neighbor searches. Despite using a kd-tree to speed up these searches, using a constant search radius to determine the point neighborhoods proves fatal for highly unevenly distributed point clouds (such as LiDAR data). Using a large radius on heavily dense areas kills performance, as the number of points to retrieve will be huge. However, using a small radius to make these neighborhoods treatable will instead lead to empty neighborhoods for most of the volume. Hence, we propose an adaptive neighborhood based on a fixed amount of neighbors ($k$), namely a $k$-neighborhood.

Considering a specific point $\bp$, we first study how to approximate its cost $w_\bp$ (which results  from applying \Cref{eq:c4e3} on its $r$-neighborhood) if we only had access to its $k$-neighborhood. We start by splitting $w_\bp$ into a term $w_\bp^{r_{min}}$ (which approximates the contribution of the points within $r_{min}$ to the cost) and $w_\bp^{r}$ (which approximates the contribution of the points between  $r_{min}$ and $r$ to the cost):

\begin{equation}
    w_\bp = w_\bp^{r_{min}} + w_\bp^{r}
\end{equation}

Then we would compute an estimate of the local point density $\rho_\bp$ using the $k$-neighborhood. Assuming $\rho_\bp$ to be constant within $r$, we can approximate the previous terms as:

\begin{equation}
\label{eq:c4e8}
	w_\bp^{r_{min}} = \int_{0}^{r_{min}}\rho_\bp 4\pi x^2 (1 - \frac{r_{min}}{r})^\alpha dx = \rho_\bp \cfrac{4}{3}\pi r_{min}^3 (1 - \frac{r_{min}}{r})^\alpha
\end{equation}

\begin{equation}
\label{eq:c4e9}
    w_\bp^{r} = \int_{r_{min}}^{r} \rho_\bp 4\pi x^2 (1 - \frac{x}{r})^\alpha dx = \rho_\bp 4\pi \int_{r_{min}}^{r}  x^2 (1 - \frac{x}{r})^\alpha dx
\end{equation}

Then, we can rewrite $w_\bp$ as:

\begin{equation}
\begin{split}
w_\bp = & K\rho_\bp \\
K = & \cfrac{4}{3}\pi r_{min}^3 (1 - \frac{r_{min}}{r})^\alpha + 4\pi \int_{r_{min}}^{r}  x^2 (1 - \frac{x}{r})^\alpha dx
\end{split}
\end{equation}

Notice that $K$ is a constant for any point $\bp \in C$ because it only depends on $r$, $r_{min}$ and $\alpha$, which are fixed during the algorithm execution and independent of $\bp$. Consequently, $K$ can be neglected when comparing the cost of two points to rank them. Therefore, using the local point density $\rho_\bp$ estimated using the $k$-neighborhood as a rank criterion will approximately have similar behavior as to using $w_\bp$ resulting from applying \Cref{eq:c4e3} to the $r$-neighborhood. This makes sense because by removing the points with the largest densities, we are minimizing the maximum density, which is the objective of our first problem formulation in \Cref{eq:c4e1}.  

\subsection{Improving sample distribution}

\begin{figure*}[!htb]
\centering
\begin{tabular*}{\linewidth}{cccccc}
    & Original Cloud & Yuksel ($\beta=1$) & Yuksel ($\beta=0.65$) & Yuksel ($\beta=0.35$) & Yuksel ($\beta=0$)
    \\
    \rotatebox[origin=c]{90}{Top Face} &
    \raisebox{-0.5\height}{\includegraphics[width=0.18\linewidth]{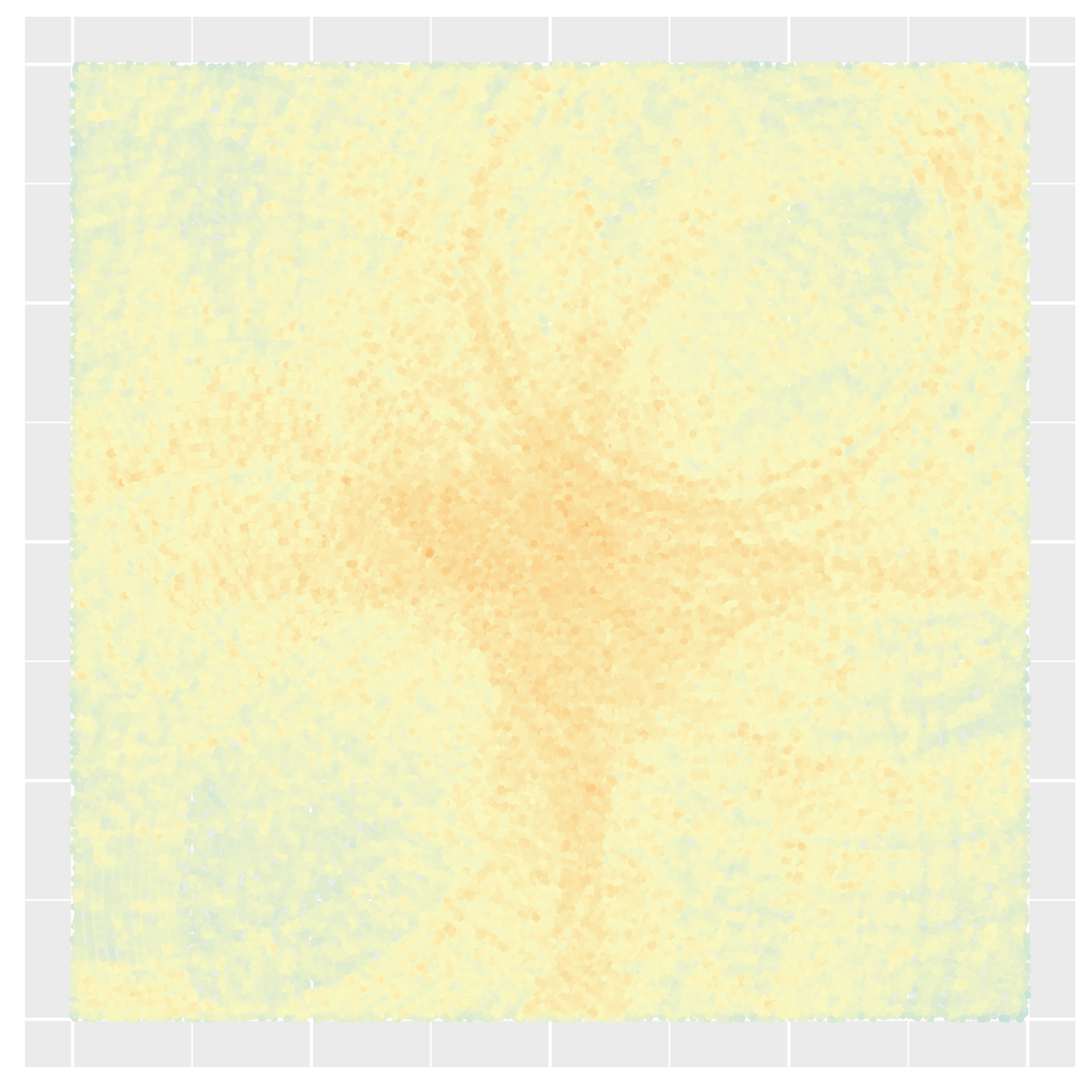}} &
    \raisebox{-0.5\height}{\includegraphics[width=0.18\linewidth]{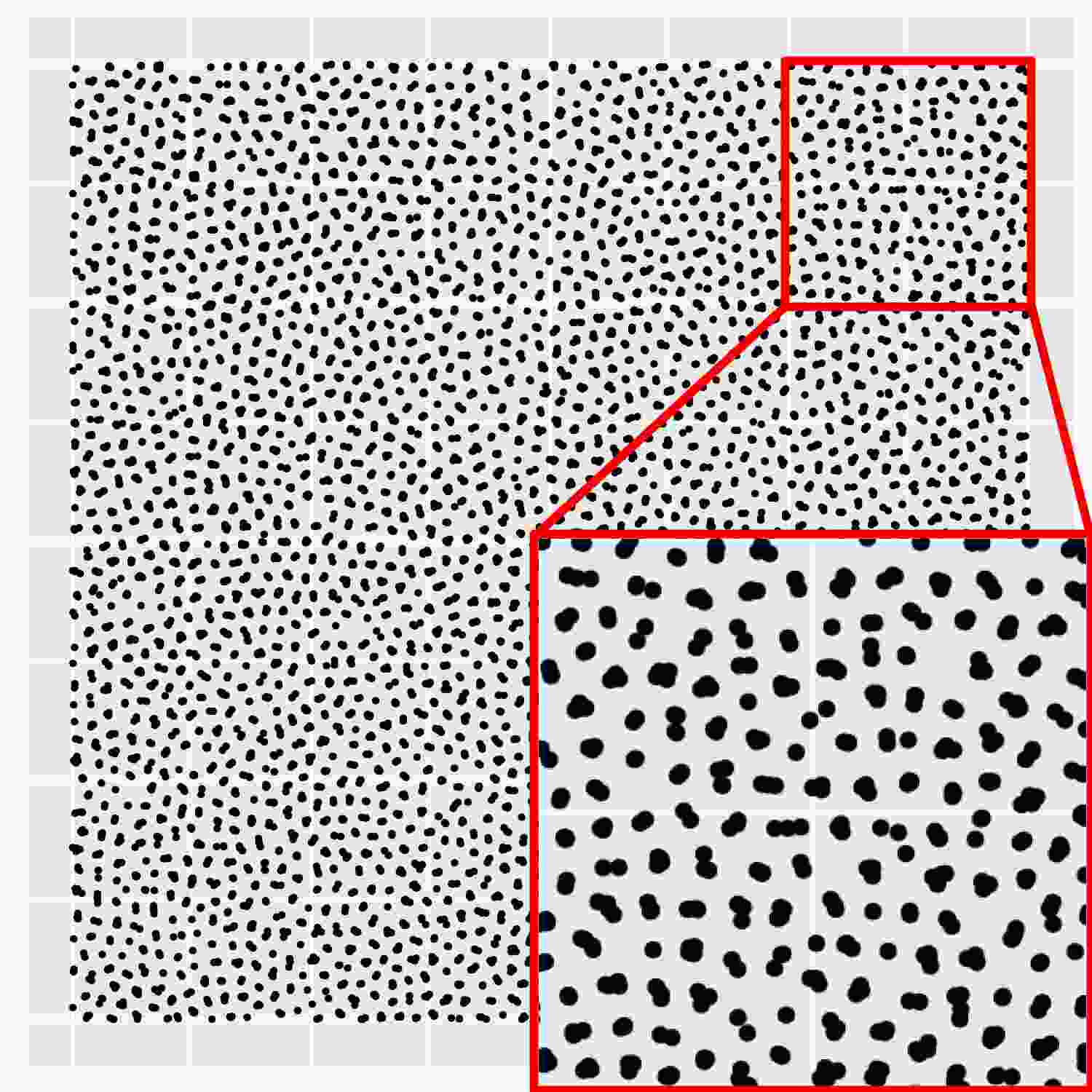}} &
    \raisebox{-0.5\height}{\includegraphics[width=0.18\linewidth]{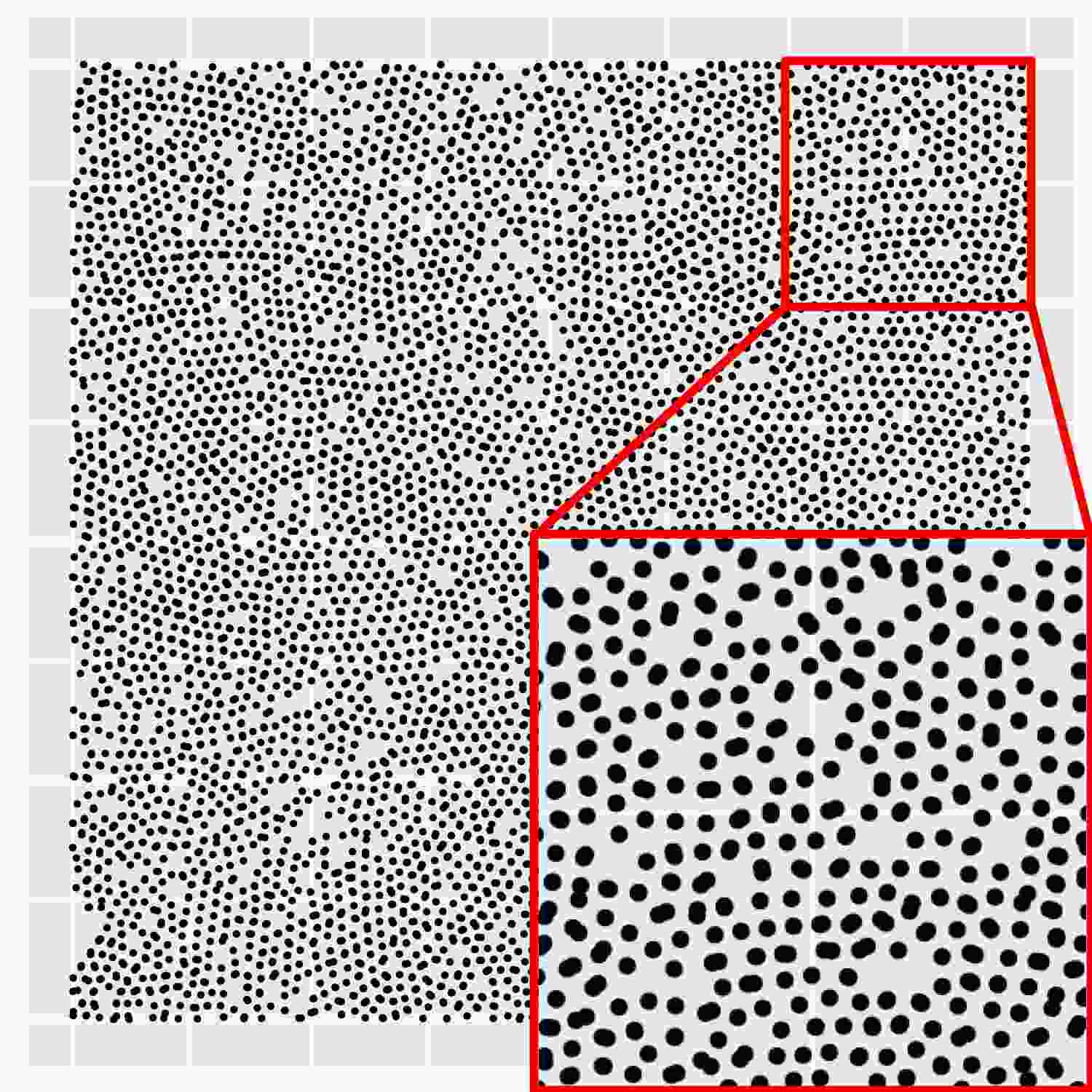}} &
    \raisebox{-0.5\height}{\includegraphics[width=0.18\linewidth]{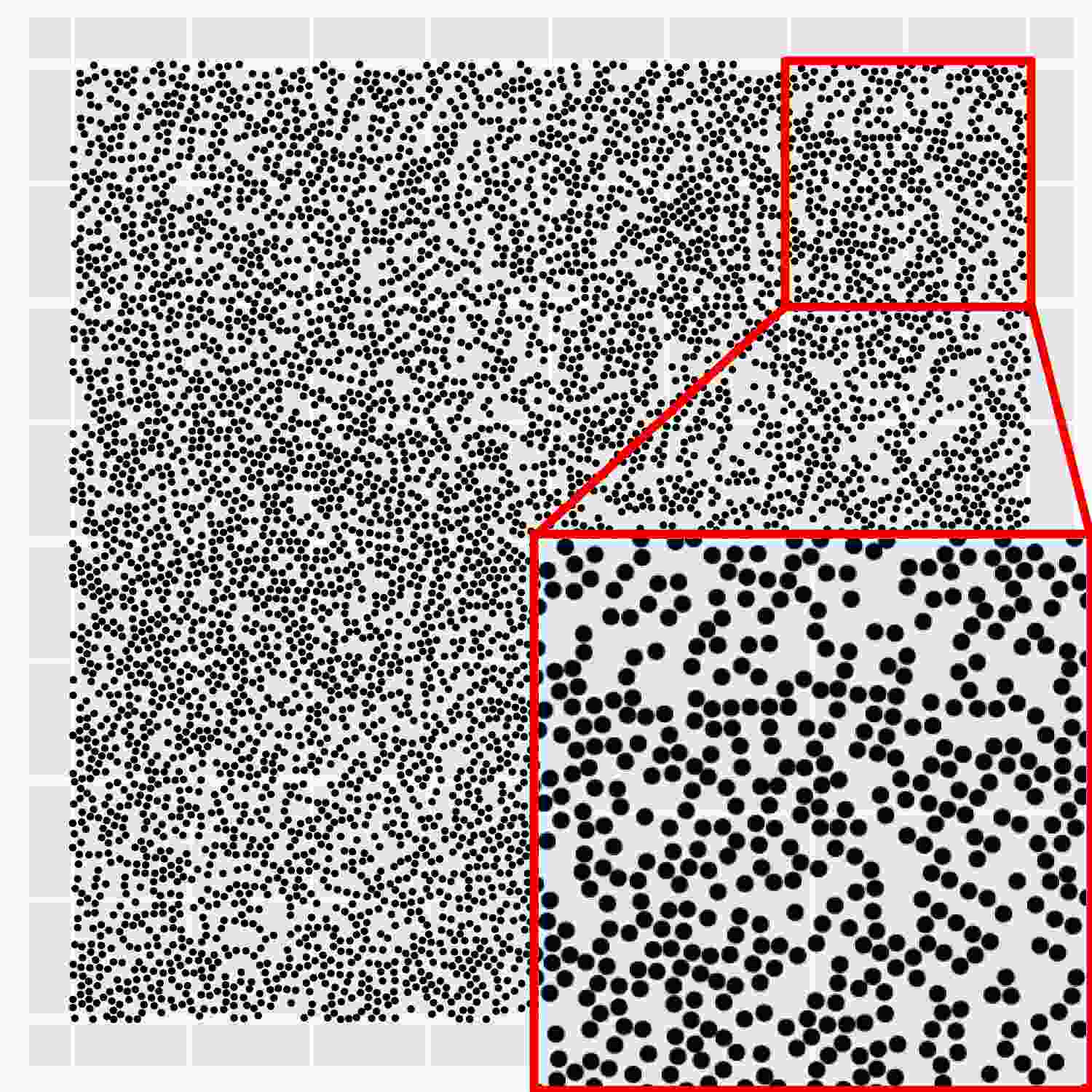}} &
    \raisebox{-0.5\height}{\includegraphics[width=0.18\linewidth]{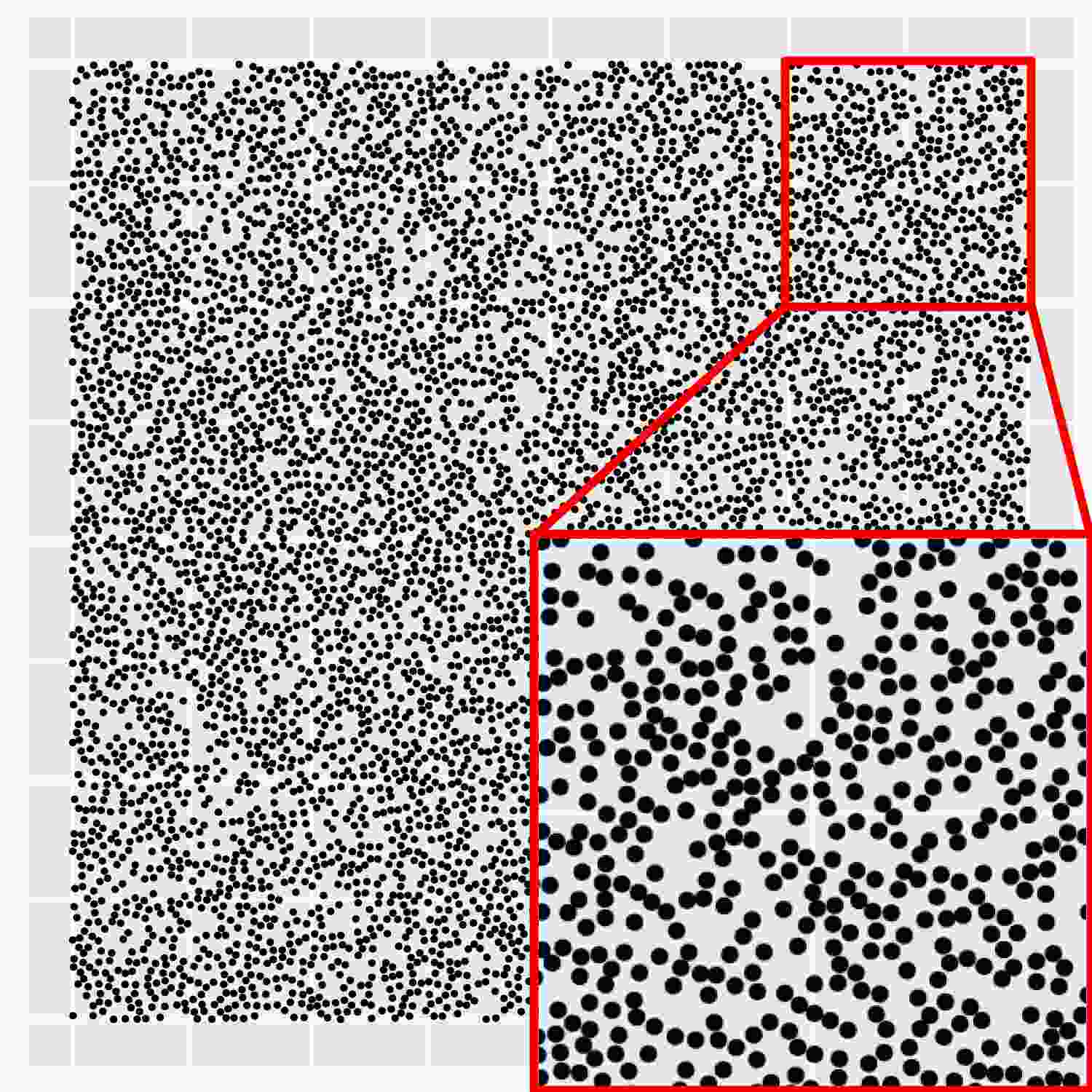}} 
    \vspace{0.2mm}
    \\
    & Dart Throwing & Corsini & Ours ($k=1$) & Ours ($k=3$) & Ours ($k=6$) 
    \\
    \rotatebox[origin=c]{90}{Top Face} &
    \raisebox{-0.5\height}{\includegraphics[width=0.18\linewidth]{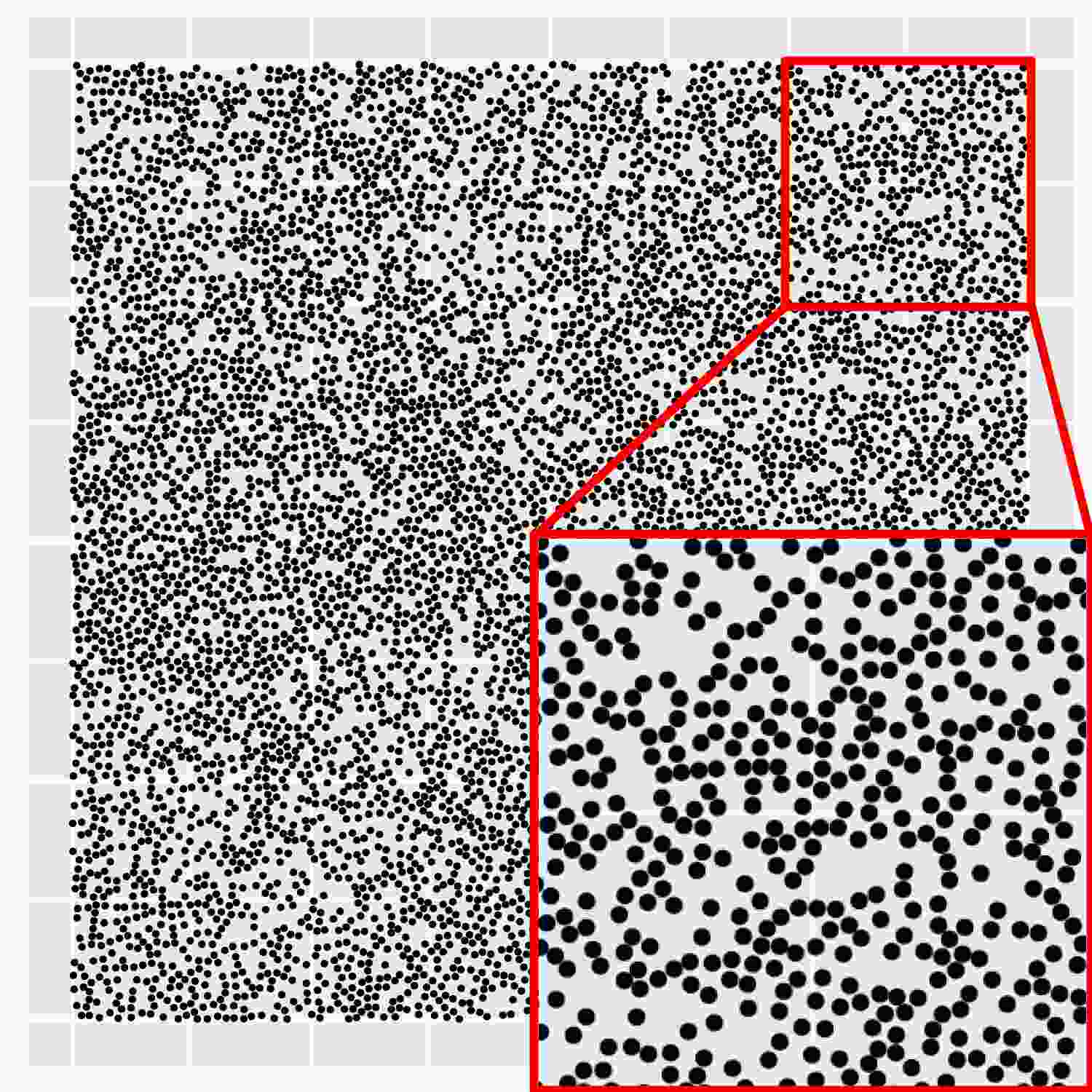}} &
    \raisebox{-0.5\height}{\includegraphics[width=0.18\linewidth]{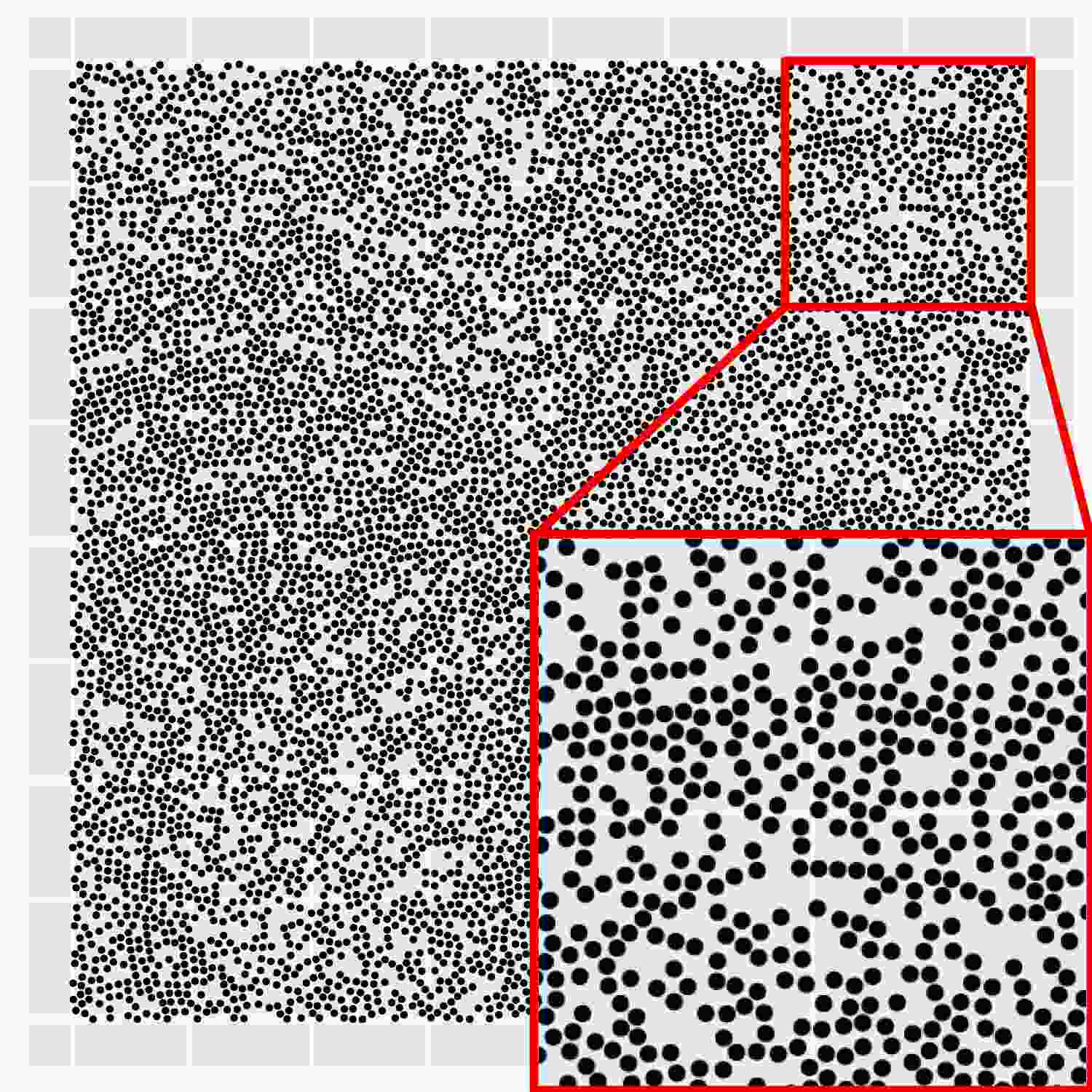}} &
    \raisebox{-0.5\height}{\includegraphics[width=0.18\linewidth]{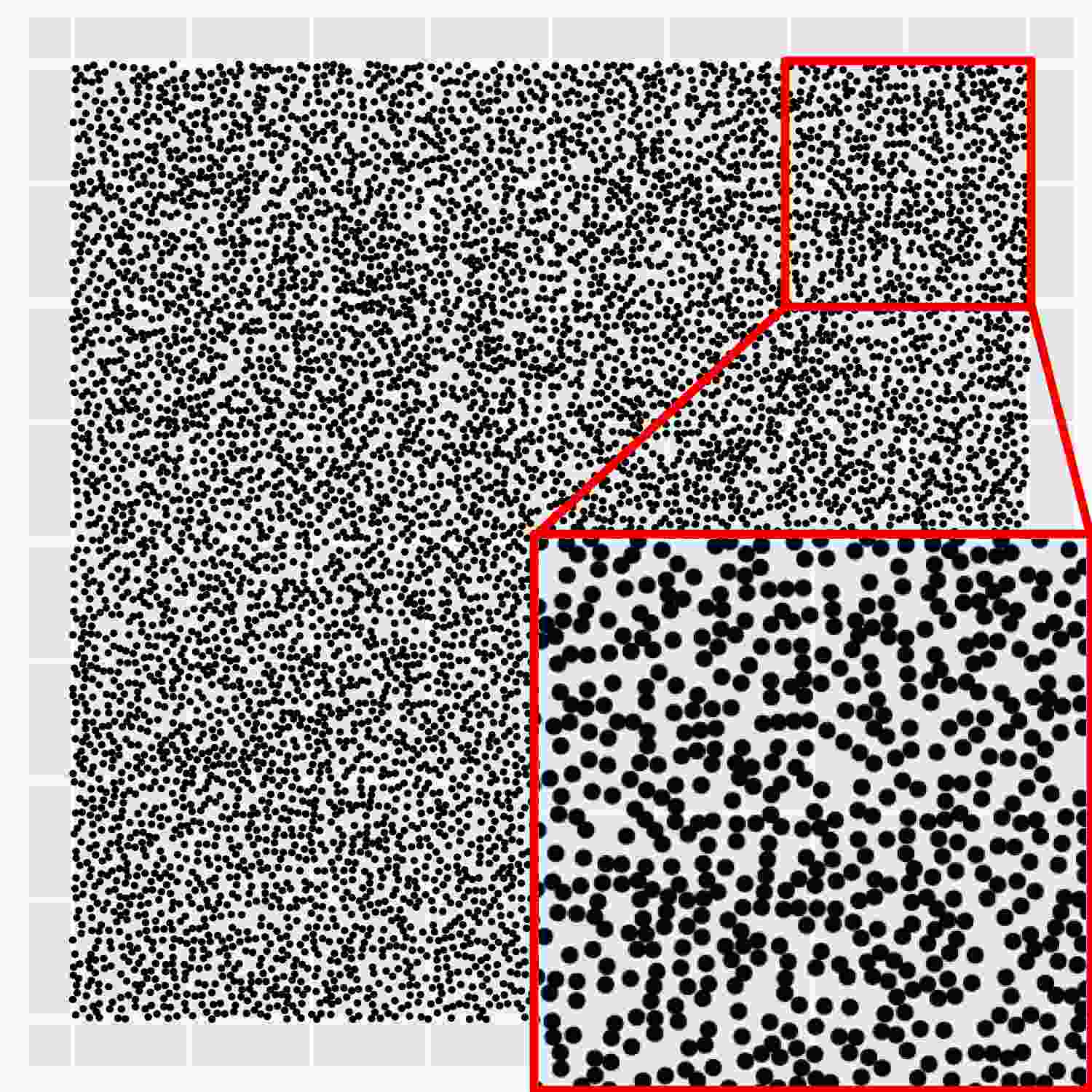}} &
    \raisebox{-0.5\height}{\includegraphics[width=0.18\linewidth]{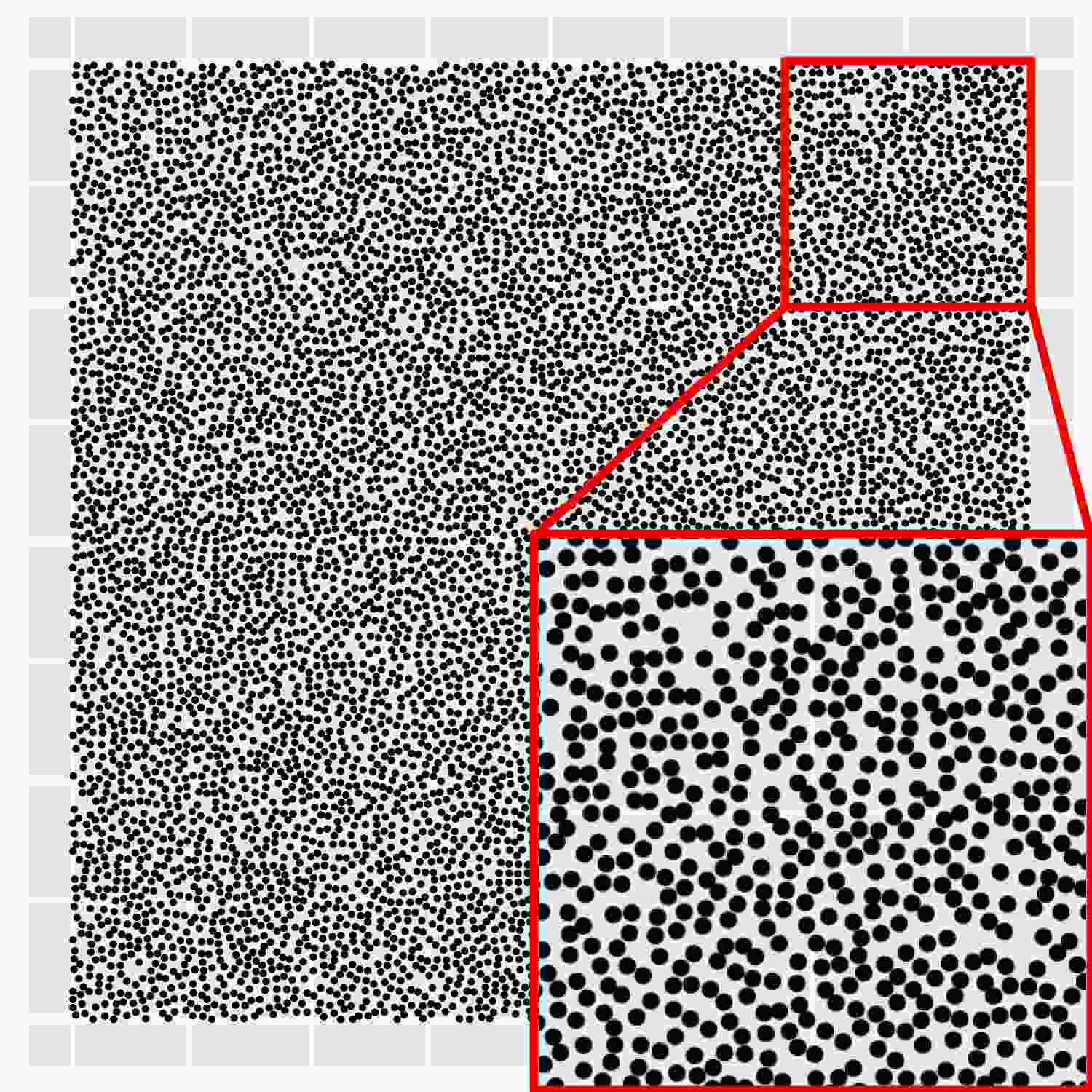}} &
    \raisebox{-0.5\height}{\includegraphics[width=0.18\linewidth]{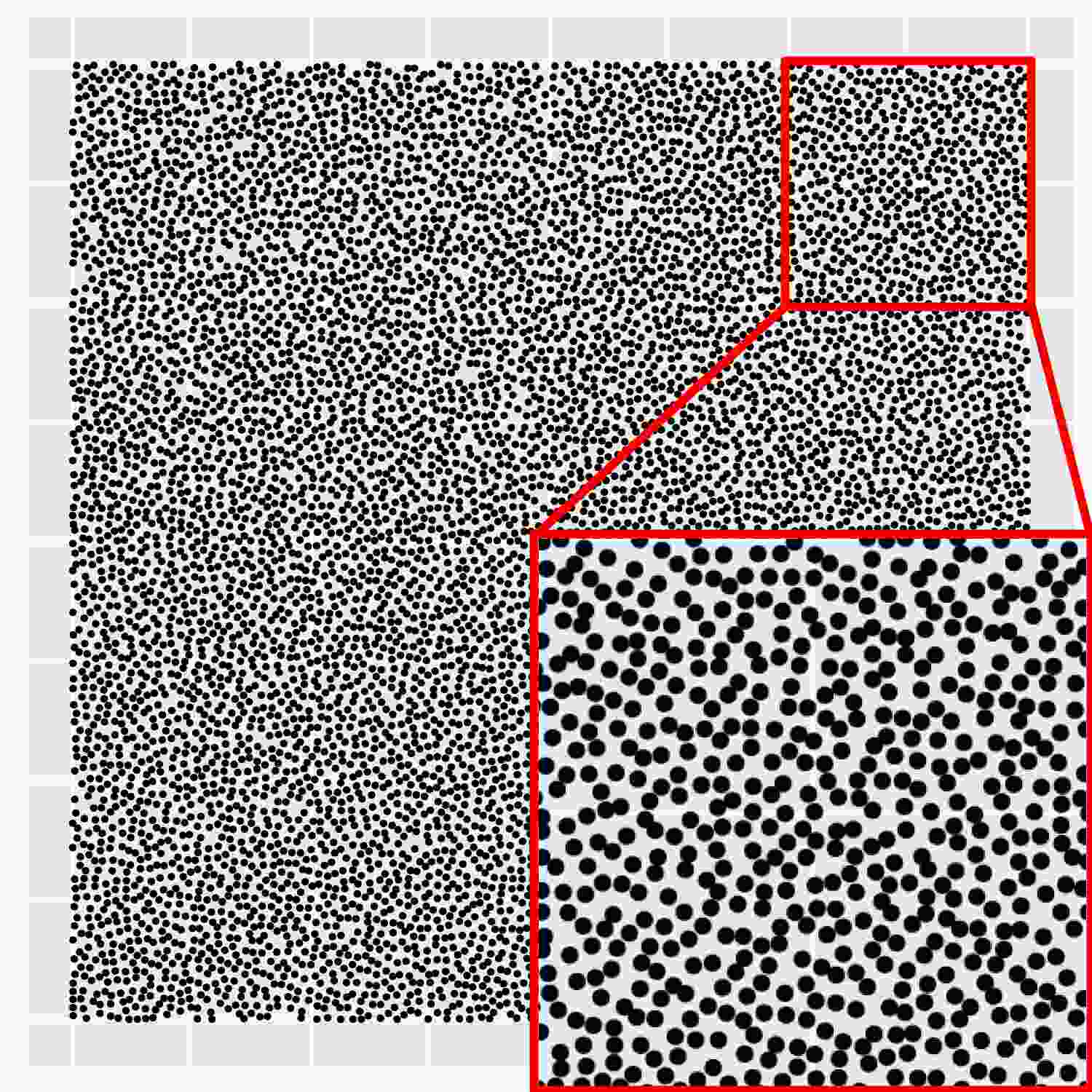}} 
    \vspace{0.2mm}
    \\
    & Original Cloud & Yuksel ($\beta=1$) & Yuksel ($\beta=0.65$) & Yuksel ($\beta=0.35$) & Yuksel ($\beta=0$)
    \\
    \rotatebox[origin=c]{90}{Right Face} &
    \raisebox{-0.5\height}{\includegraphics[width=0.18\linewidth]{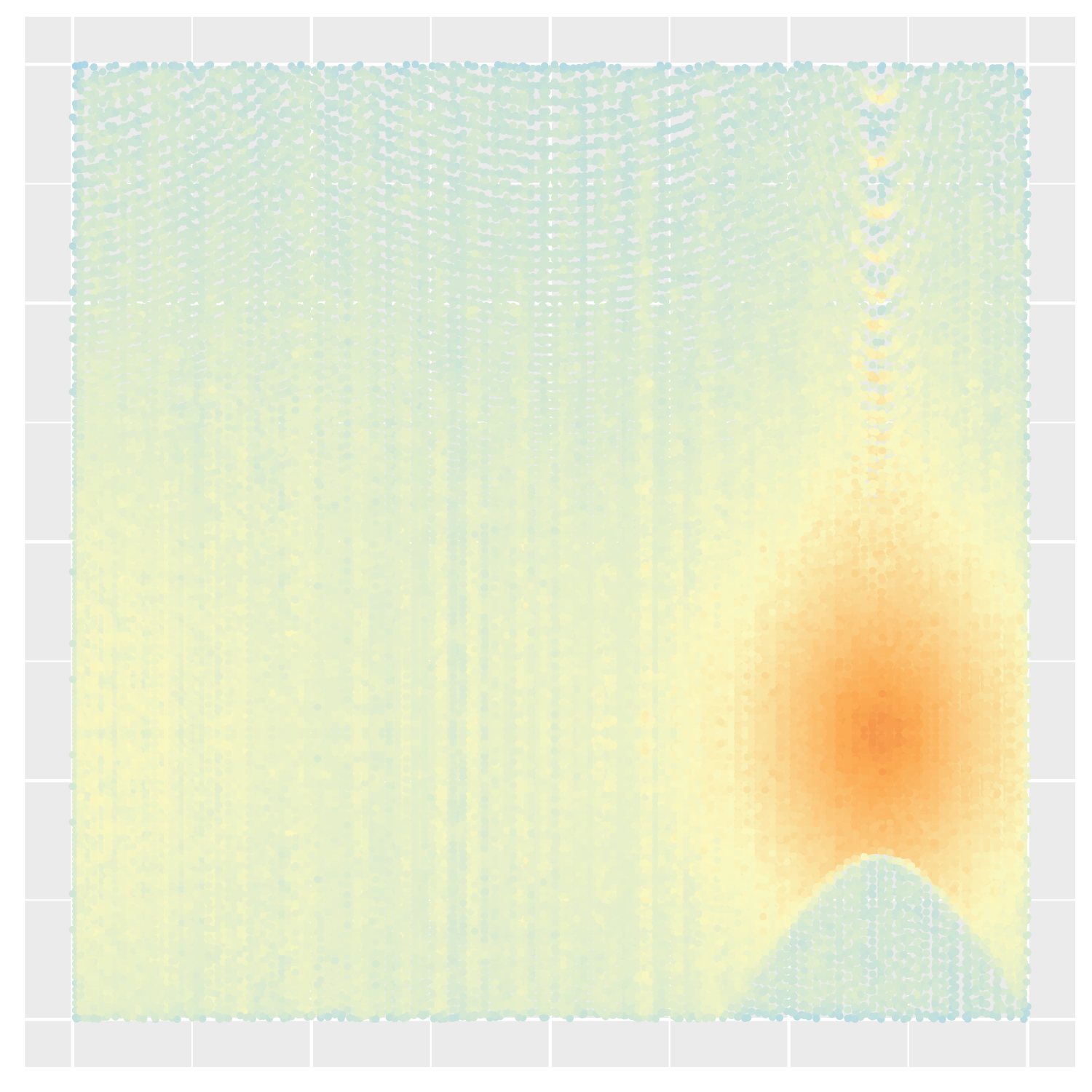}} &
    \raisebox{-0.5\height}{\includegraphics[width=0.18\linewidth]{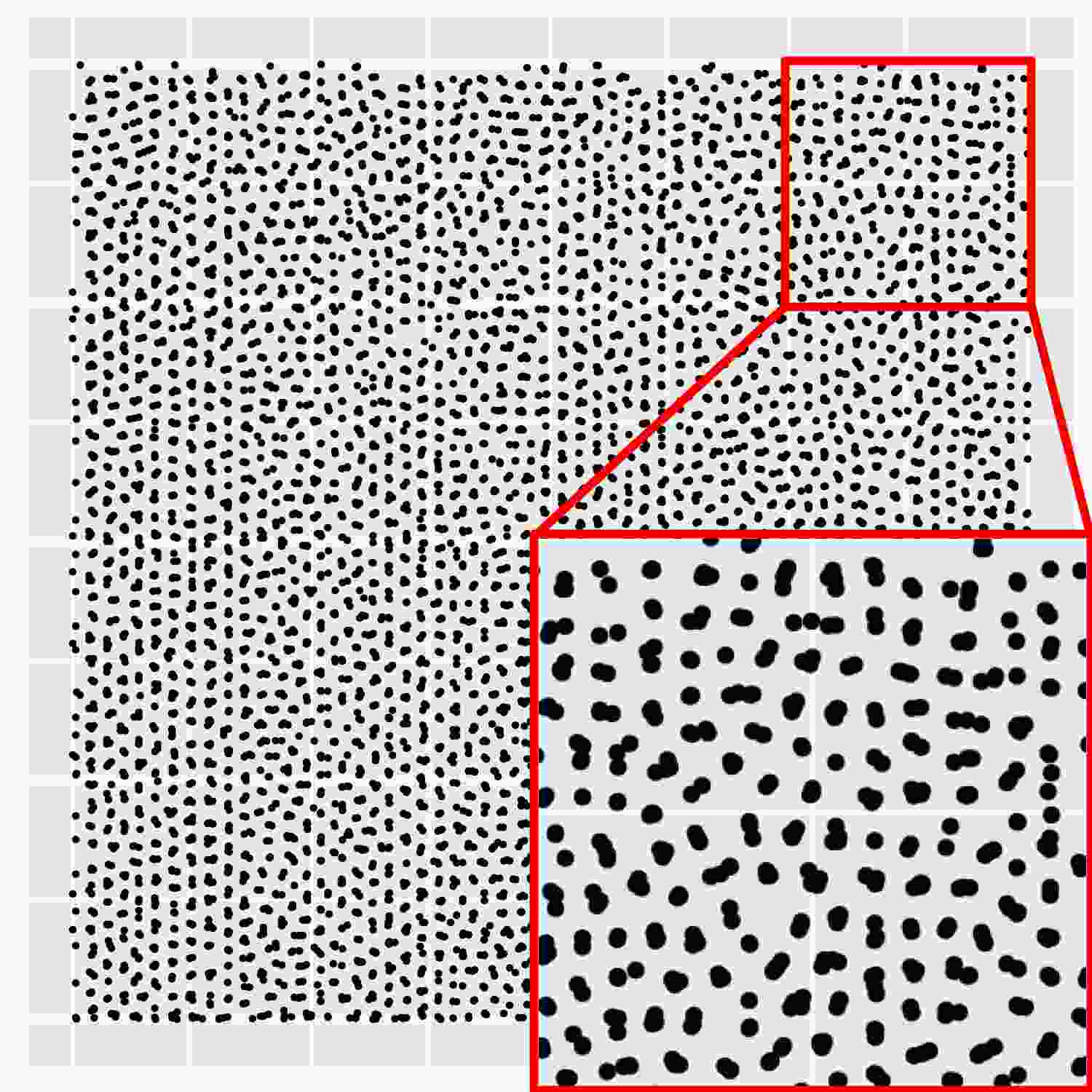}} &
    \raisebox{-0.5\height}{\includegraphics[width=0.18\linewidth]{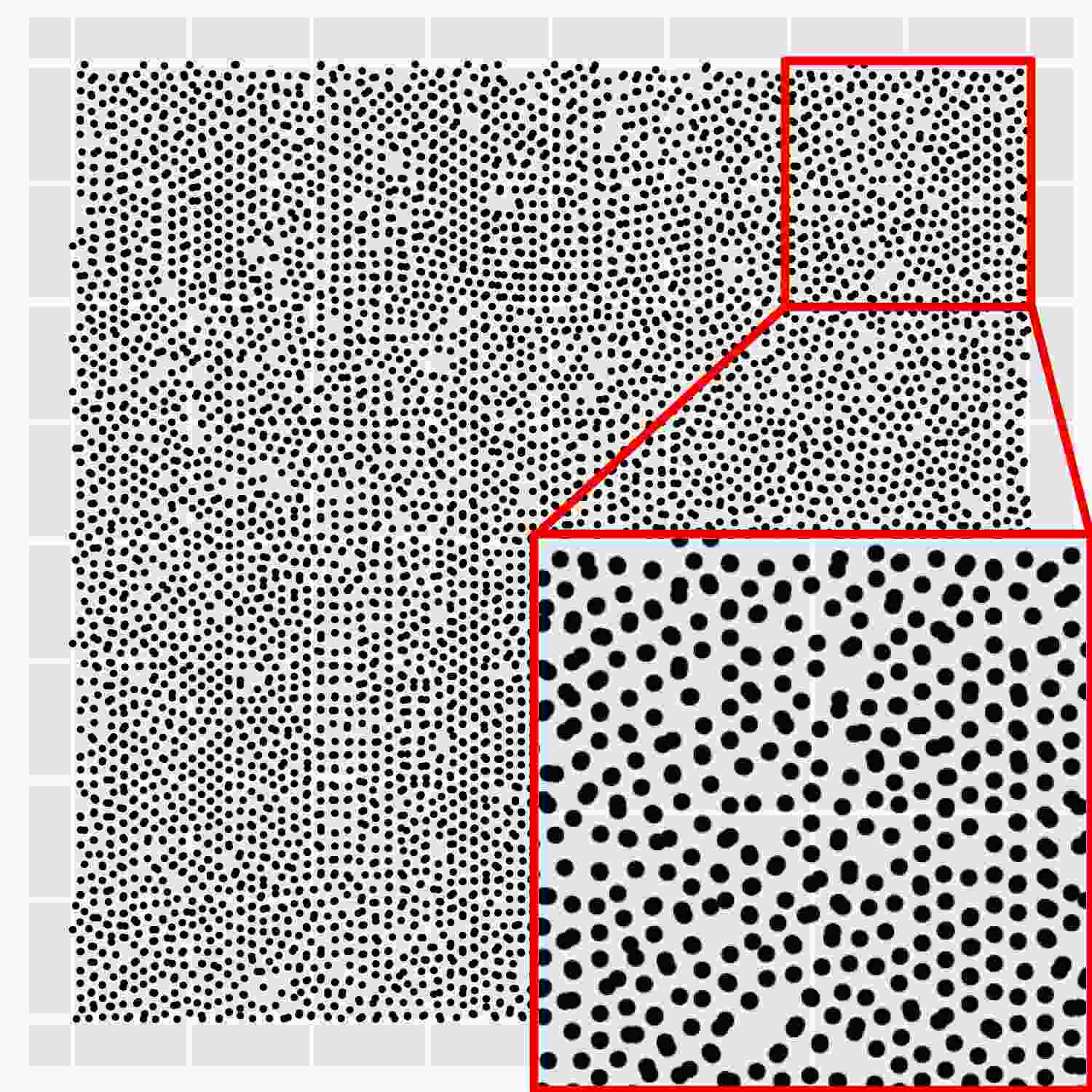}} &
    \raisebox{-0.5\height}{\includegraphics[width=0.18\linewidth]{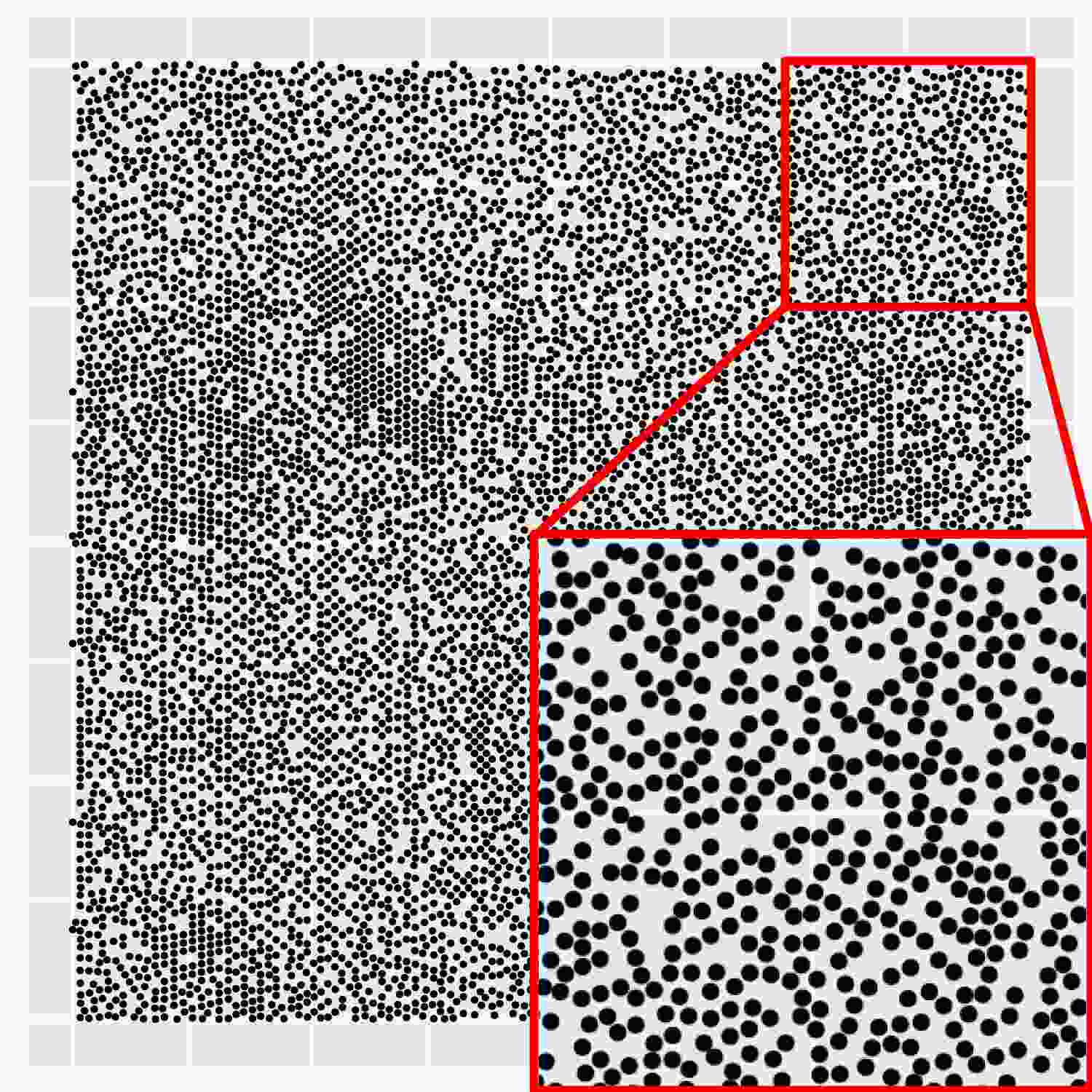}} &
    \raisebox{-0.5\height}{\includegraphics[width=0.18\linewidth]{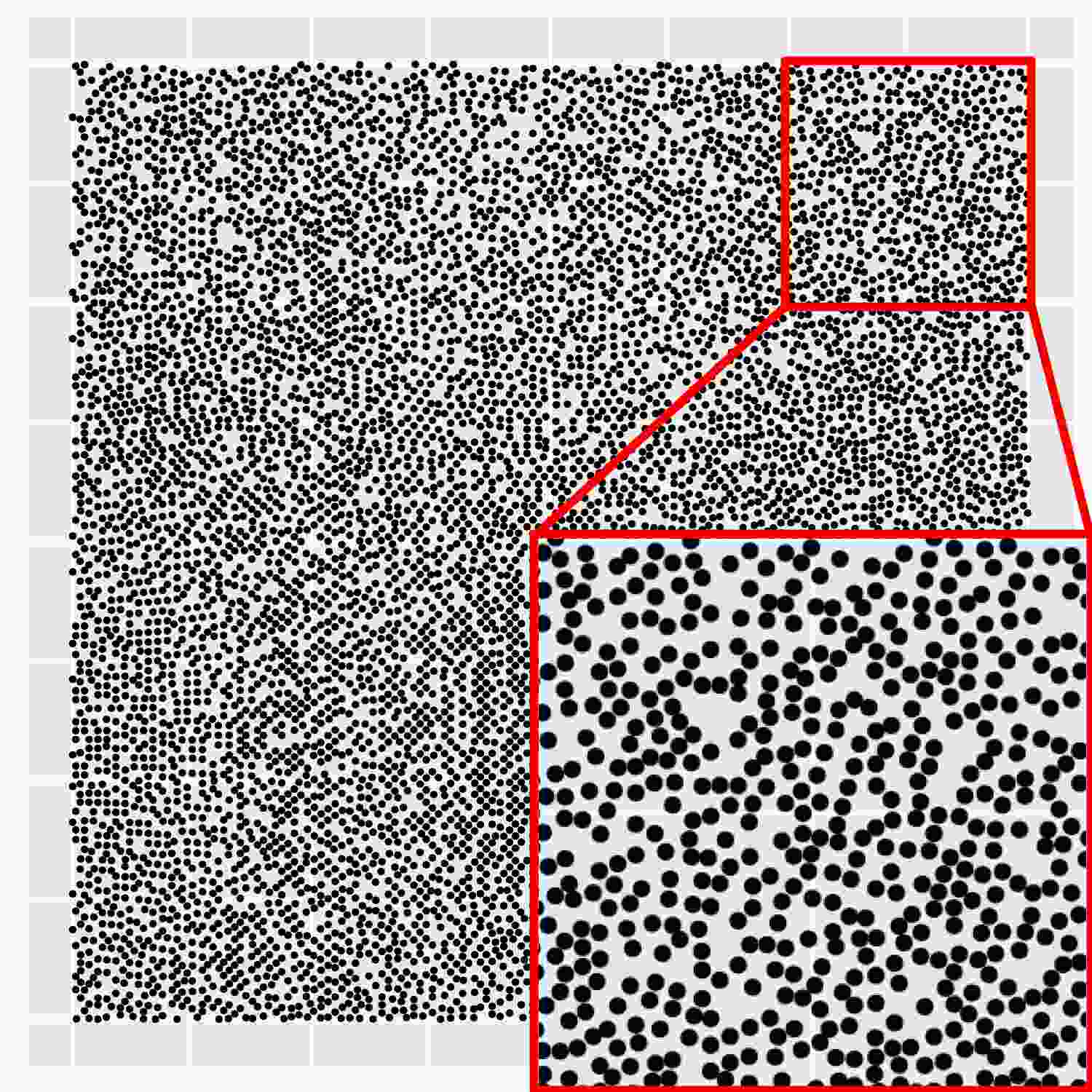}} 
    \vspace{0.2mm}
    \\
    & Dart Throwing & Corsini & Ours ($k=1$) & Ours ($k=3$) & Ours ($k=6$)
    \\
    \rotatebox[origin=c]{90}{Right Face} &
    \raisebox{-0.5\height}{\includegraphics[width=0.18\linewidth]{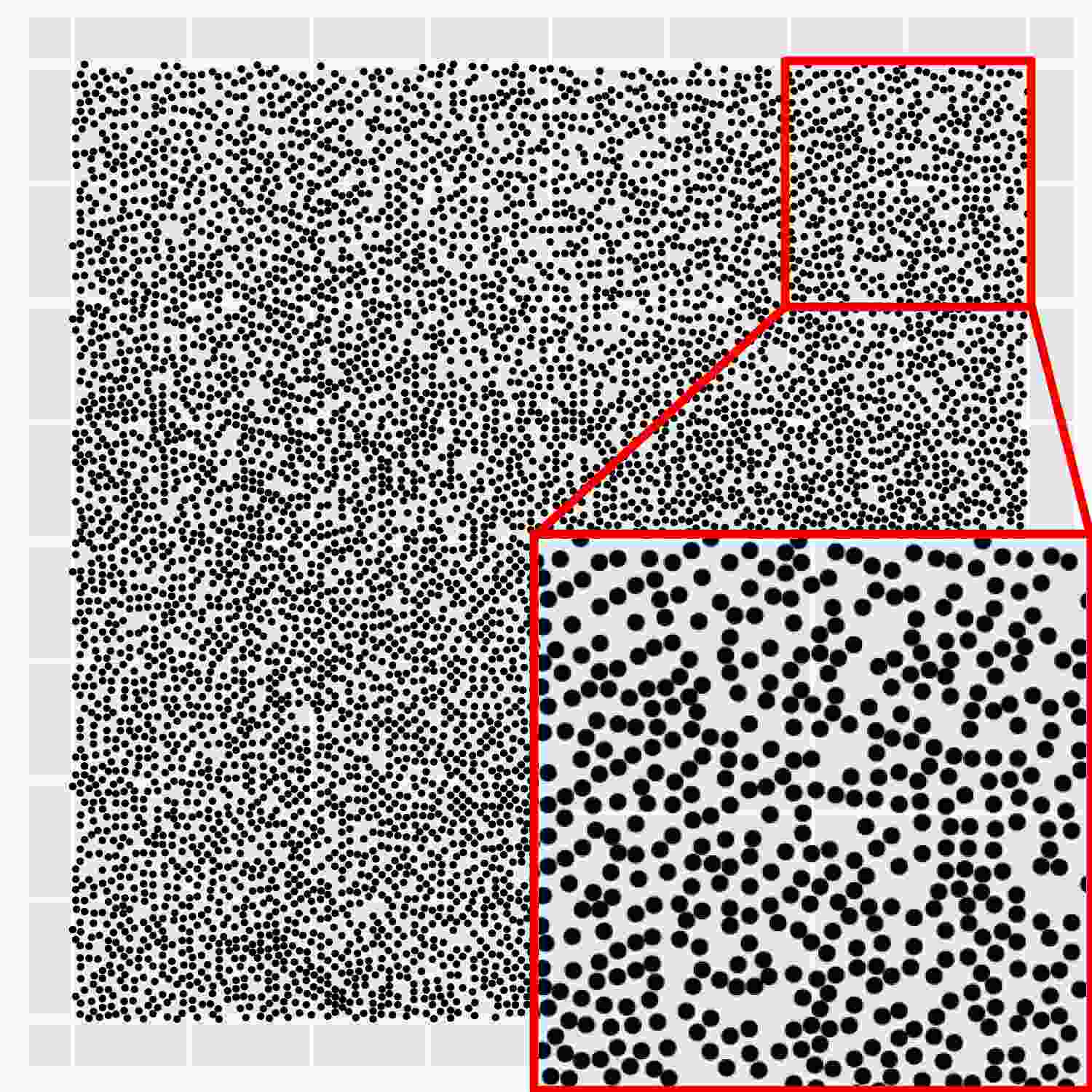}} &
    \raisebox{-0.5\height}{\includegraphics[width=0.18\linewidth]{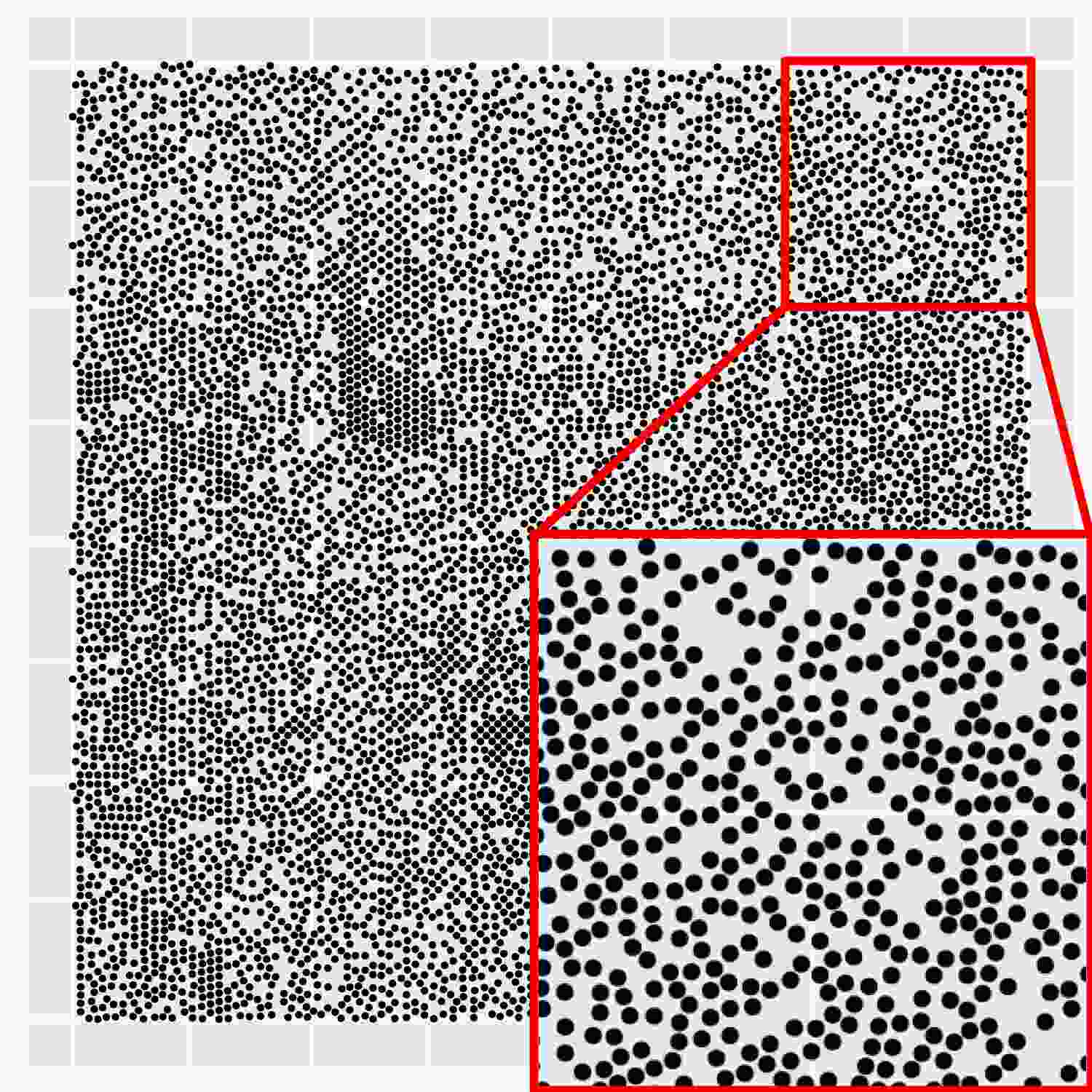}} &
    \raisebox{-0.5\height}{\includegraphics[width=0.18\linewidth]{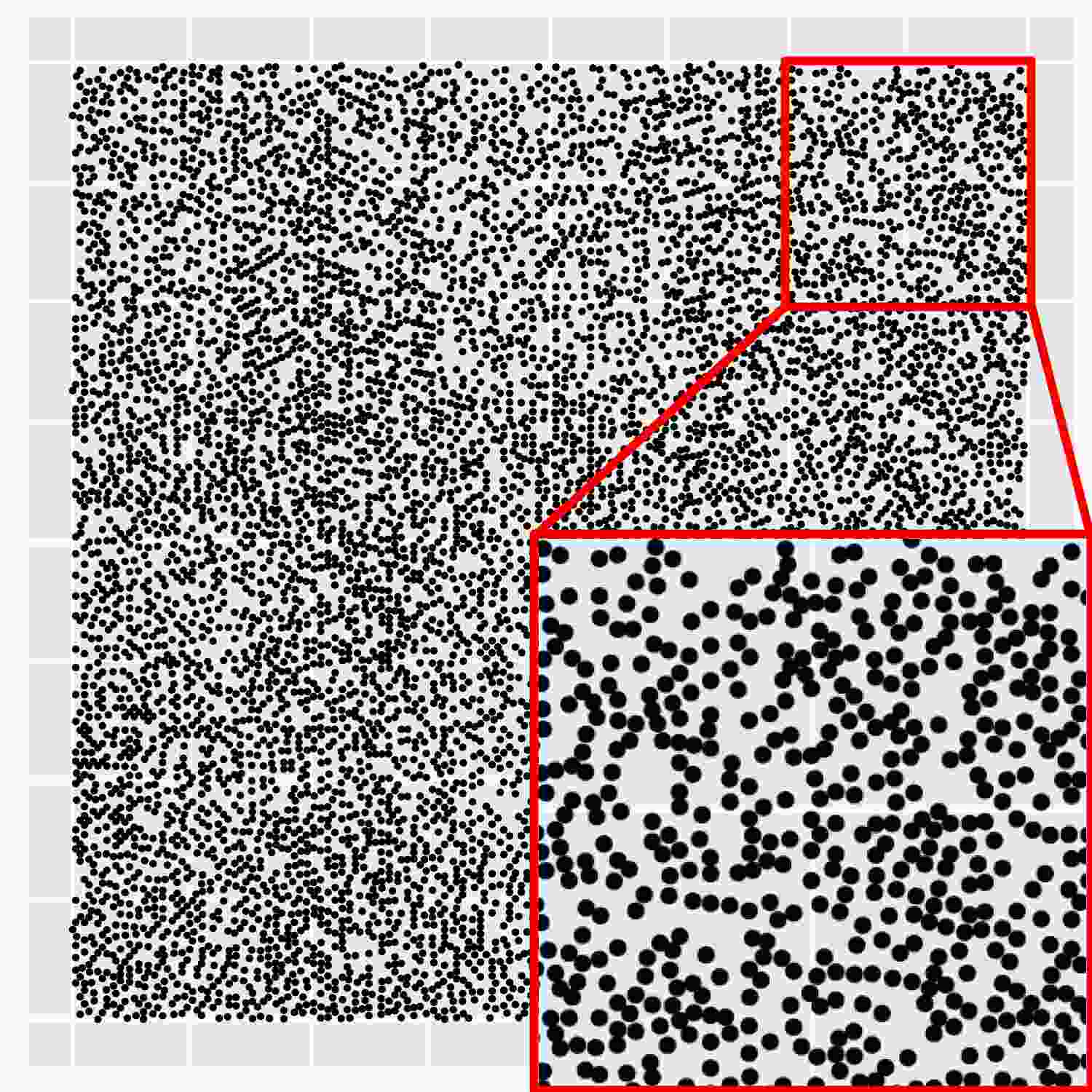}} &
    \raisebox{-0.5\height}{\includegraphics[width=0.18\linewidth]{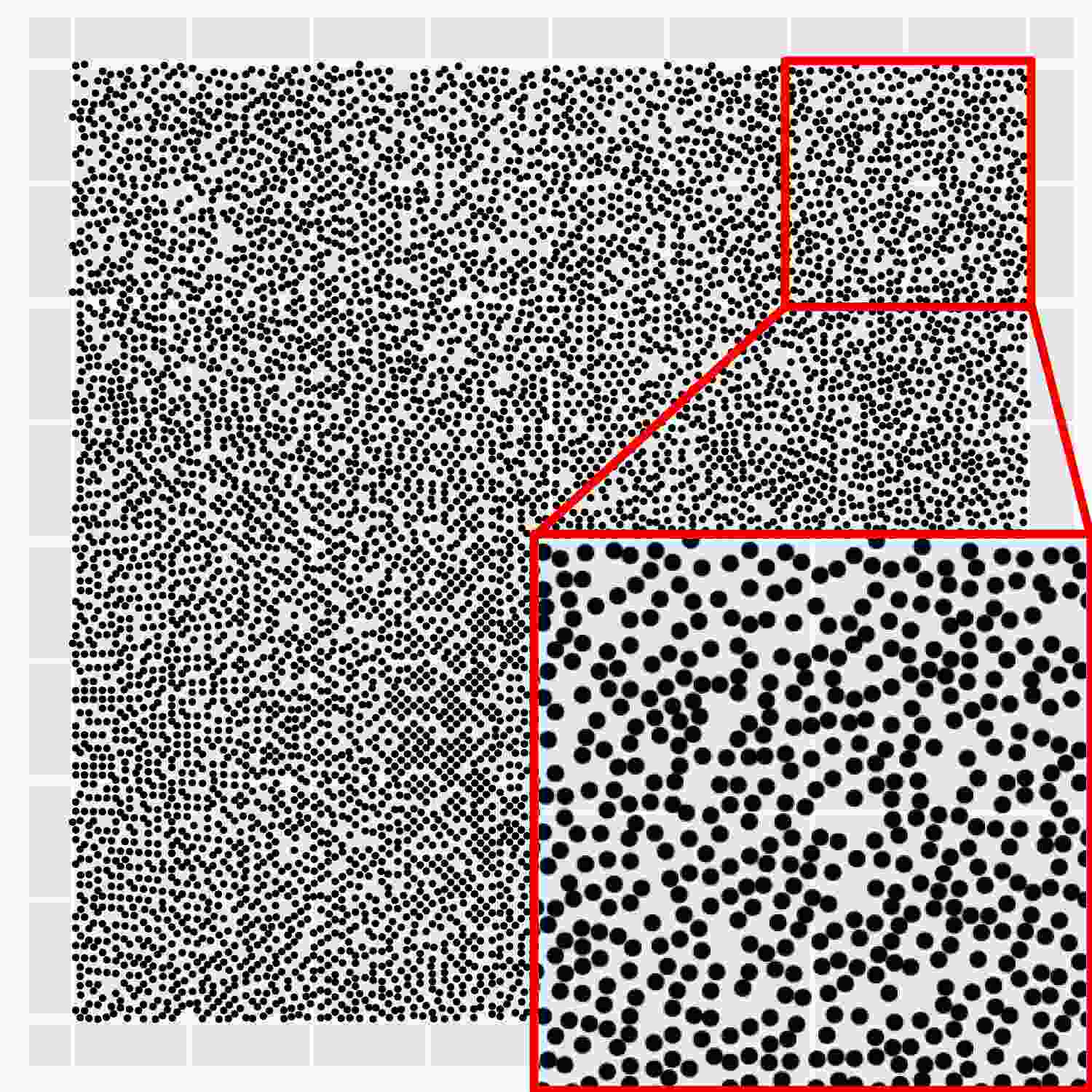}} &
    \raisebox{-0.5\height}{\includegraphics[width=0.18\linewidth]{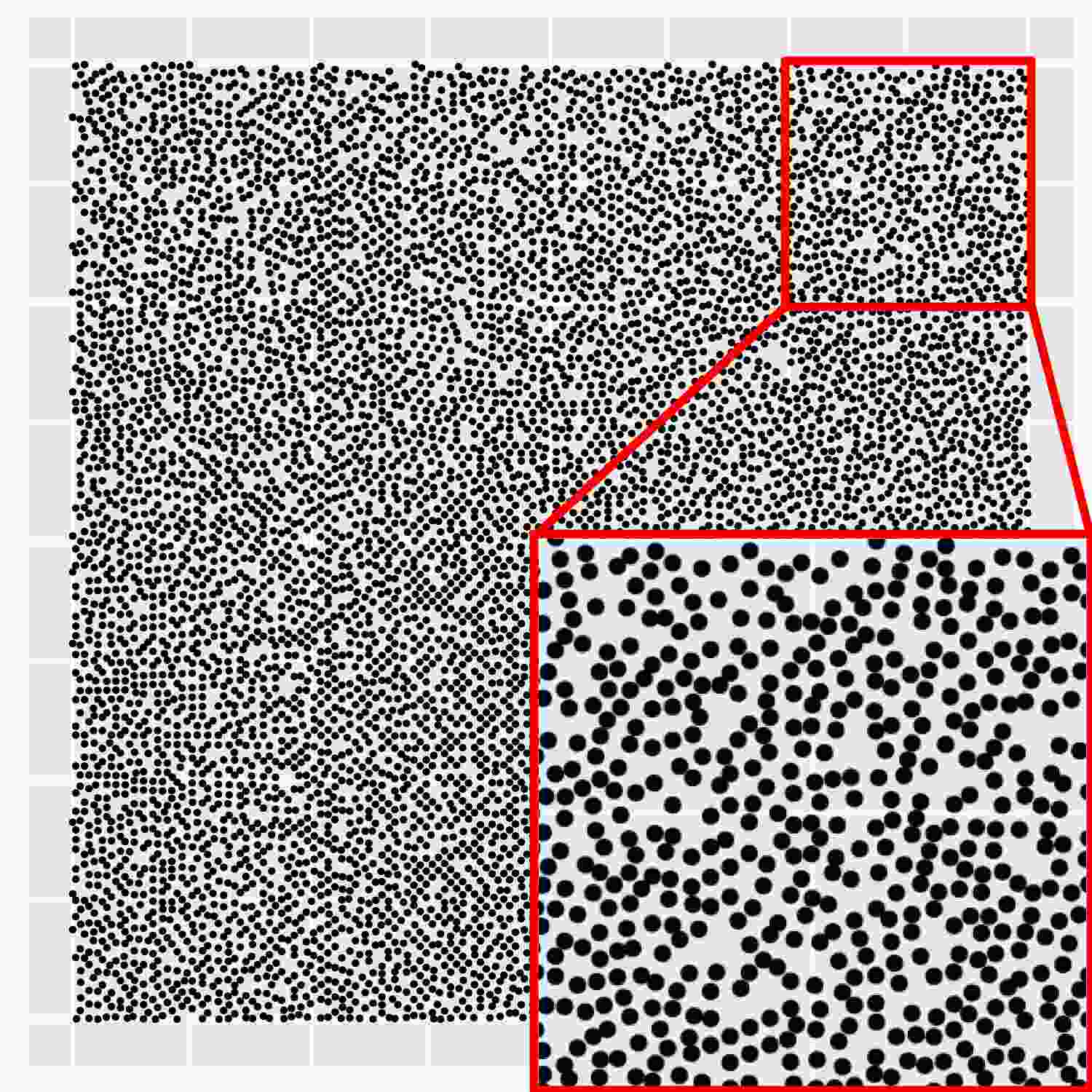}} 
    \vspace{0.2mm}
    \\
    & Legend & Yuksel ($\beta=1$) & Yuksel ($\beta=0.65$) & Yuksel ($\beta=0.35$) & Yuksel ($\beta=0$)
    \\
    \rotatebox[origin=c]{90}{Statistics} &
    \raisebox{-0.5\height}{\includegraphics[width=0.18\linewidth]{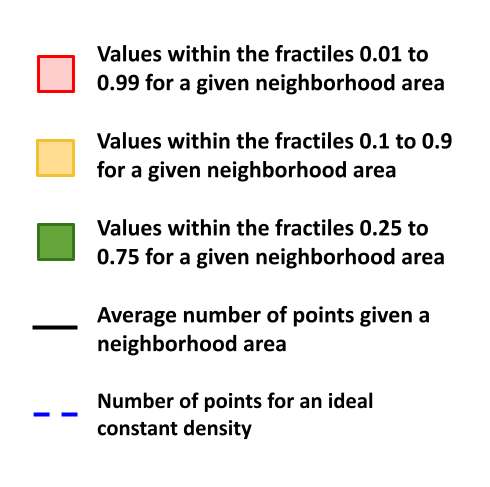}} &
    \raisebox{-0.5\height}{\includegraphics[width=0.18\linewidth]{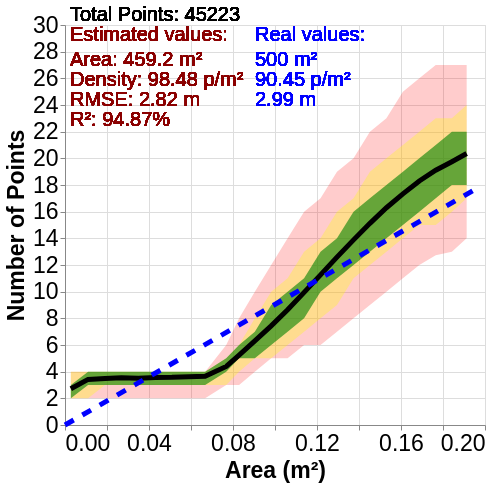}} &
    \raisebox{-0.5\height}{\includegraphics[width=0.18\linewidth]{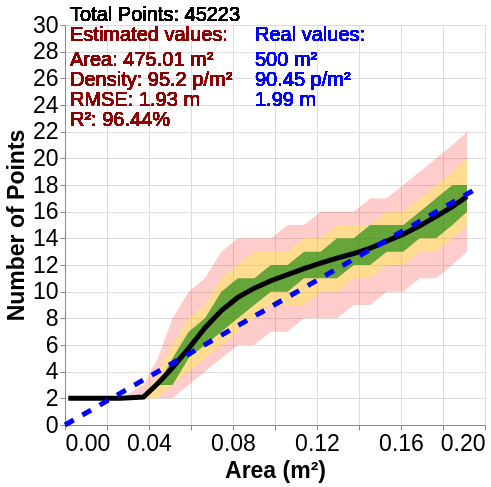}} &
    \raisebox{-0.5\height}{\includegraphics[width=0.18\linewidth]{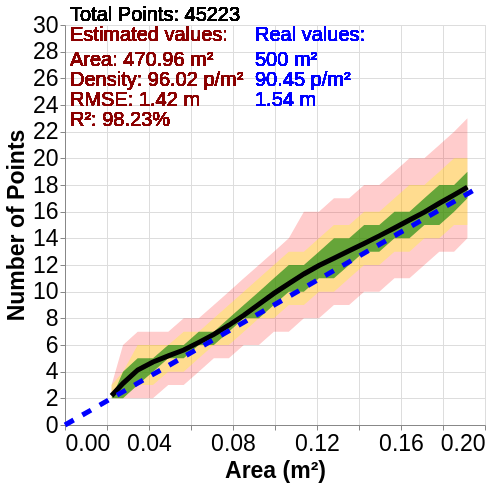}} &
    \raisebox{-0.5\height}{\includegraphics[width=0.18\linewidth]{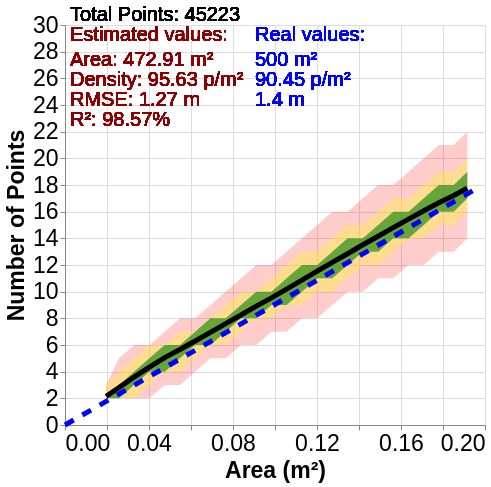}} 
    \\
    & Dart Throwing & Corsini & Ours ($k=1$) & Ours ($k=3$) & Ours ($k=6$)
    \\
    \rotatebox[origin=c]{90}{Statistics} &
    \raisebox{-0.5\height}{\includegraphics[width=0.18\linewidth]{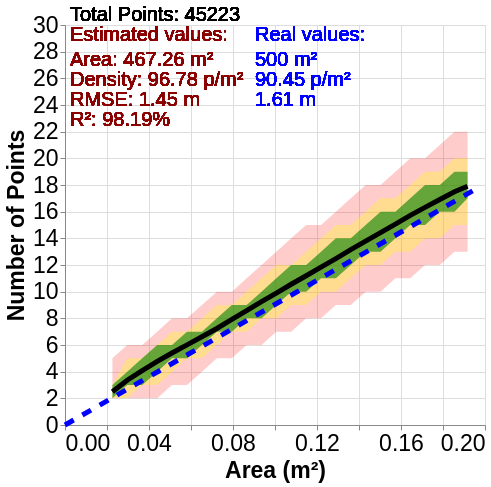}} &
    \raisebox{-0.5\height}{\includegraphics[width=0.18\linewidth]{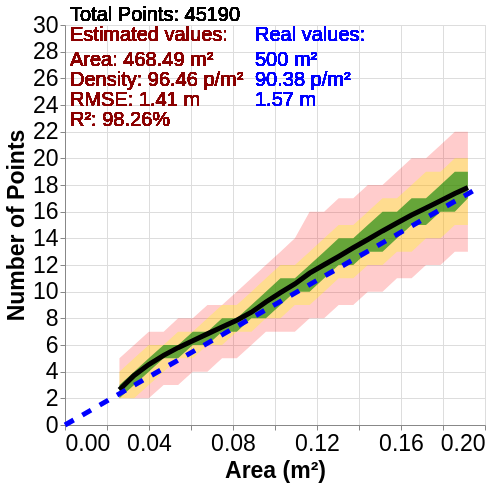}} &
    \raisebox{-0.5\height}{\includegraphics[width=0.18\linewidth]{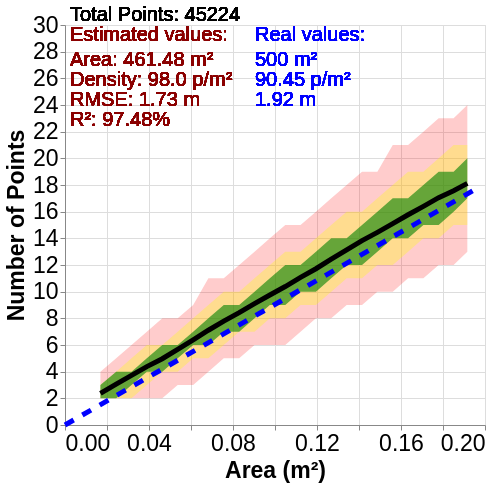}} &
    \raisebox{-0.5\height}{\includegraphics[width=0.18\linewidth]{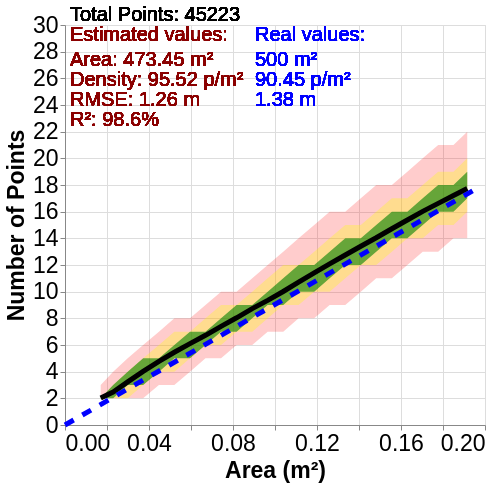}} &
    \raisebox{-0.5\height}{\includegraphics[width=0.18\linewidth]{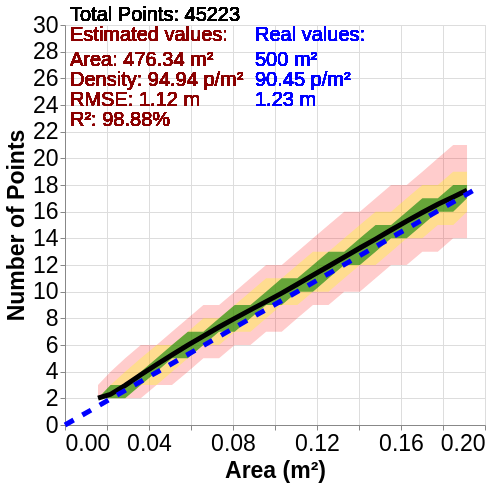}} 
\end{tabular*}
\caption{\label{fig:c4f2} Analysis of the simplifications ($\lambda=0.1$) the cube point cloud using different algorithms. 
The first four rows show the results on two faces of the cloud (top, right) compared to the original cloud (colored according to the local point density). The last two rows show the distribution of neighborhood sizes at increasing area. Our method ($k=6$) yields the most homogeneous point distribution and the least local density variability at different neighborhood areas.}
\end{figure*}

Another shortcoming of Yuksel's method is that their ranking criterion does not contain enough information to ensure good point distributions within $r_{min}$. The authors note that better results are obtained when ``samples with many relatively close neighbors are removed earlier than samples with fewer but very close neighbors'' and they introduce $r_{min}$ in order to induce this. However, they also note, ``The resulting Poisson disk radius $r$ of the final sample set improves with increasing $r_{min}$ until $r_{min}$ gets close to $r$, at which point the output includes pairs of adjacent samples''. We experimentally found that their strategy (tested using their implementation~\cite{cyCodeBase}) produces many tight clusters (3 to 7 points) depending on the choice of $\beta$. In \Cref{fig:c4f2} we can observe that clusters appear for $\beta=0.65$ (which corresponds to the conservative estimate proposed by the authors) and $\beta=1$. We argue that this happens because the method does not optimize for the distribution of points within $r_{min}$. In fact, we observed that the best results are achieved when $r_{min}$ is not used ($\beta=0$). 

Using the local density $\rho_\bp$ as a ranking criterion still inherits the problem of generating small clusters of points. This is because this value disregards how points are distributed within the $k$-neighborhood.

So, instead of optimizing for equal densities, we first tried to maximize the distance between any two closest points (as we introduced in our second problem formulation \Cref{eq:c4e2}). This is equivalent to estimating the local point density at $\bp$ as a value inversely proportional to the square of the distance to its closest point ($d_\bp)$:

\begin{equation}
\label{eq:c4e10}
    w_\bp^{k=1} = \frac{1}{d_\bp^2} 
\end{equation}

In~\Cref{fig:c4f2} we show the result of our method using $w_\bp^{k=1}$ (labeled as "ours $k=1$") as ranking criterion. Notice that this method does not produce small tight clusters of points at the expense of sometimes leaving empty gaps. Let $\bn{k}{\bp}$ be the $k$-neighborhood of $\bp$; we introduce $w_\bp^{k}$ to further improve the sample distribution: 

\begin{equation}
\label{eq:c4e11}
    w_\bp^{k} = \sum_{\bp' \in \bn{k}{\bp}} \frac{1}{\bdist{\bp}{\bp'}^2} 
\end{equation}

\begin{figure}[!htb]
\centering
    \includegraphics[width=0.49\linewidth]{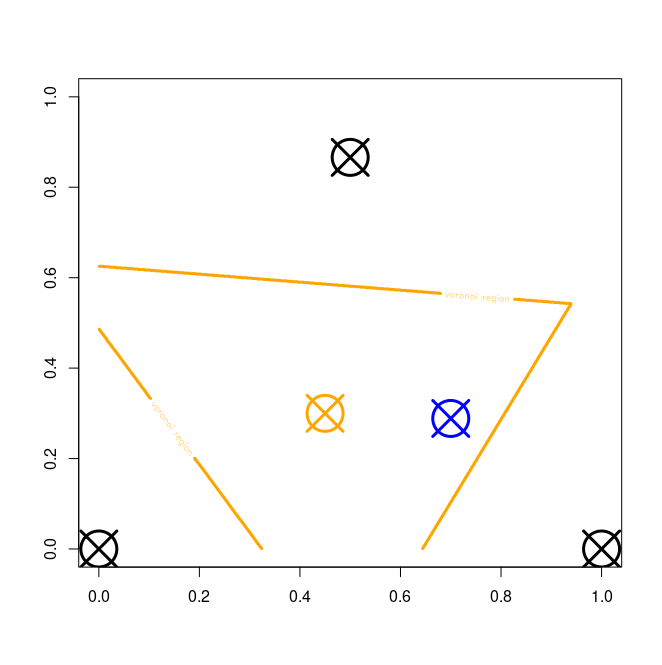}
    \includegraphics[width=0.49\linewidth]{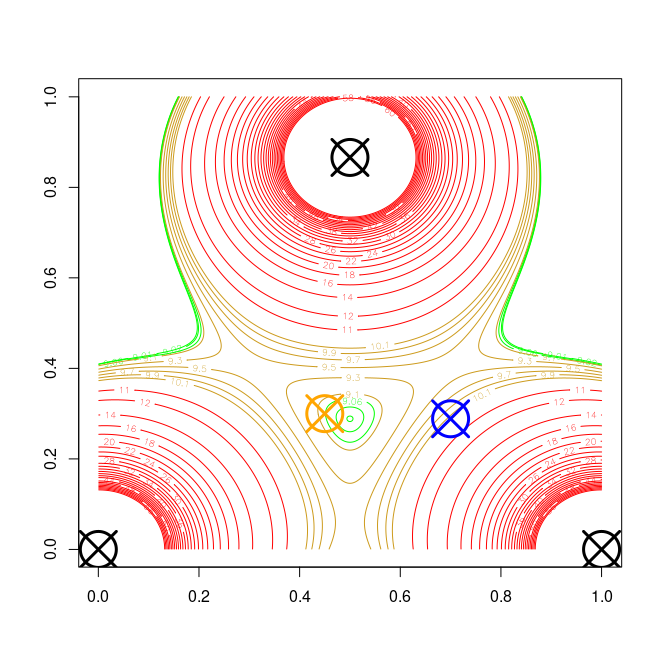}
\caption{Left: In this example, we can see three well-distributed points (in black) and two closest points to each other (blue and orange), which have the same $w_\bp^{k=1}$. In fact, if we move the blue point within the Voronoi region of the orange point (considering the black points), it would still have the same $w_\bp^{k=1}$ as the orange point. Hence, the greedy algorithm could arbitrarily choose to remove either of them.
Right: Visualization of the discerning cost ($w_\bp^{k=4}$ - $w_\bp^{k=1}$). Points within the green region would have a smaller cost than the orange one's cost, and those within the red region would be higher than the blue one's cost. Using the $w_\bp^{k=4}$, the greedy algorithm would choose to remove the blue point. } 
\label{fig:c4f3}
\end{figure}

Recall that we iteratively remove samples with the highest cost in a greedy-like fashion. Hence the likelihood of finding a suboptimal solution is high. The intuition behind $w_\bp^{k}$ is that it will usually be similar to $w_\bp^{k=1}$ (since the inverse of the smallest distance will have the largest contribution to $w_\bp^{k}$), but will help the algorithm discern which samples to remove in cases such as the one presented in~\Cref{fig:c4f3}.

\setlength{\tabcolsep}{3.57pt}
\renewcommand{\arraystretch}{0}

\section{Sub-sampling algorithm}

Maintaining a priority queue for a whole cloud in memory is not feasible when handling point clouds with billions of points. Instead, we have designed a strategy that uses a voxelized out-of-core representation and produces high-quality results.

We start by generating a voxelization of the point cloud and storing it out-of-core. In our experiments with architectural models, we have tested voxels covering $1^3m^3, 3^3m^3$ or $5^3m^3$. The voxel size must also be small enough to ensure all the voxels fit individually in main memory. Then we iterate the following two steps until we reach the desired numbers of points $\lambda |C|$:

\begin{enumerate}
    \item \textbf{Cost updates}: We individually read the voxels from the disk and, for each point, we compute its cost $w_\bp^{k}$. The neighbors in $\bn{k}{\bp}$ must also be searched in the surrounding 6, 18, or 26 voxels for correctness (we search the 26 adjacent voxels in our implementation). We keep an updated max priority queue (global, not per-voxel) to keep the computed costs. As we shall see, we only keep the $\lambda |C|$ points with the lowest cost in the queue. Points with a cost higher than those in the queue are not stored in the priority queue.
    \item \textbf{Point removal:} Now, at the top of the queue, we have an upper bound $w_{up}$ of the cost the $\lambda |C|$-th sample will have (because removing a point preserves or decreases the costs $w_\bp^{k}$). Thus, if we only remove points with a cost above $w_{up}$, we ensure we will still keep the $\lambda |C|$ samples with the lowest costs. We individually read the voxels from the disk and remove any point whose cost is higher than $w_{up}$. Let $\bni{\bp} = \{\bp' | \bp \in \bn{k}{\bp'}\}$ i.e., the set of points which contain $\bp$ in their neighborhood. Upon removing $\bp$, we update the cost for each point in $\bni{\bp}$.

\end{enumerate}

These two steps are repeated until $|C'| = \lambda |C|$.

We implemented different strategies to improve the efficiency of this algorithm. During the cost updating step, we keep the maximum cost of the samples contained in each voxel. A voxel whose maximum cost is smaller than $w_{up}$ does not require processing during the second step. During the first step, we also construct and store a list of the points in $\bni{\bp}$ for each $\bp$, which allows us to find the points which require cost updating quickly. Moreover, to avoid performing neighborhood searches each time a point requires a cost update, during the first step, we also compute and store an extended $k+b$-neighborhood for each point (in our experiments, we use $k+b = 14$). When less than $k$ valid neighbors remain on the stored $k+b$-neighborhood of a point, the point is flagged as ``dirty'', which means we will avoid removing it until we recompute its cost on the following iteration of the algorithm.

\section{Results and Discussion}

We implemented the proposed sub-sampling algorithm using non-parallel C++ code. The program was forced to run out-of-core regardless of the model size, i.e., voxels were kept on disk for all test models. The test hardware was a commodity PC equipped with an Intel Core i7-10700K CPU, 32GB of RAM, and a 500GB KIOXIA-EXCERIA PLUS SSD running Ubuntu 20.04.

We tested the algorithm with both synthetic and real point clouds representing architectural models. The size of the point clouds varied from about 400K points to about 1.2 billion points. The number of scans within each point cloud also varied from 2 to 38 scans per dataset.  Real models were taken using actual LiDAR equipment, including a Leica P20 ScanStation and Leica RTC 360. 

For real datasets, scanner locations were restricted to be about 1.5\,m above the ground (the scanner device was mounted on a tripod). In contrast, for synthetic datasets, we could test more varied scanner locations within the scanning volume.

\setlength{\tabcolsep}{5pt}
\renewcommand{\arraystretch}{1}

\begin{table*}[!htb]
\centering
\begin{tabular}{rccccccccc}
   & \textbf{Points} & \multicolumn{2}{c}{\textbf{Ours}} & \multicolumn{2}{c}{\textbf{Ours}} & \textbf{Yuksel} & \textbf{Yuksel} & \textbf{Corsini} & \textbf{Dart} \\
   & & \multicolumn{2}{c}{($k=1$)} & \multicolumn{2}{c}{($k=6$)} & ($\beta=0$) & ($\beta = 0.65$) & & \textbf{Throwing} \\
   & & \textbf{b=0} & \textbf{k+b=14} & \textbf{b=0} & \textbf{k+b=14} & & & &   \\
\textbf{Cathedral}     & 315k   & 9.64s   & 9.99s   & 17.08s  & 8.82s   & 15.55s  & 15.51s   & 0.56s & 0.14s  \\ 
\textbf{Cottage}       & 1424k  & 32.08s  & 35.13s  & 59.11s  & 37.15s  & 32.58s  & 32.42s   & 2.30s & 0.83s  \\ 
\textbf{Japanese}      & 540k   & 15.73s  & 17.33s  & 28.40s  & 16.33s  & 14.26s  & 14.54s   & 0.84s & 0.27s  \\ 
\textbf{Mansion}       & 1887k  & 64.07s  & 60.91s  & 118.56s & 58.39s  & 133.89s & 128.65s  & 2.82s & 1.05s  \\ 
\textbf{Monastery}     & 779k   & 20.56s  & 21.75s  & 34.28s  & 20.52s  & 21.47s  & 21.17s   & 1.12s & 0.39s  \\ 
\textbf{Old House}     & 1653k  & 62.01s  & 63.01s  & 103.41s & 57.36s  & 40.24s  & 40.12s   & 2.20s & 0.96s  \\ 
\textbf{San Francisco} & 2831k  & 96.30s  & 105.28s & 165.07s & 94.48s  & 64.60s  & 65.22s   & 3.90s & 1.72s  
\end{tabular}
\caption{\label{tab:c4t1} Execution times for the different simplification algorithms on the different synthetic models. Using the a neighbor buffer we are able to achieve similar execution times for $k=6$ and $k=1$. The fastest methods are Corsini's and Dart Throwing while our method performs similar to Yuksel's. }

\vspace{-4mm}
\end{table*}

\paragraph*{Scanning a Cube} 
We start by discussing a simple example of a cube of $10^3m^3$ scanned from 80 different locations, resulting in 452k points. These scans were simulated by imitating the sampling pattern and range noise distribution of a high-end LiDAR scanner (Leica P20). We compared our algorithm using $k \in \{1,3,6\}$ with Yuksel's method~\cite{yuksel2015, cyCodeBase} with $\beta \in \{0, 0.35, 0.65, 1\}$, MeshLab's implementation of Corsini's method~\cite{corsini2012, meshlab} and a simple implementation of Dart Throwing where we perform bisection search on the radius until obtaining the target number of samples. In all these cases, we decimated the cloud to a $10\%$ of its original size.

Schl\"omer et al.~\cite{schlomer2011, schlomer2011spectral} and Lagae et al.~\cite{lagae2008} propose evaluating 2D point distributions using spectral measures. Although we could apply this analysis to the cube's faces separately, we found this evaluation method inadequate for different reasons. For instance, 3D simplification methods use 3D neighborhoods which behave differently near edges. Moreover, Schl\"omer et al.~\cite{schlomer2011spectral} emphasize that ``The FT will be contaminated by boundary artifacts if the points are defined on the bounded unit square instead of the unit torus (...) For this reason, spectral analysis should be performed only on point sets that are periodic in space'', which is not our case.

In~\Cref{fig:c4f2} we show the simplified clouds on two faces of the cube (top and right). Performing a visual evaluation, we can see that our method with $k=6$ yields the most well-distributed clouds without significant gaps or clusters of points.

Next, we will discuss the charts on the last two rows of~\Cref{fig:c4f2}. Our clouds contain 3D points with an underlying 2D structure (the surface of the scanned elements). Therefore, when samples are well-distributed, we expect the number of points in a given neighborhood to increase linearly to its area. To check this property, we plot the number of neighbors at increasing neighborhood area (i.e., at increasing search radii) for each point on the simplified cubes. We bin these values and show colored bands enclosing the fractiles 0.01 to 0.99 (red), 0.1 to 0.9 (yellow), and 0.25 to 0.75 (green), and the average number of points (black line) to convey their variability. Finally, we estimate the average point density as the slope of the regression line (dotted dark red line). These charts contain different pieces of information that can help us compare the different methods:

\begin{enumerate}
    \item \textbf{Goodness of fit:} If the data does not show a linear behavior, this can hint at heterogeneous local densities. For instance, Yuksel's method with $\beta \in \{0.65, 1\}$ yields small clusters that make the density oscillate. This behavior can be detected by a small value of $R^2$ or a large RMSE and can be clearly observed by looking at the charts.
    \item \textbf{Variability:} The spread of the colored bands shows the variability in the number of neighbors for a given neighborhood area. Ideally, the less variability, the better (we want homogeneous densities across the cloud). Similarly, the $R^2$ value is the proportion of the variation in the number of points predicted by the neighborhood area (the bigger, the better). Finally, the RMSE shows how far the points are from the regression line (the smaller, the better). Looking at these three measures, our algorithm with $k=6$ appears to be the one with the smallest variability.
    \item \textbf{Covered Area:} For the cube case, we know that the real scanned area is $500m^2$, and knowing the number of points after the simplification ($\approx 45,223$), we can also compute the real density ($\approx 90.45 p/m^2$). However, the estimated densities (slopes of the regression lines) seem larger than the real density. Since the number of points is fixed, this can only mean that the decimated clouds cover a smaller area than the original one (probably due to the behavior on the model edges). The closer the estimated density is to the real one, the better the decimation process preserves the scanned area (this only applies if the density estimate is reliable, which does not happen for Yuksel with $\beta \in \{0.65, 1\}$). In this regard, our algorithm with $k=6$ also outperforms the rest.
\end{enumerate}

\paragraph*{Synthetic models} 

Next, we discuss our results on synthetic models. These were simulated similarly to the cube but are much more complex (resembling real buildings) and contain a larger amount of points (although they still fit in memory, thus enabling a comparison with competing approaches). However, unlike the cube example, we no longer have an accurate estimate on the real scanned area, as part of the model's surface could be occluded or scanned twice (front and back sides of faces).

\setlength{\tabcolsep}{2pt}
\renewcommand{\arraystretch}{0}

\begin{figure*}[!htb]
\centering
\begin{tabular*}{\linewidth}{ccccccc}
    & Ours ($k=1$) & Ours ($k=6$) & Yuksel ($\beta=0$) & Yuksel ($\beta=0.65$) & Corsini & Dart Throwing \\
    \rotatebox[origin=c]{90}{Cathedral} &
    \raisebox{-0.5\height}{\includegraphics[width=0.15\linewidth]{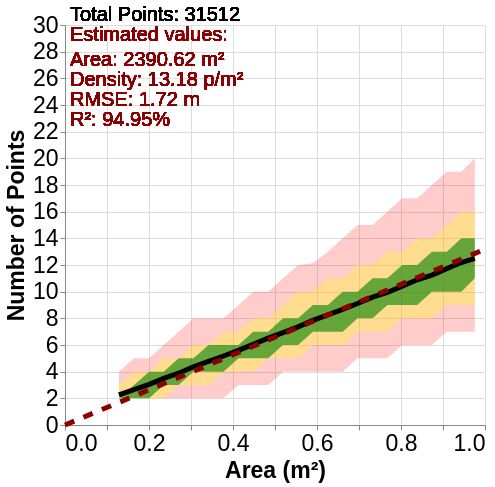}} &
    \raisebox{-0.5\height}{\includegraphics[width=0.15\linewidth]{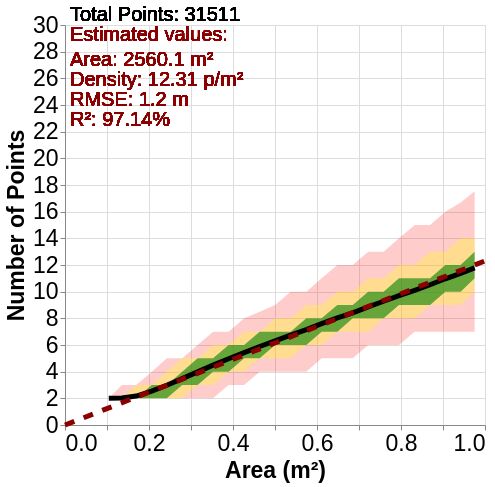}} &
    \raisebox{-0.5\height}{\textcolor{green}{\fbox{\includegraphics[width=0.15\linewidth]{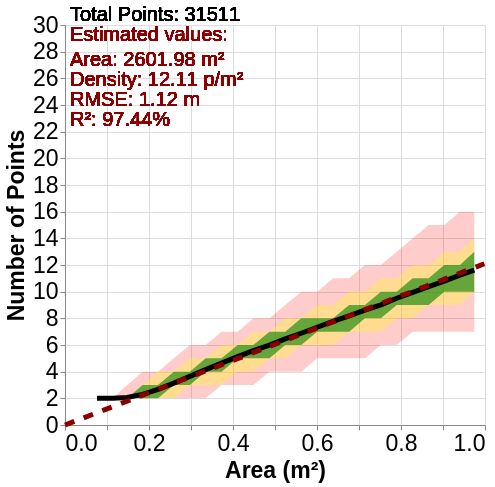}}}} &
    \raisebox{-0.5\height}{\includegraphics[width=0.15\linewidth]{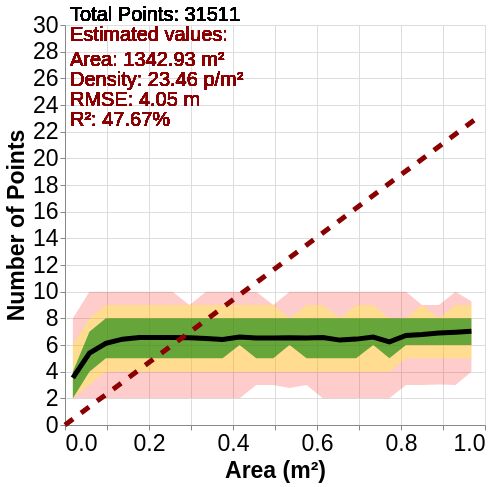}} &
    \raisebox{-0.5\height}{\includegraphics[width=0.15\linewidth]{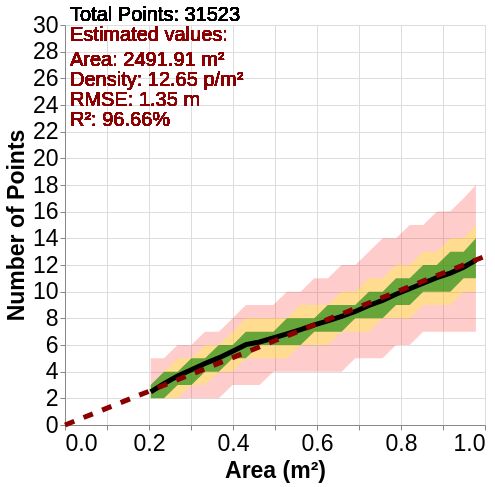}} &
    \raisebox{-0.5\height}{\includegraphics[width=0.15\linewidth]{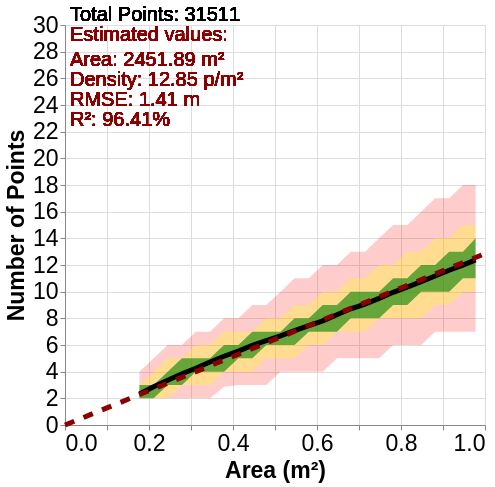}}
    \\
    \rotatebox[origin=c]{90}{Cottage} &
    \raisebox{-0.5\height}{\includegraphics[width=0.15\linewidth]{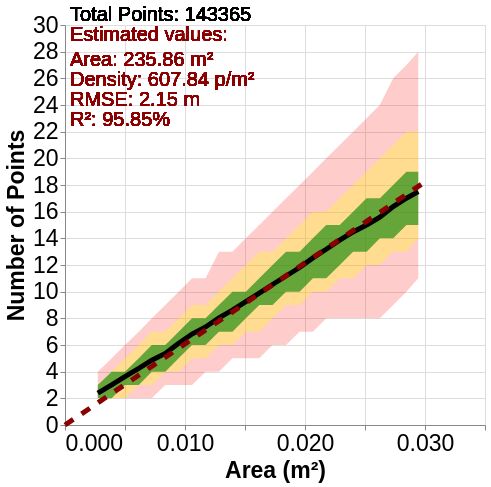}} &
    \raisebox{-0.5\height}{\textcolor{green}{\fbox{\includegraphics[width=0.15\linewidth]{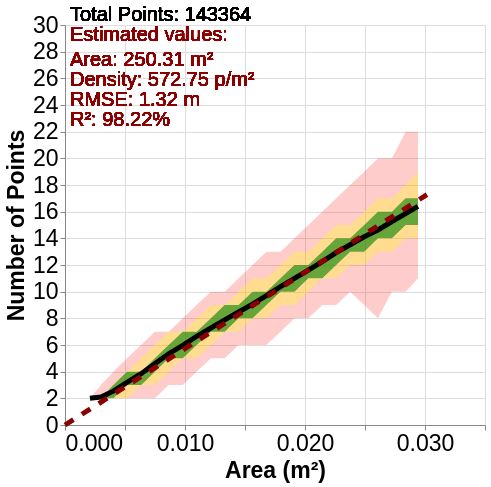}}}} &
    \raisebox{-0.5\height}{\includegraphics[width=0.15\linewidth]{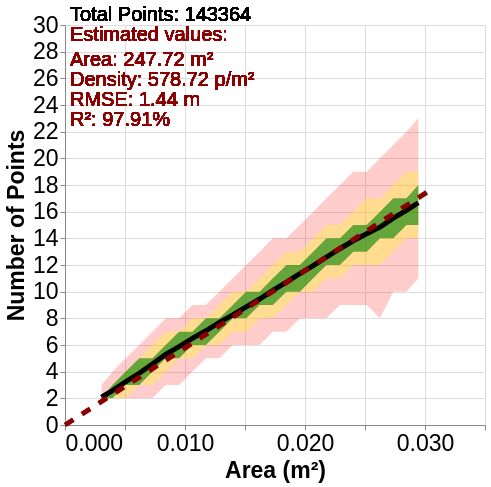}} &
    \raisebox{-0.5\height}{\includegraphics[width=0.15\linewidth]{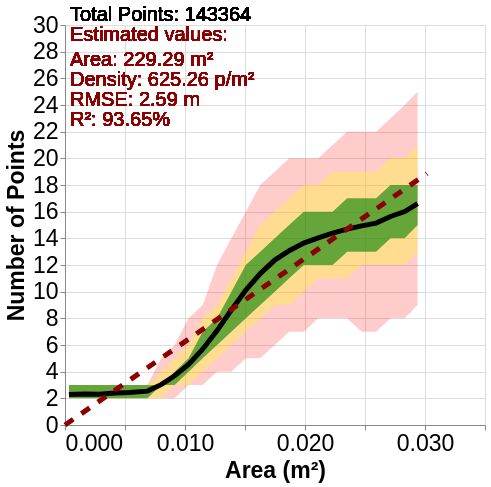}} &
    \raisebox{-0.5\height}{\includegraphics[width=0.15\linewidth]{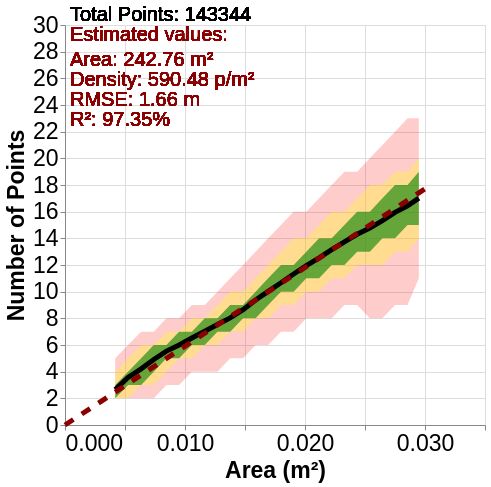}} &
    \raisebox{-0.5\height}{\includegraphics[width=0.15\linewidth]{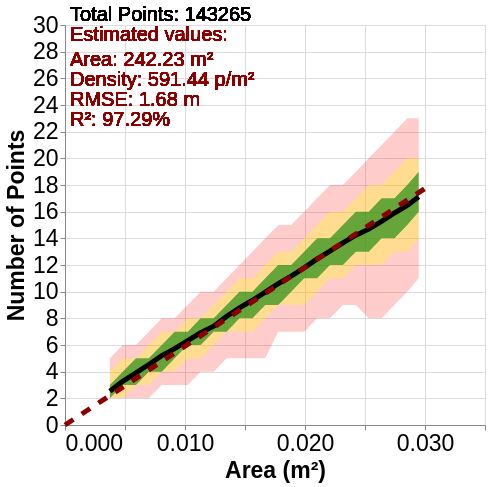}}
    \\
    \rotatebox[origin=c]{90}{Japanese} &
    \raisebox{-0.5\height}{\includegraphics[width=0.15\linewidth]{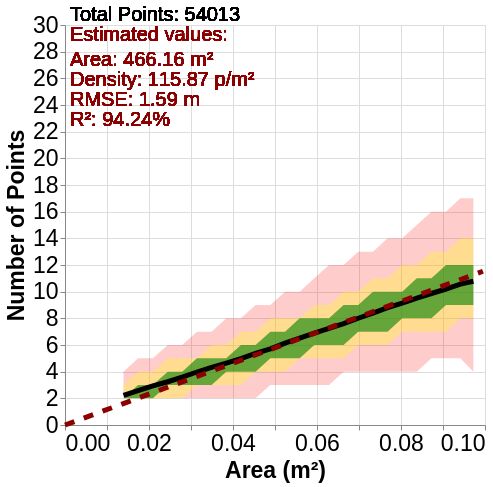}} &
    \raisebox{-0.5\height}{\textcolor{green}{\fbox{\includegraphics[width=0.15\linewidth]{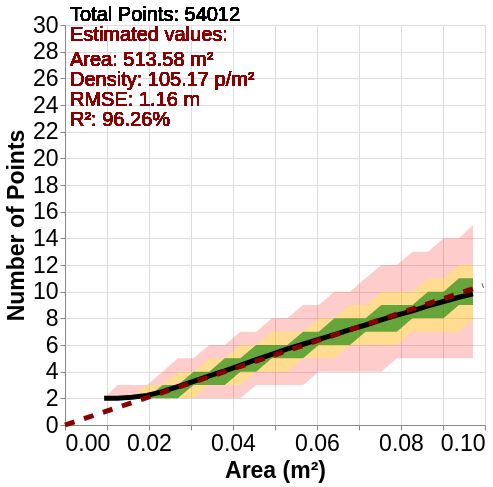}}}} &
    \raisebox{-0.5\height}{\includegraphics[width=0.15\linewidth]{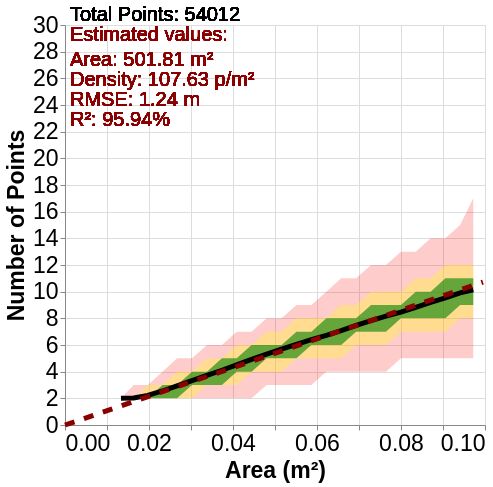}} &
    \raisebox{-0.5\height}{\includegraphics[width=0.15\linewidth]{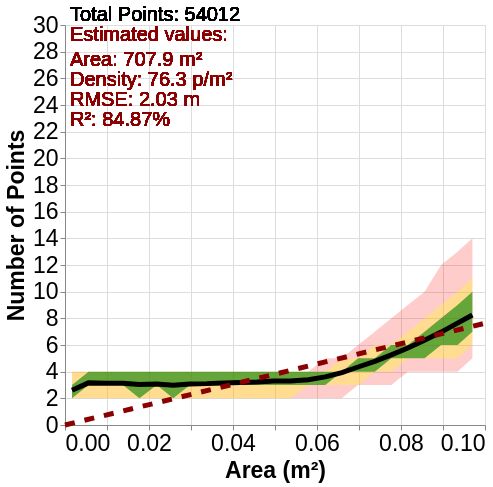}} &
    \raisebox{-0.5\height}{\includegraphics[width=0.15\linewidth]{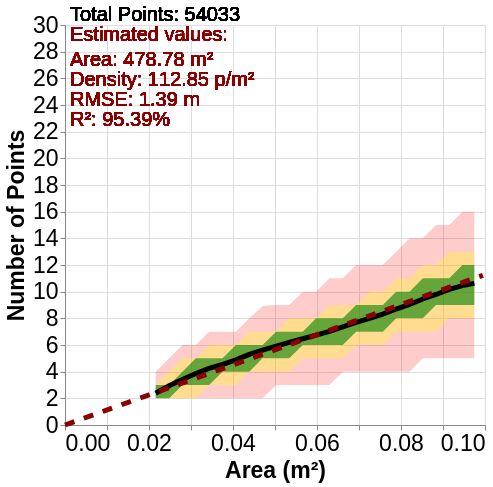}} &
    \raisebox{-0.5\height}{\includegraphics[width=0.15\linewidth]{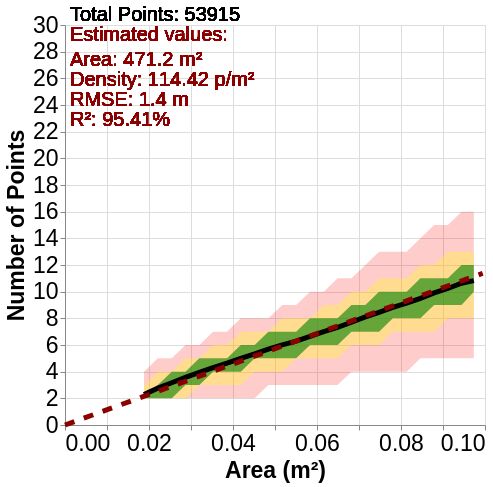}}
    \\
    \rotatebox[origin=c]{90}{Mansion} &
    \raisebox{-0.5\height}{\includegraphics[width=0.15\linewidth]{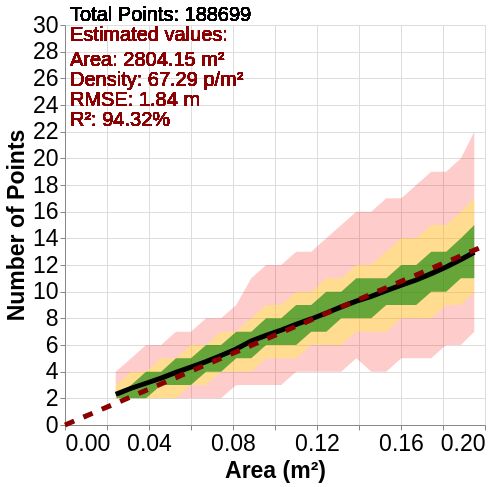}} &
    \raisebox{-0.5\height}{\textcolor{green}{\fbox{\includegraphics[width=0.15\linewidth]{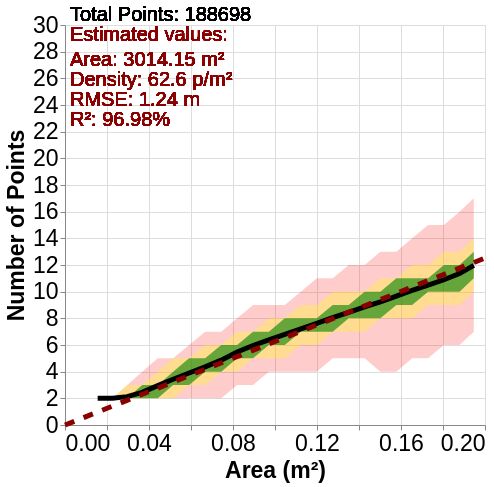}}}} &
    \raisebox{-0.5\height}{\includegraphics[width=0.15\linewidth]{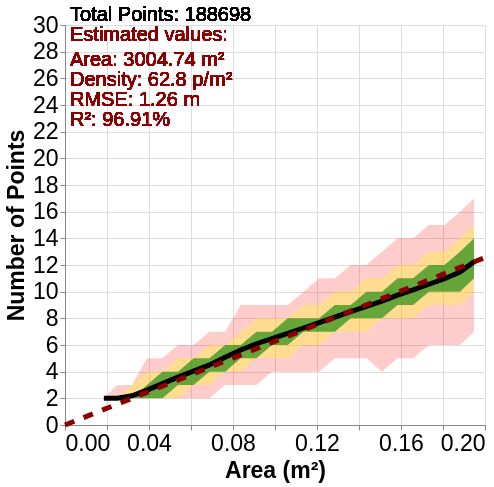}} &
    \raisebox{-0.5\height}{\includegraphics[width=0.15\linewidth]{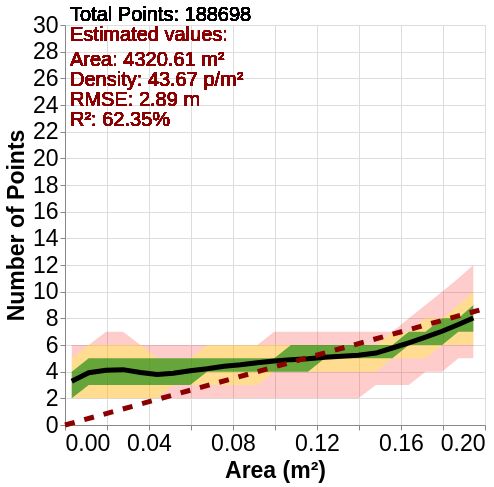}} &
    \raisebox{-0.5\height}{\includegraphics[width=0.15\linewidth]{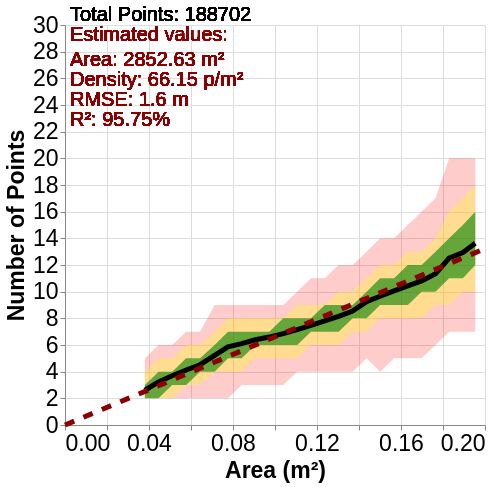}} &
    \raisebox{-0.5\height}{\includegraphics[width=0.15\linewidth]{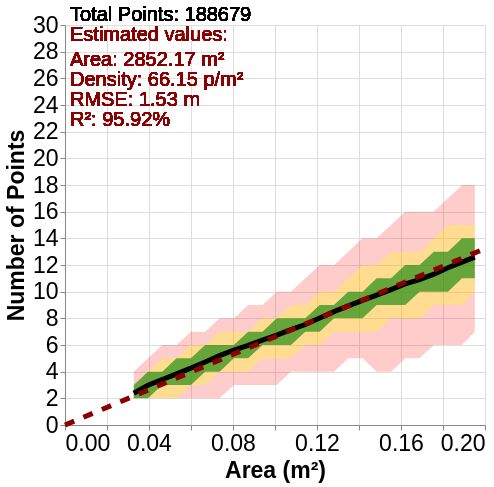}}
    \\
    \rotatebox[origin=c]{90}{Monastery} &
    \raisebox{-0.5\height}{\includegraphics[width=0.15\linewidth]{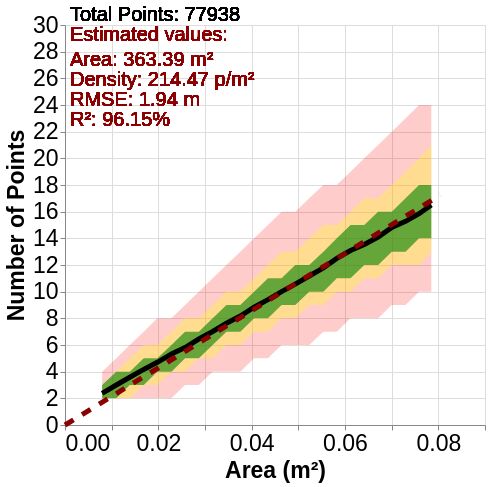}} &
    \raisebox{-0.5\height}{\textcolor{green}{\fbox{\includegraphics[width=0.15\linewidth]{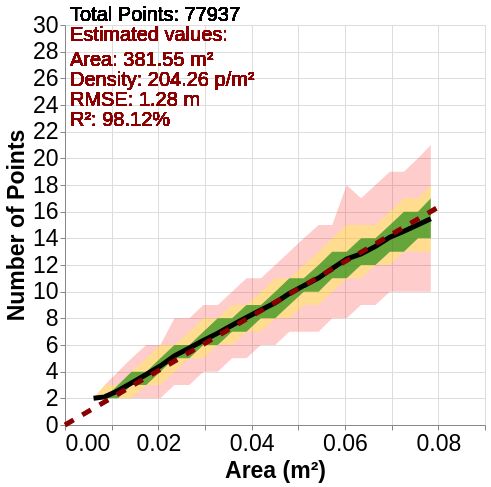}}}} &
    \raisebox{-0.5\height}{\includegraphics[width=0.15\linewidth]{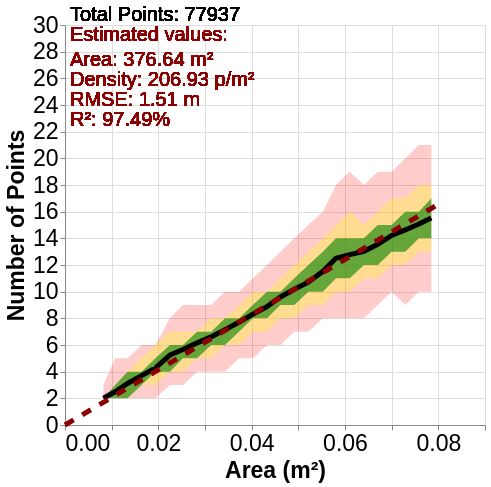}} &
    \raisebox{-0.5\height}{\includegraphics[width=0.15\linewidth]{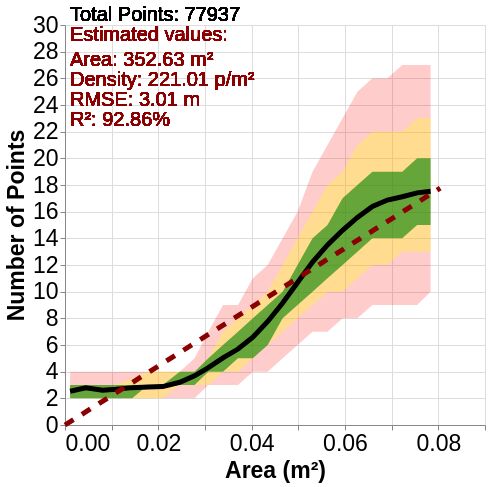}} &
    \raisebox{-0.5\height}{\includegraphics[width=0.15\linewidth]{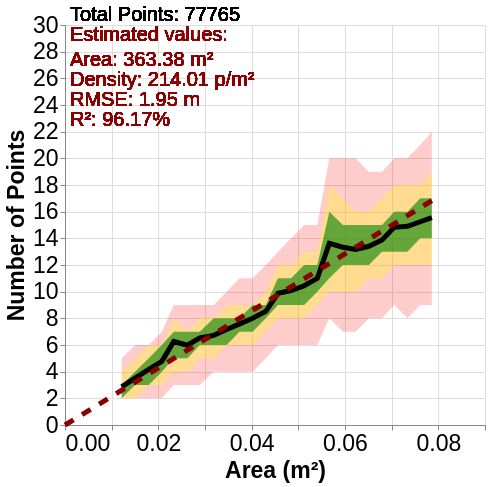}} &
    \raisebox{-0.5\height}{\includegraphics[width=0.15\linewidth]{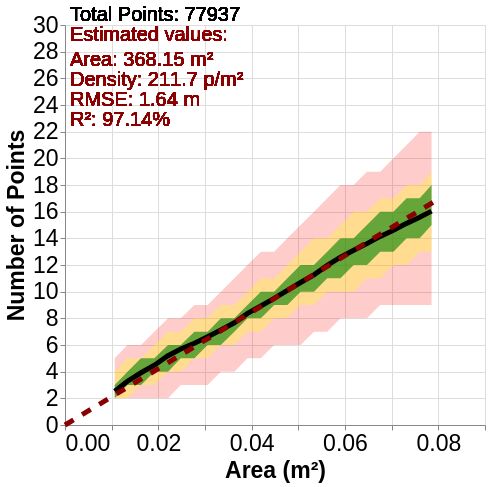}}
    \\
    \rotatebox[origin=c]{90}{Old House} &
    \raisebox{-0.5\height}{\includegraphics[width=0.15\linewidth]{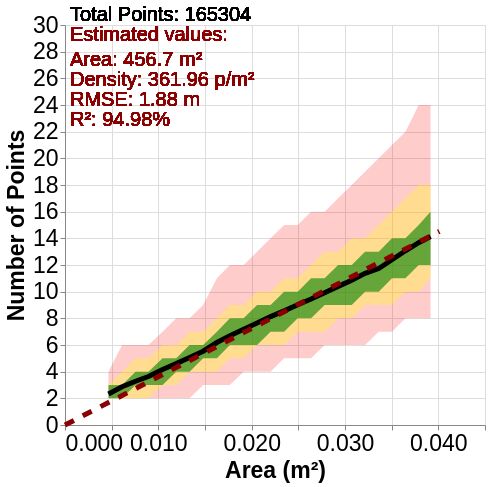}} &
    \raisebox{-0.5\height}{\textcolor{green}{\fbox{\includegraphics[width=0.15\linewidth]{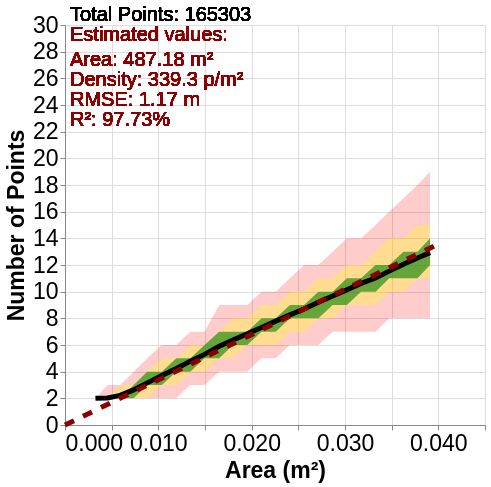}}}} &
    \raisebox{-0.5\height}{\includegraphics[width=0.15\linewidth]{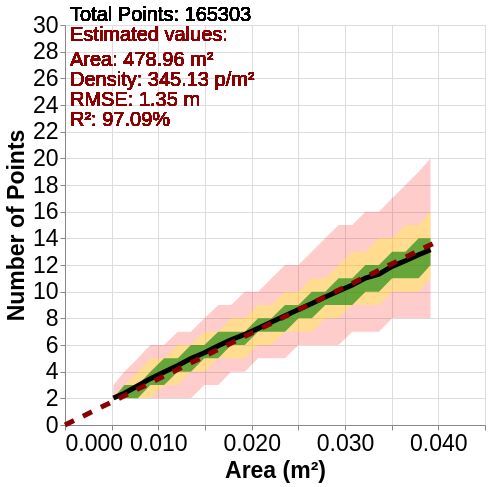}} &
    \raisebox{-0.5\height}{\includegraphics[width=0.15\linewidth]{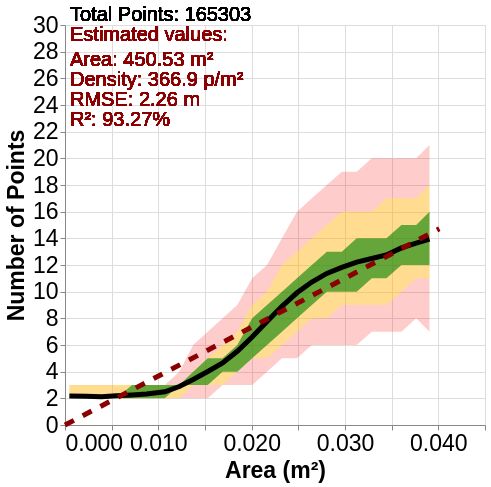}} &
    \raisebox{-0.5\height}{\includegraphics[width=0.15\linewidth]{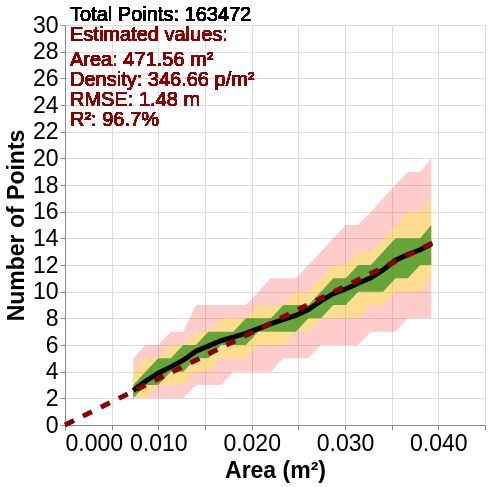}} &
    \raisebox{-0.5\height}{\includegraphics[width=0.15\linewidth]{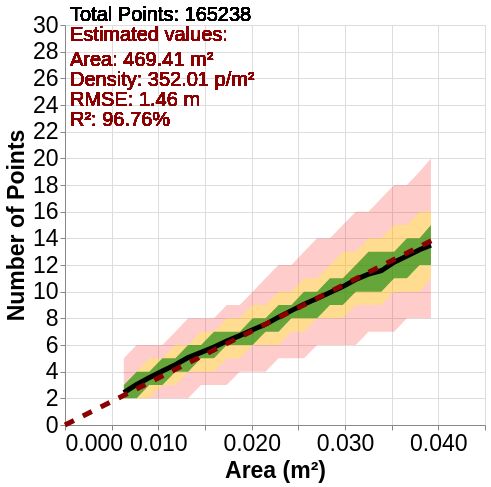}}
    \\
    \rotatebox[origin=c]{90}{San Francisco} &
    \raisebox{-0.5\height}{\includegraphics[width=0.15\linewidth]{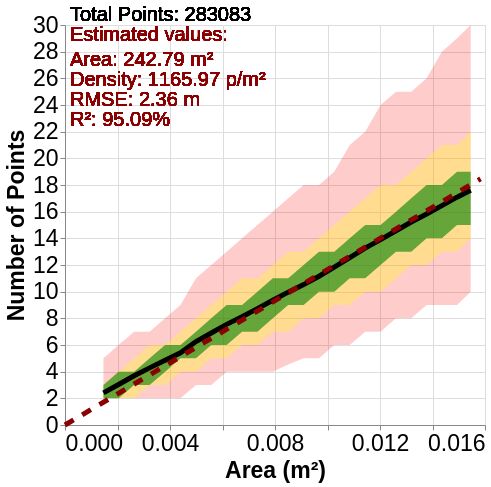}} &
    \raisebox{-0.5\height}{\textcolor{green}{\fbox{\includegraphics[width=0.15\linewidth]{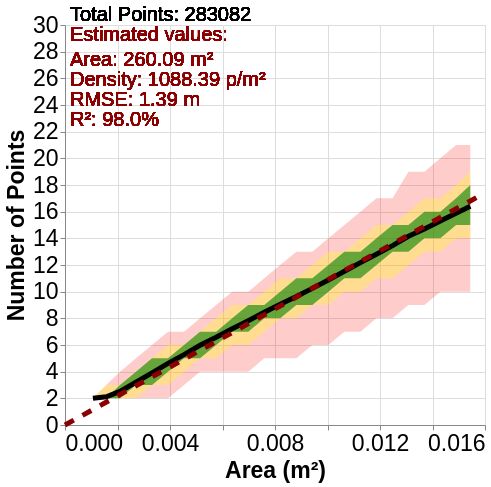}}}} &
    \raisebox{-0.5\height}{\includegraphics[width=0.15\linewidth]{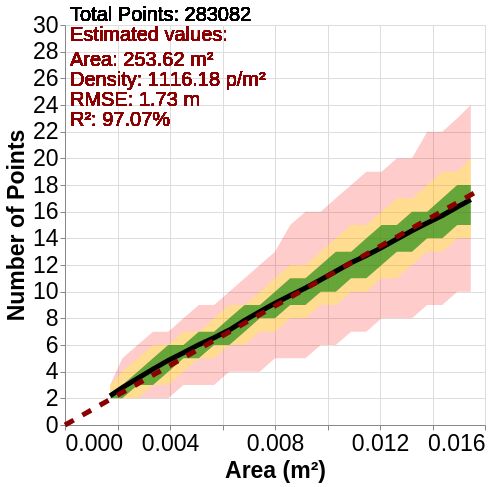}} &
    \raisebox{-0.5\height}{\includegraphics[width=0.15\linewidth]{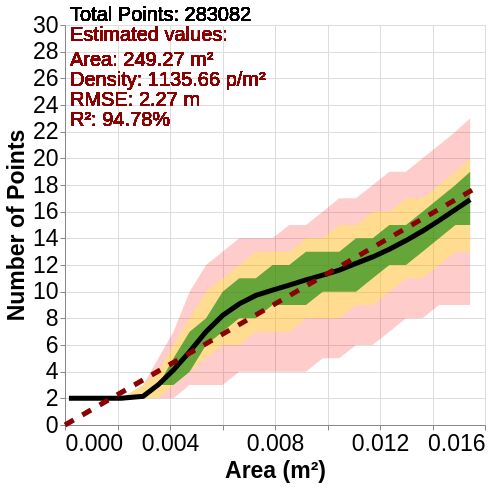}} &
    \raisebox{-0.5\height}{\includegraphics[width=0.15\linewidth]{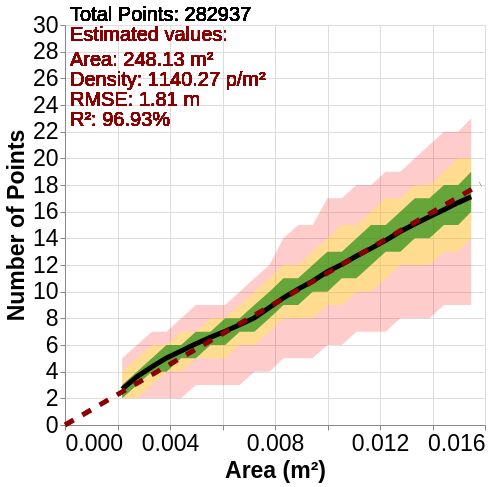}} &
    \raisebox{-0.5\height}{\includegraphics[width=0.15\linewidth]{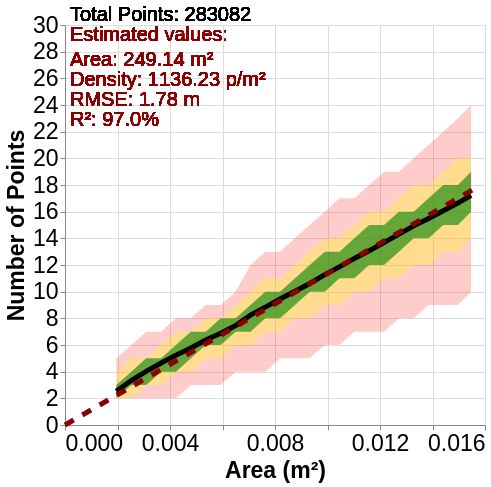}}
    \\
\end{tabular*}
\caption{\label{fig:c4f4} Distribution of neighborhood sizes (number of points) at increasing area. Results the clouds simplified with our method and competing approaches. The red band contains points between the fractiles 0.01 and 0.99, the yellow one contains points between the fractiles 0.1 and 0.9, and the green one contains points between the quartiles 0.25 and 0.75. The best results in terms of smallest variability and highest covered area are highlighted in green. Area values for Yuksel with $\beta=0.65$ are not reliable because the number of points clearly does not follow a linear relationship with respect to the neighborhood area.}
\end{figure*}

\setlength{\tabcolsep}{4.9pt}
\renewcommand{\arraystretch}{0}

\newcommand{\figfull}[5]{
\begin{overpic}[width={#5}cm,height=0.3cm,tics=10]{#1}
	\put(0, 0){{\makebox(0,0)[lb]{\raisebox{0.5mm}{\hspace{0.5mm}\scriptsize\contour{black}{\color{#4}{#2}}}}}}
	\put(98, 0){{\makebox(0,0)[rb]{\raisebox{0.5mm}{\hspace{0.5mm}\scriptsize\contour{black}{\color{#4}{#3}}}}}}
\end{overpic}
}

\begin{figure*}[!htb]
\centering
\begin{tabular*}{\linewidth}{ccccccc}
    & Mesh & Scanned Cloud & Ours ($k=6$) & Yuksel ($\beta=0$) & Corsini & Dart Throwing \\
    \rotatebox[origin=c]{90}{Cathedral} &
    \raisebox{-0.5\height}{\includegraphics[width=0.14\linewidth]{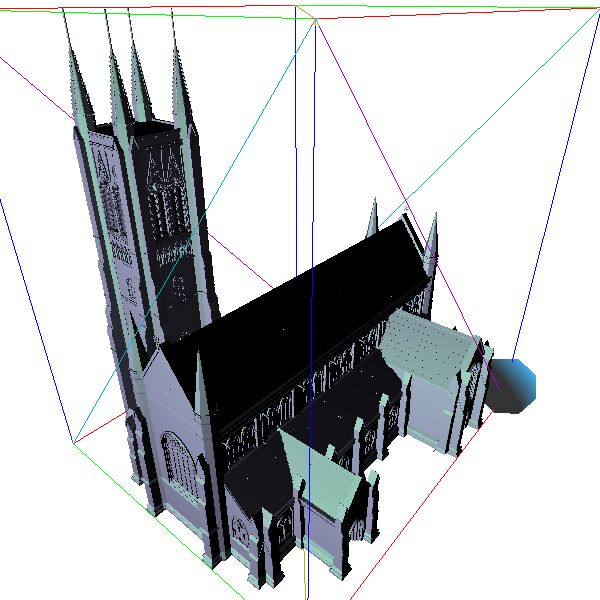}} &
    \raisebox{-0.5\height}{\includegraphics[width=0.14\linewidth]{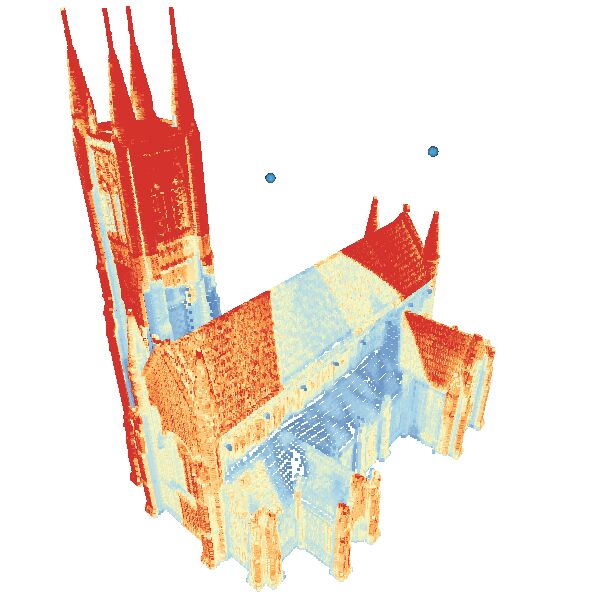}} &
    \raisebox{-0.5\height}{\includegraphics[width=0.14\linewidth]{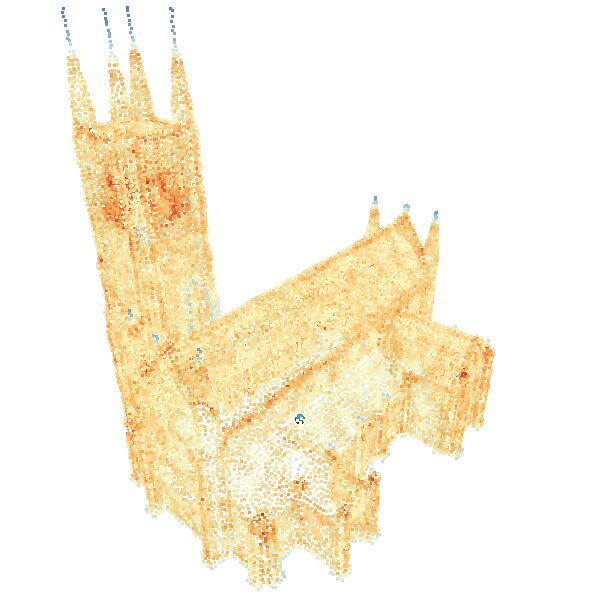}} &
    \raisebox{-0.5\height}{\includegraphics[width=0.14\linewidth]{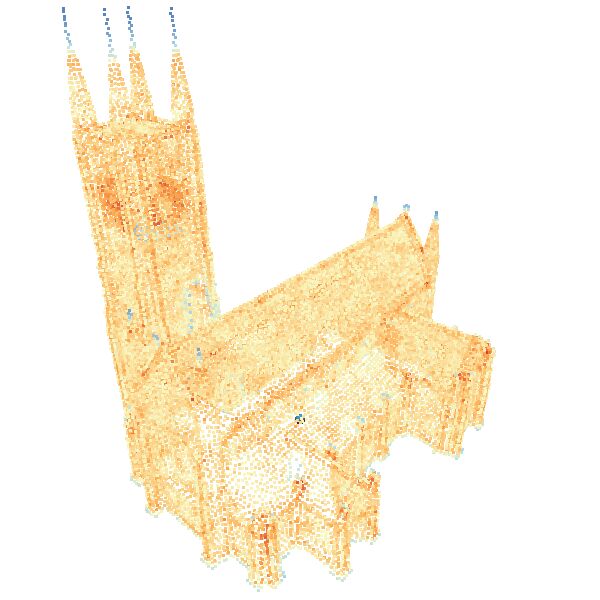}} &
    \raisebox{-0.5\height}{\includegraphics[width=0.14\linewidth]{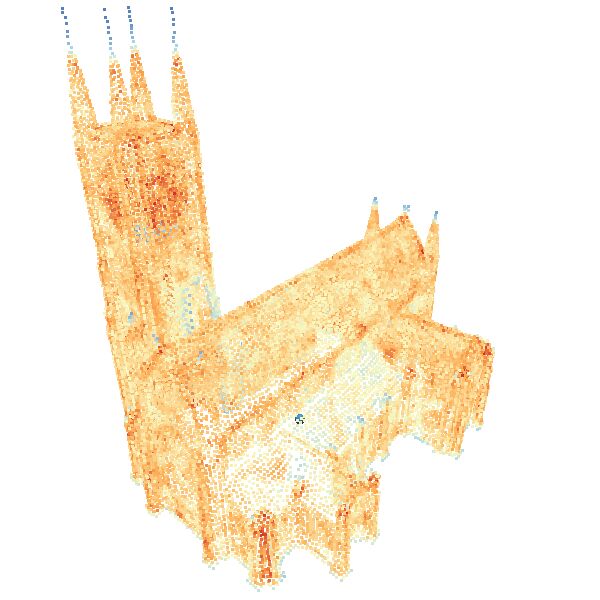}} &
    \raisebox{-0.5\height}{\includegraphics[width=0.14\linewidth]{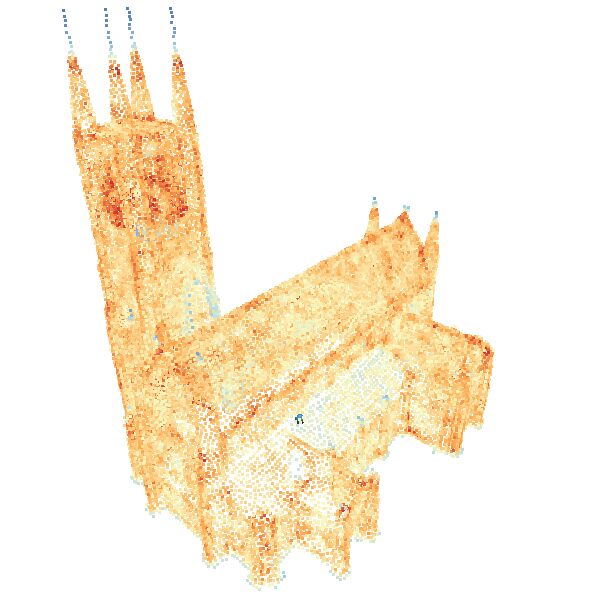}}
    \\
    & &
    \figfull{synthetic_ss/legend}{$0\,p/m^2$}{$2e2\,p/m^2$}{white}{2.5} &
    \multicolumn{4}{c}{\figfull{synthetic_ss/legend}{$0\,p/m^2$}{$20\,p/m^2$}{white}{11}}
    \\
    \vspace{2mm}
    \\
    \rotatebox[origin=c]{90}{Cottage} &
    \raisebox{-0.5\height}{\includegraphics[width=0.14\linewidth]{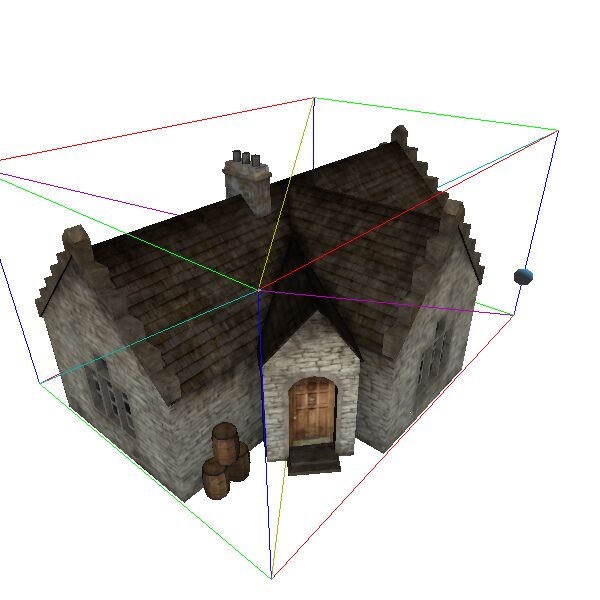}} &
    \raisebox{-0.5\height}{\includegraphics[width=0.14\linewidth]{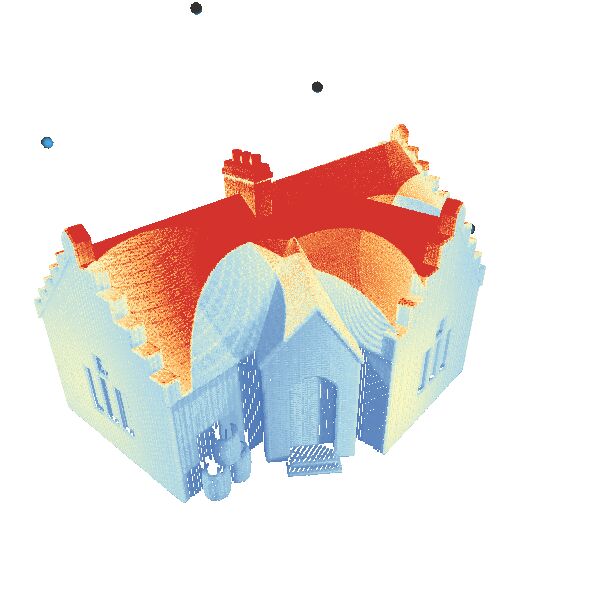}} &
    \raisebox{-0.5\height}{\includegraphics[width=0.14\linewidth]{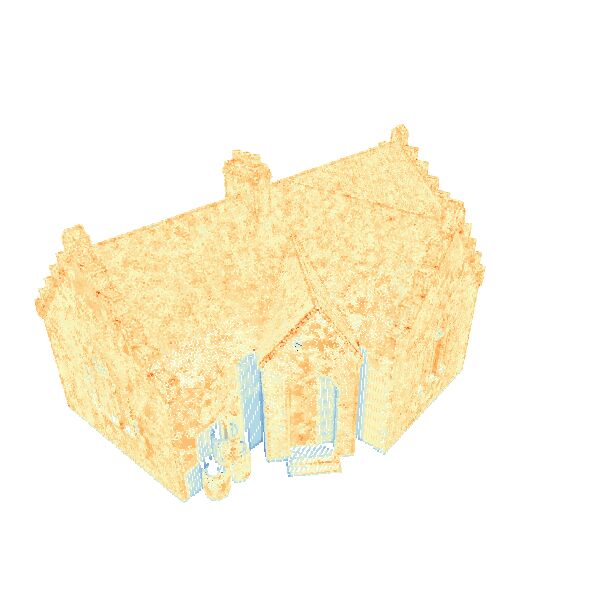}} &
    \raisebox{-0.5\height}{\includegraphics[width=0.14\linewidth]{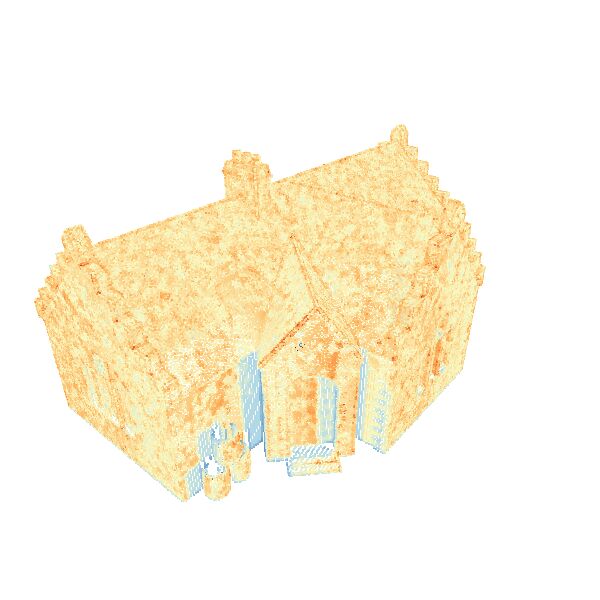}} &
    \raisebox{-0.5\height}{\includegraphics[width=0.14\linewidth]{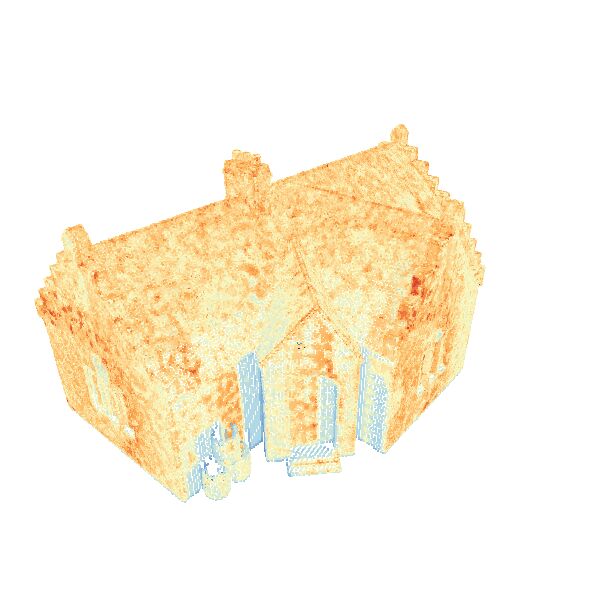}} &
    \raisebox{-0.5\height}{\includegraphics[width=0.14\linewidth]{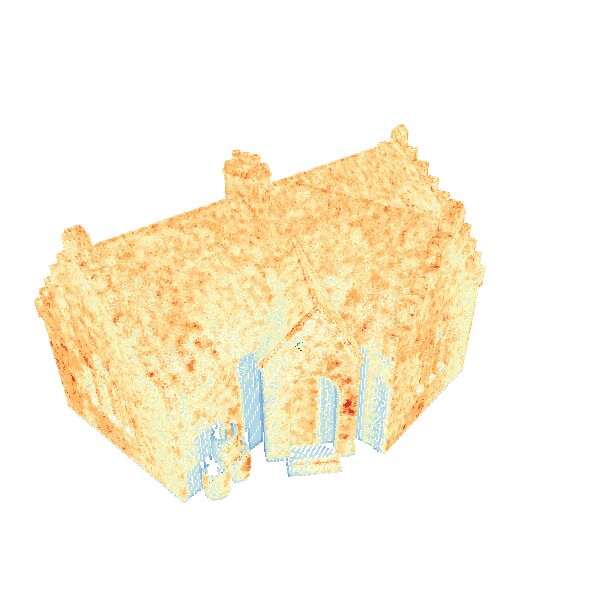}}
    \\
    & &
    \figfull{synthetic_ss/legend}{$0\,p/m^2$}{$1e4\,p/m^2$}{white}{2.5} &
    \multicolumn{4}{c}{\figfull{synthetic_ss/legend}{$0\,p/m^2$}{$1e3\,p/m^2$}{white}{11}}
    \\
    \vspace{2mm}
    \\
    \rotatebox[origin=c]{90}{Japanese} &
    \raisebox{-0.5\height}{\includegraphics[width=0.14\linewidth]{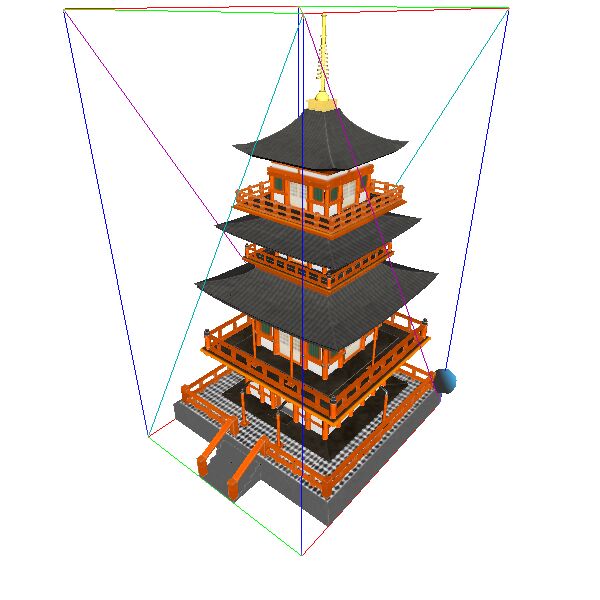}} &
    \raisebox{-0.5\height}{\includegraphics[width=0.14\linewidth]{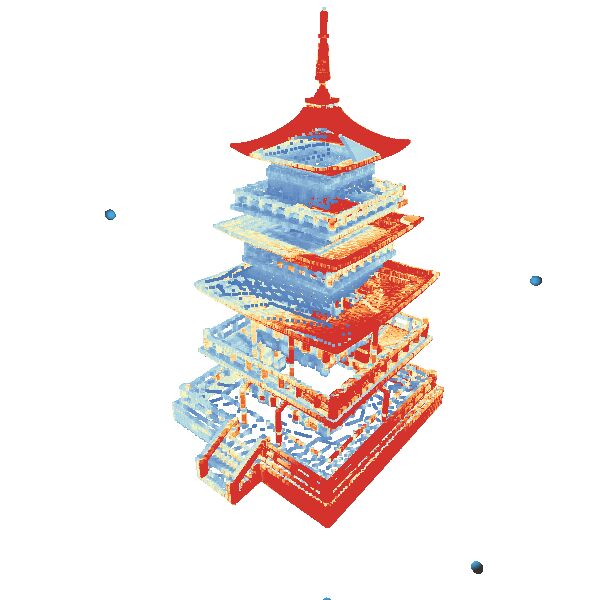}} &
    \raisebox{-0.5\height}{\includegraphics[width=0.14\linewidth]{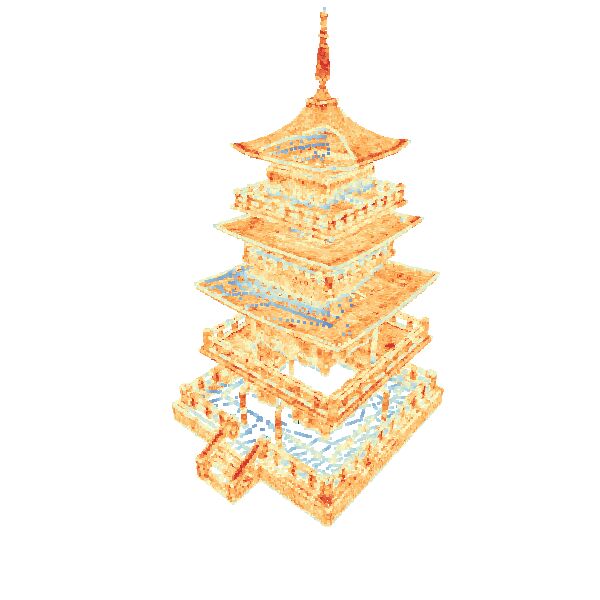}} &
    \raisebox{-0.5\height}{\includegraphics[width=0.14\linewidth]{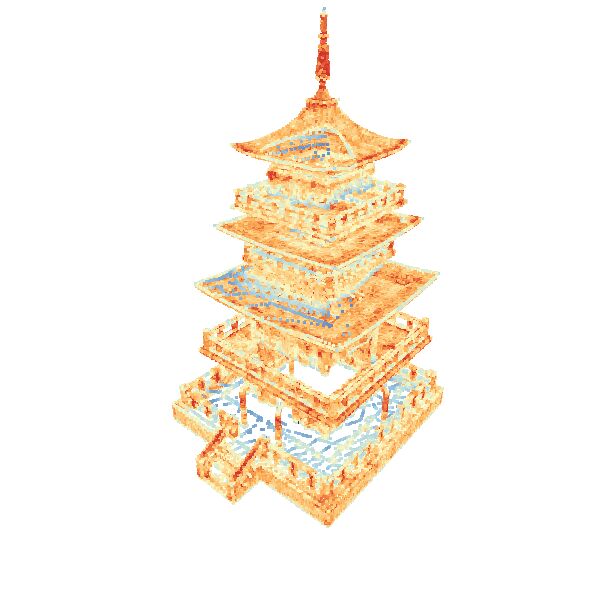}} &
    \raisebox{-0.5\height}{\includegraphics[width=0.14\linewidth]{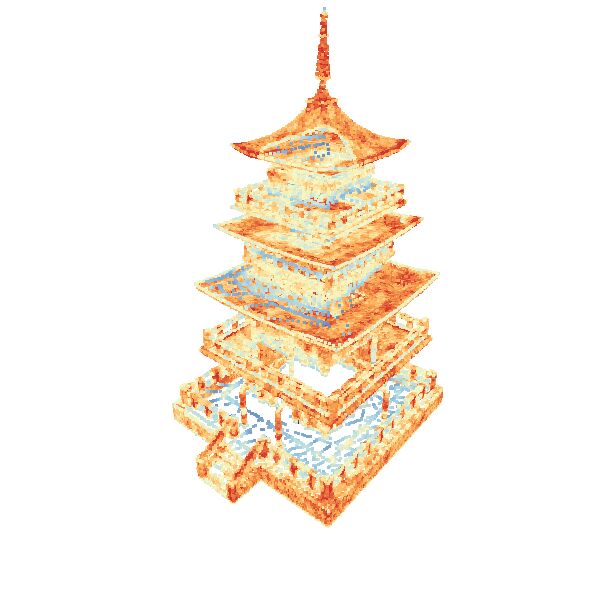}} &
    \raisebox{-0.5\height}{\includegraphics[width=0.14\linewidth]{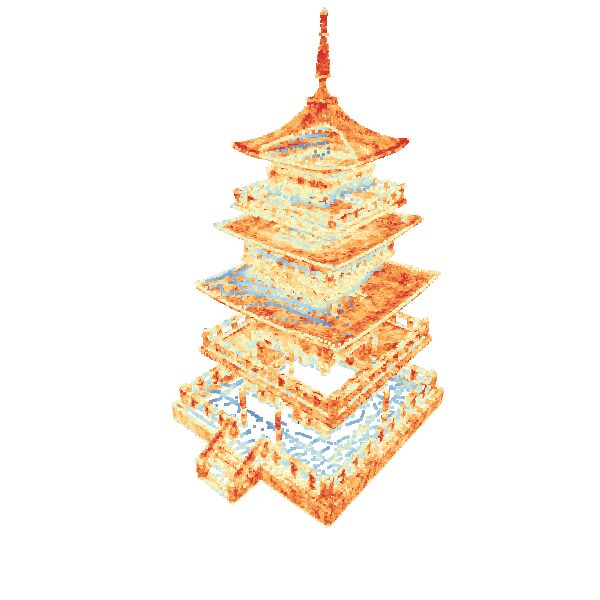}}
    \\
    & &
    \figfull{synthetic_ss/legend}{$0\,p/m^2$}{$1.5e3\,p/m^2$}{white}{2.5} &
    \multicolumn{4}{c}{\figfull{synthetic_ss/legend}{$0\,p/m^2$}{$1.5e2\,p/m^2$}{white}{11}}
    \\
    \rotatebox[origin=c]{90}{Mansion} &
    \raisebox{-0.5\height}{\includegraphics[width=0.14\linewidth]{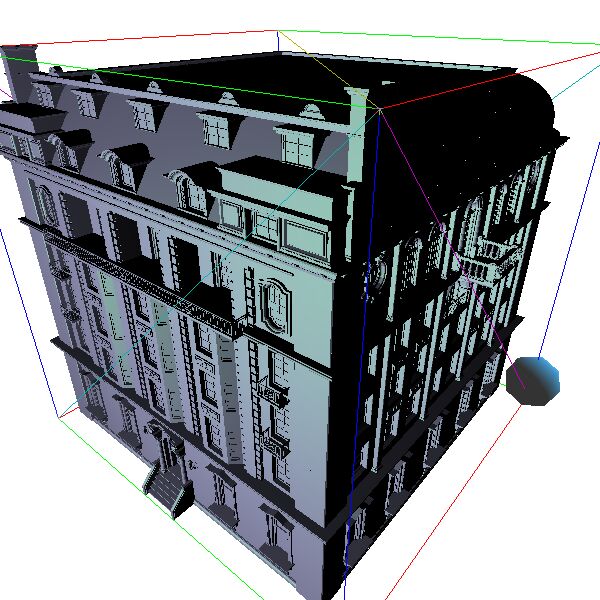}} &
    \raisebox{-0.5\height}{\includegraphics[width=0.14\linewidth]{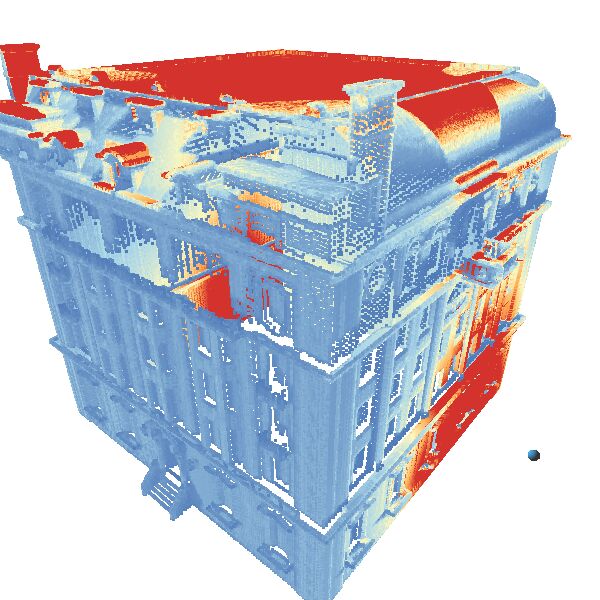}} &
    \raisebox{-0.5\height}{\includegraphics[width=0.14\linewidth]{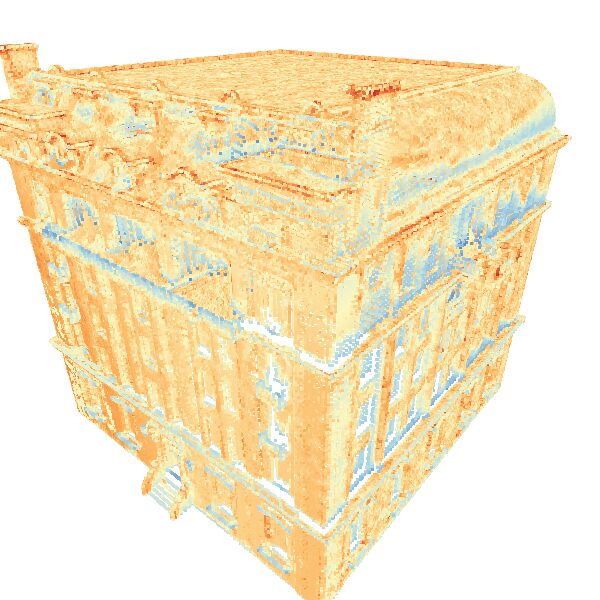}} &
    \raisebox{-0.5\height}{\includegraphics[width=0.14\linewidth]{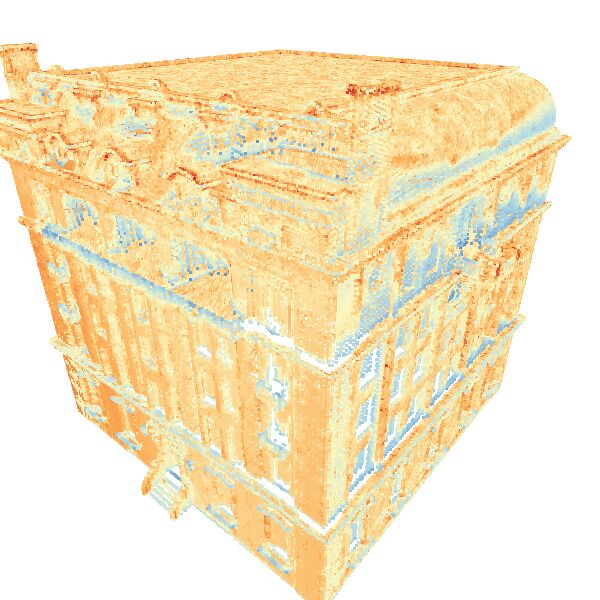}} &
    \raisebox{-0.5\height}{\includegraphics[width=0.14\linewidth]{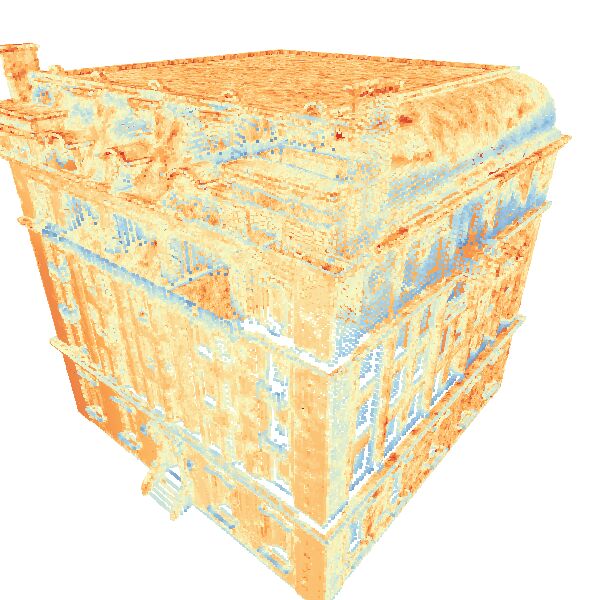}} &
    \raisebox{-0.5\height}{\includegraphics[width=0.14\linewidth]{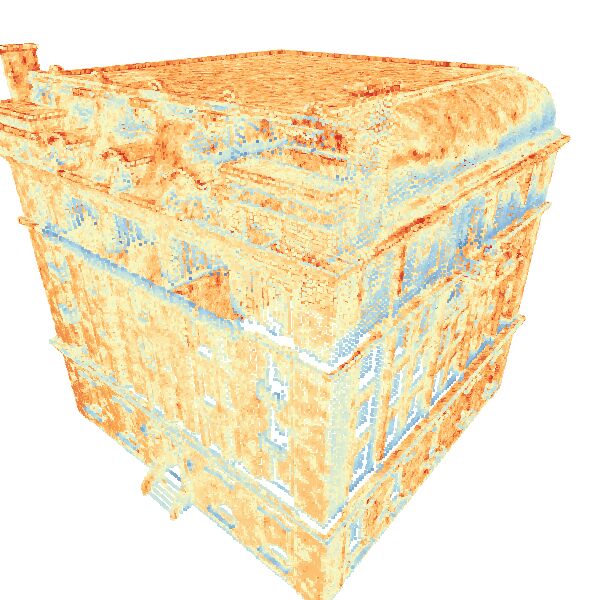}}
    \\
    & &
    \figfull{synthetic_ss/legend}{$0\,p/m^2$}{$1e3\,p/m^2$}{white}{2.5} &
    \multicolumn{4}{c}{\figfull{synthetic_ss/legend}{$0\,p/m^2$}{$1.5e2\,p/m^2$}{white}{11}}
    \\
    \vspace{2mm}
    \\
    \rotatebox[origin=c]{90}{Monastery} &
    \raisebox{-0.5\height}{\includegraphics[width=0.14\linewidth]{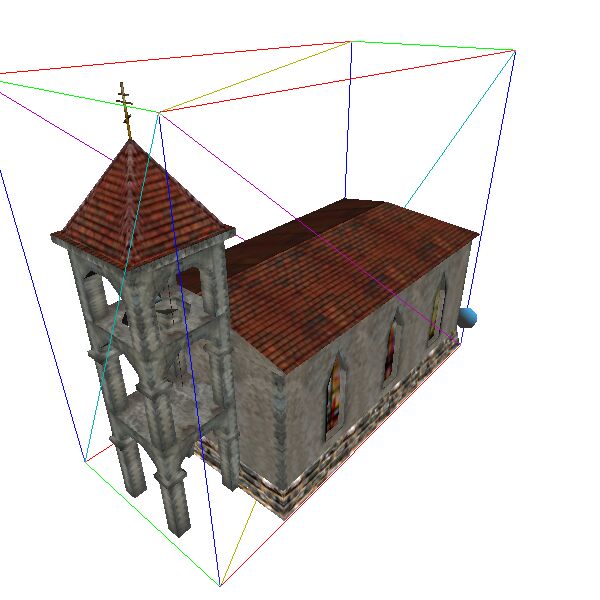}} &
    \raisebox{-0.5\height}{\includegraphics[width=0.14\linewidth]{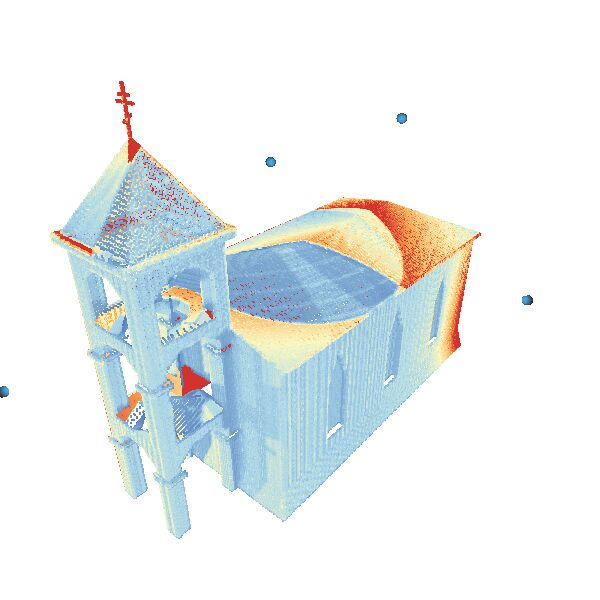}} &
    \raisebox{-0.5\height}{\includegraphics[width=0.14\linewidth]{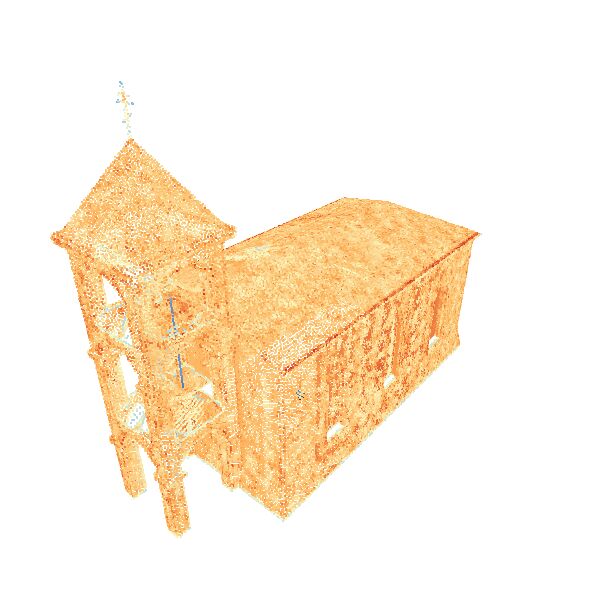}} &
    \raisebox{-0.5\height}{\includegraphics[width=0.14\linewidth]{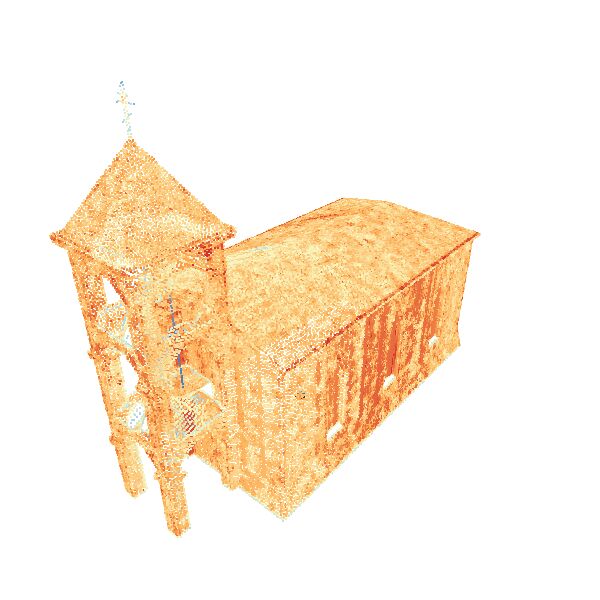}} &
    \raisebox{-0.5\height}{\includegraphics[width=0.14\linewidth]{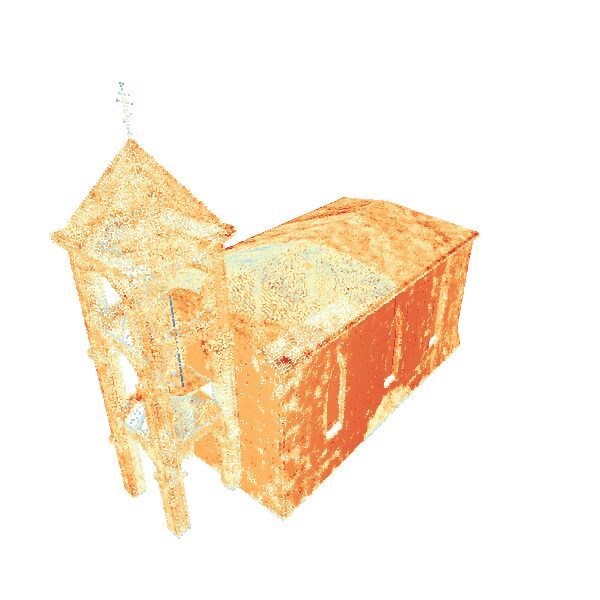}} &
    \raisebox{-0.5\height}{\includegraphics[width=0.14\linewidth]{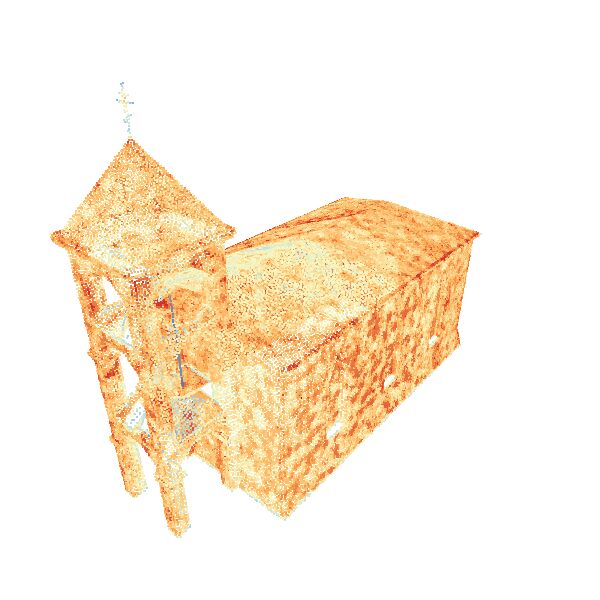}}
    \\
    & &
    \figfull{synthetic_ss/legend}{$0\,p/m^2$}{$3e3\,p/m^2$}{white}{2.5} &
    \multicolumn{4}{c}{\figfull{synthetic_ss/legend}{$0\,p/m^2$}{$3e2\,p/m^2$}{white}{11}}
    \\
    \vspace{2mm}
    \\
    \rotatebox[origin=c]{90}{Old House} &
    \raisebox{-0.5\height}{\includegraphics[width=0.14\linewidth]{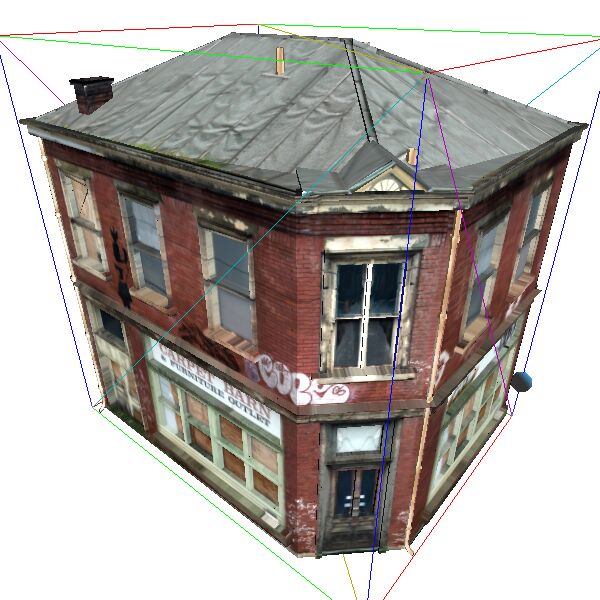}} &
    \raisebox{-0.5\height}{\includegraphics[width=0.14\linewidth]{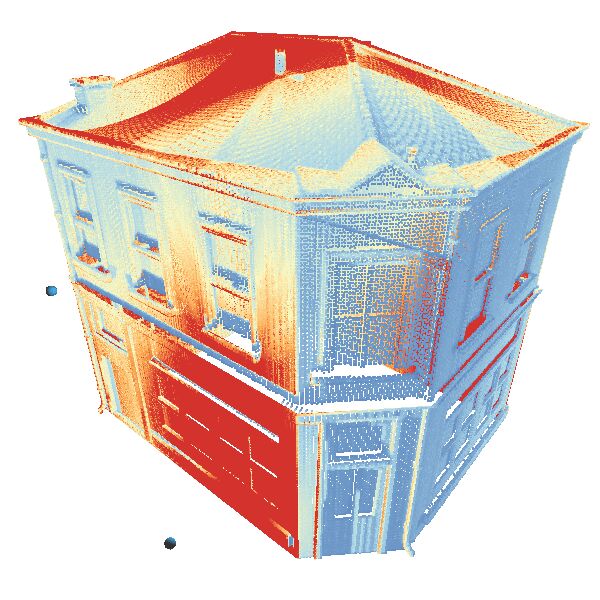}} &
    \raisebox{-0.5\height}{\includegraphics[width=0.14\linewidth]{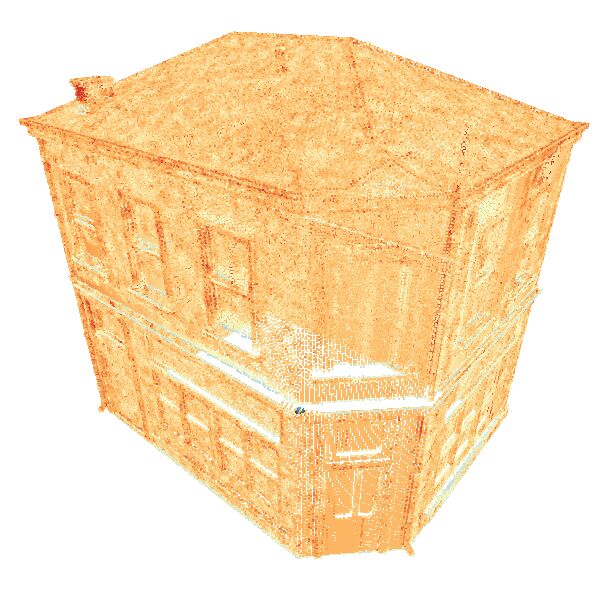}} &
    \raisebox{-0.5\height}{\includegraphics[width=0.14\linewidth]{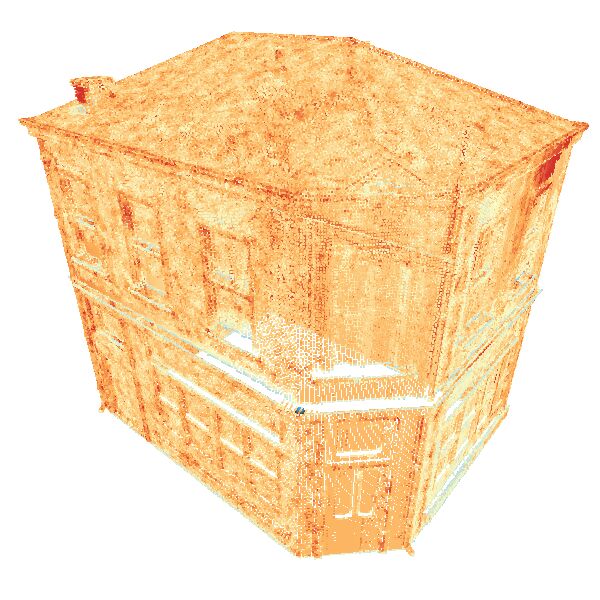}} &
    \raisebox{-0.5\height}{\includegraphics[width=0.14\linewidth]{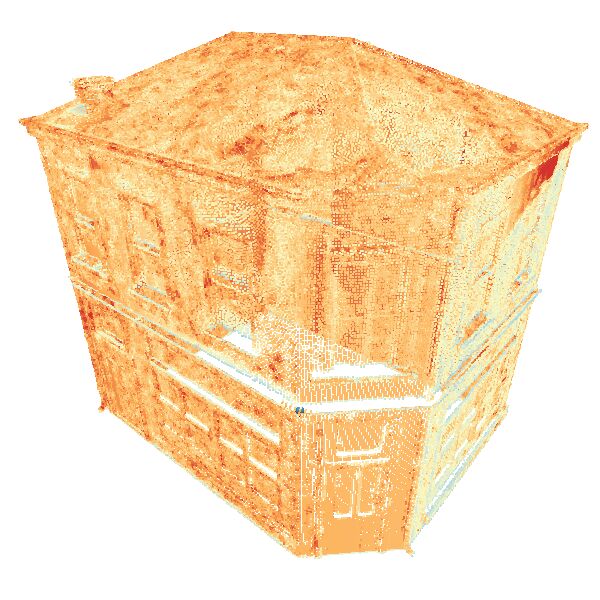}} &
    \raisebox{-0.5\height}{\includegraphics[width=0.14\linewidth]{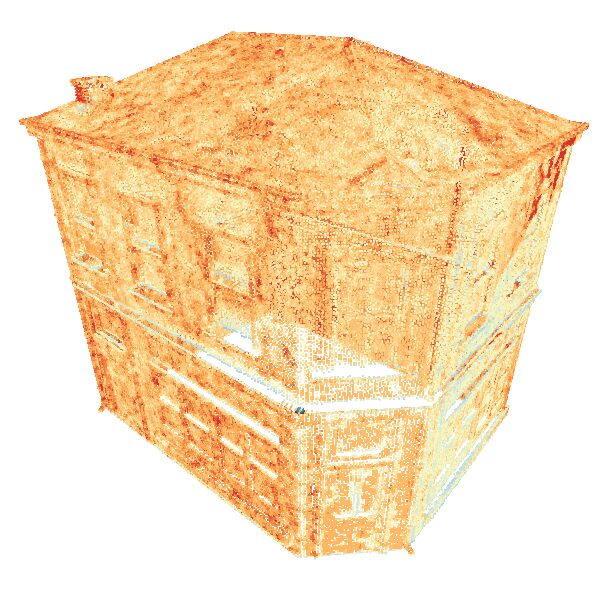}}
    \\
    & &
    \figfull{synthetic_ss/legend}{$0\,p/m^2$}{$5e3\,p/m^2$}{white}{2.5} &
    \multicolumn{4}{c}{\figfull{synthetic_ss/legend}{$0\,p/m^2$}{$5e2\,p/m^2$}{white}{11}} 
    \\
    \vspace{2mm}
    \\
    \rotatebox[origin=c]{90}{San Francisco} &
    \raisebox{-0.5\height}{\includegraphics[width=0.14\linewidth]{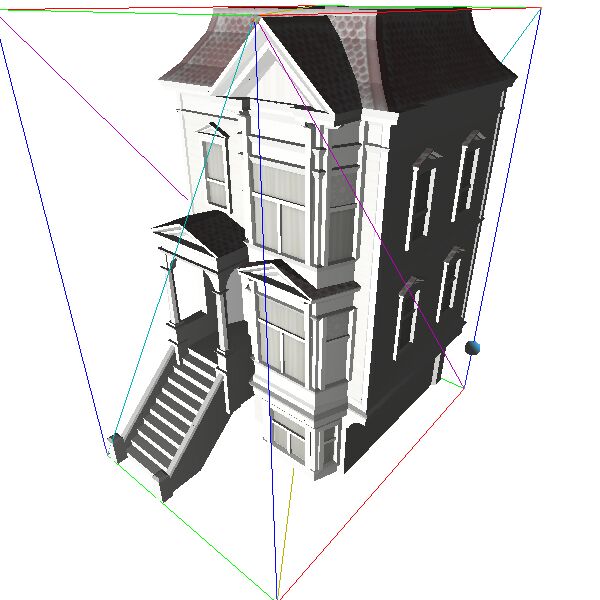}} &
    \raisebox{-0.5\height}{\includegraphics[width=0.14\linewidth]{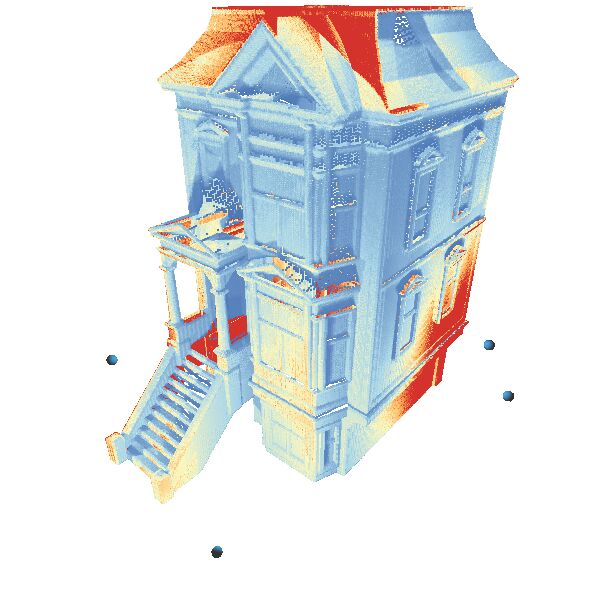}} &
    \raisebox{-0.5\height}{\includegraphics[width=0.14\linewidth]{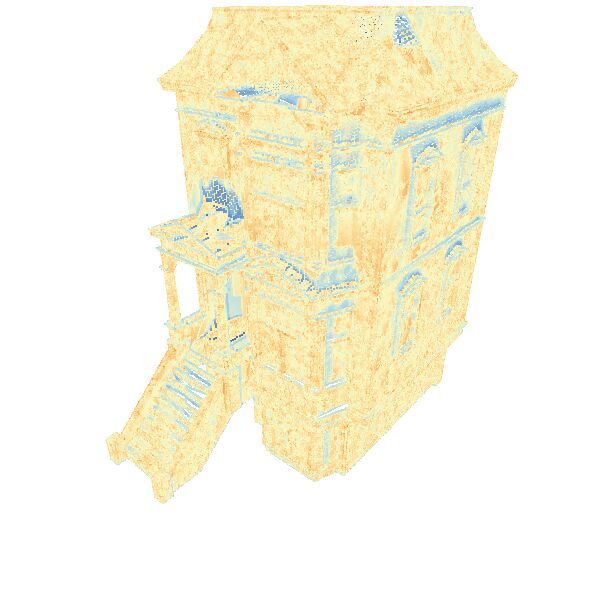}} &
    \raisebox{-0.5\height}{\includegraphics[width=0.14\linewidth]{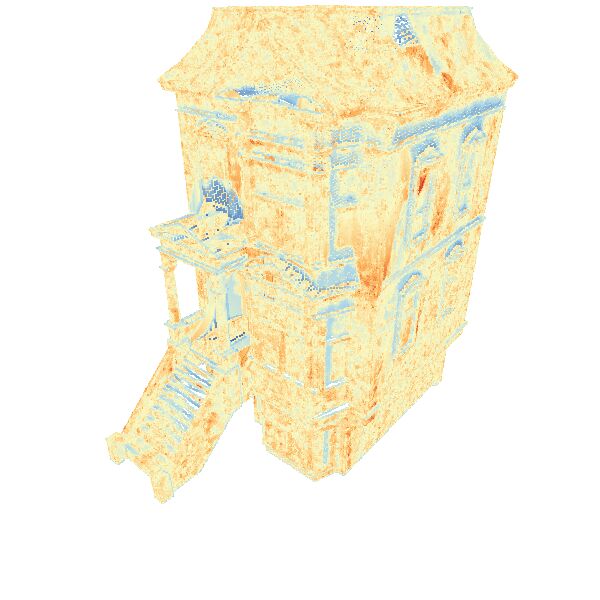}} &
    \raisebox{-0.5\height}{\includegraphics[width=0.14\linewidth]{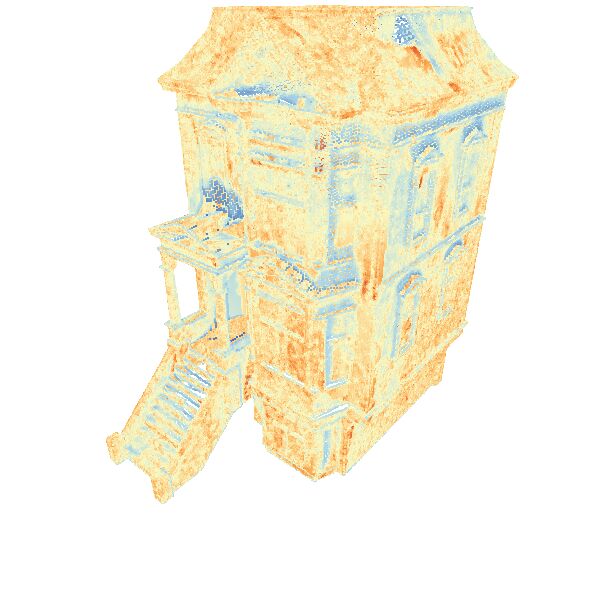}} &
    \raisebox{-0.5\height}{\includegraphics[width=0.14\linewidth]{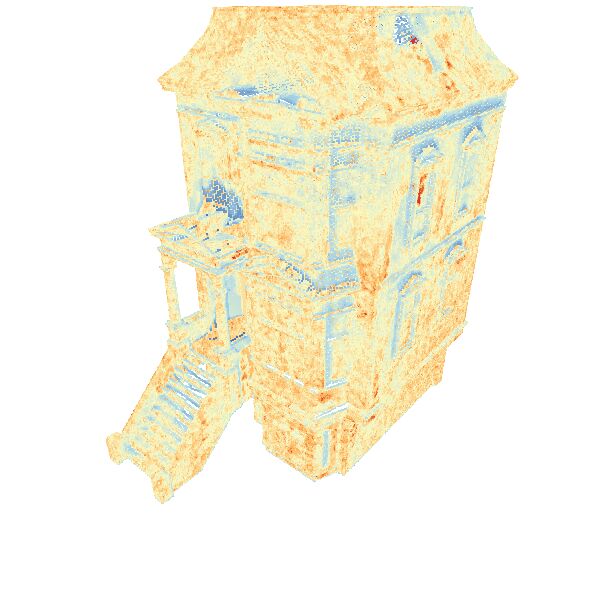}}
    \\
    & &
    \figfull{synthetic_ss/legend}{$0\,p/m^2$}{$2e4\,p/m^2$}{white}{2.5} &
    \multicolumn{4}{c}{\figfull{synthetic_ss/legend}{$0\,p/m^2$}{$2e3\,p/m^2$}{white}{11}}
    \\
\end{tabular*}
\caption{\label{fig:c4f5} Renders of the synthetic datasets. From left to right: (1) scanned mesh, (2) simulated cloud and (3-6) simplified clouds ($\lambda=0.1$). The color scale emphasizes the homogeneity/heterogeneity of the local point density across the cloud. Our method ($k=6$) shows the most homogeneous densities. }
\end{figure*}
\begin{figure*}[!htb]
\centering
    \includegraphics[width=1\linewidth]{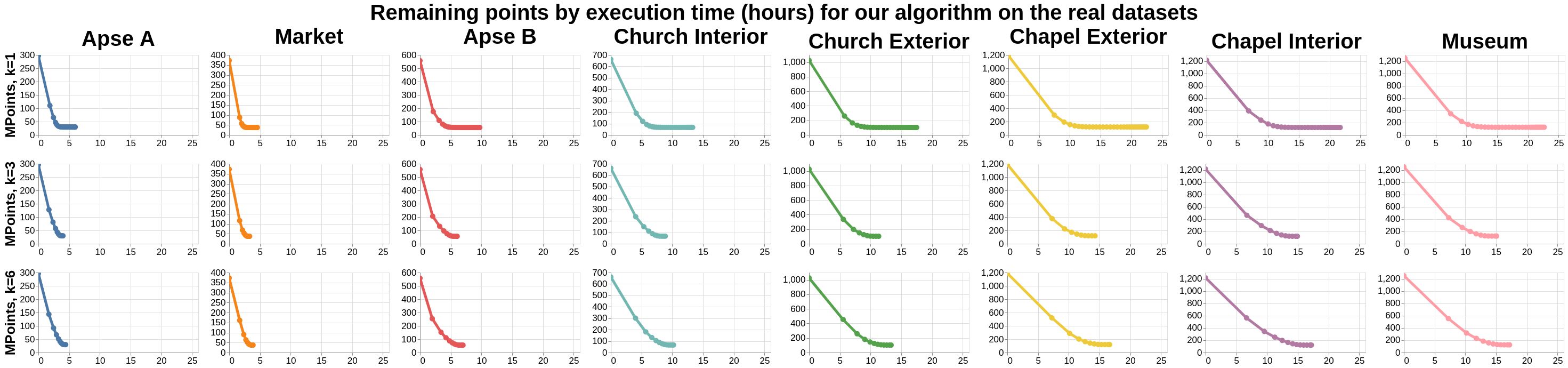}
\caption{Remaining samples on the simplified clouds as our algorithm progresses (each point represents the end of one iteration). Results for our method with $k\in\{1,3,6\}$ and $k+b=14$ for $\lambda=0.1$, on the different real models. The algorithm with $k=6$ converges in 11 to 12 iterations. For $k=1$, execution times are longer because the algorithm uses a few additional refining iterations where very few points are removed. } 
\label{fig:c4f6}
\end{figure*}
\begin{figure}[!htb]
\centering
    \includegraphics[width=0.49\linewidth]{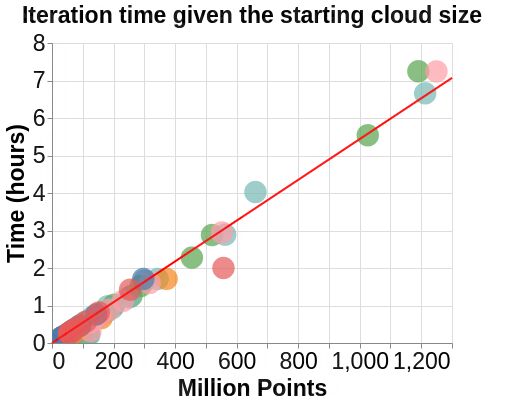}
    \includegraphics[width=0.49\linewidth]{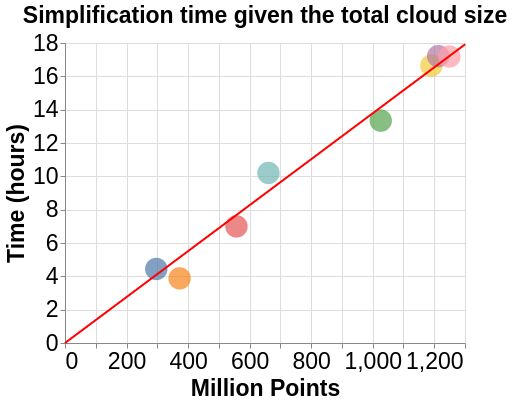}
\caption{Illustrating the relation between the size of a cloud and simplification time. Results for our method with $k=6$ and $k+b=14$ for $\lambda=0.1$, on the different real models. Left: each point represents the execution time for a single iteration, depending on the size of the cloud at the beginning of each iteration. The relation is approximately linear, taking 1 hour for every 184 million points. Right: execution time for the complete simplification process according to the original size of the cloud. The relation is approximately linear taking 1 hour for every 73 million points. } 
\label{fig:c4f7}
\end{figure}

\Cref{fig:c4f4} shows the distribution of local point densities with increasing neighborhood radius for these models. Due to space constraints, we only report results for our algorithm using $k=1$ (baseline) and $k=6$ (best result) and $10^3m^3$ voxels, for Yuksel~\cite{yuksel2015, cyCodeBase}'s method with $\beta=0.65$ (author's parameter choice) and $\beta=0$ (best result), for MeshLab's implementation of Corsini's method~\cite{corsini2012, meshlab} and for our implementation of Dart Throwing. Looking at these charts, we can extract similar conclusions as for the cube's case. For most cases, our algorithm with $k=6$ reports the highest $R^2$ and smallest RMSE, indicating a good linear fit between the neighborhood area and points (homogeneous density). We can also see that our method achieves the largest covered area if we ignore the case of Yuksel with $\beta=0.65$, for which the linear relationship does not hold.

\Cref{fig:c4f5} shows the meshes of the synthetic models, the original simulated clouds, and the simplified ones obtained with our method with $k=6$, Yuksel with $\beta=0.65$, Corsini and Dart Throwing. We used a color scale to illustrate the local point density, allowing the visual inspection of the density homogeneity across the cloud. Upon close inspection, we can see that our method with $k=6$ produces the most homogeneous results. 

Finally, \Cref{tab:c4t1} reports the execution times for the different tested algorithms. Notice that our running times include out-of-core operations, while all competing approaches were executed in-core. Although Corsini's method~\cite{corsini2012, meshlab} and Dart Throwing are the fastest ones, we have also seen they produce the worst quality results. Our method performed similarly to Yuksel's on these examples. We can also see that the performance of our method drops when increasing the number of neighbors $k$ when not using the neighbor buffer. When increasing the buffer size to $k+b=14$, the performance for $k=1$ does not significantly change. However, for $k=6$ we improve the performance and get close to the execution time for $k=1$.

\begin{figure*}[!htb]
\centering
\begin{tabular*}{\linewidth}{ccccccc}
    & Scanned Cloud & Original Density & Ours ($k=1$) & Ours ($k=3$) & Ours ($k=6$) & Ours ($k=6$) \\
    \rotatebox[origin=c]{90}{Apse A} \rotatebox[origin=c]{90}{297M Points} &
    \raisebox{-0.5\height}{\includegraphics[width=0.12\linewidth]{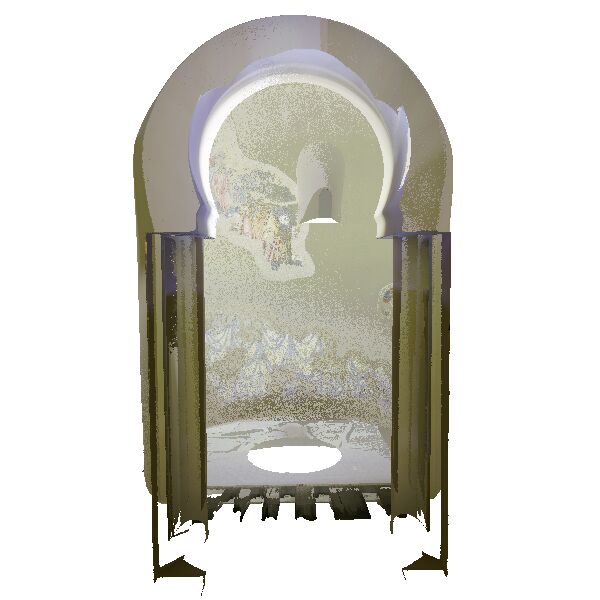}} &
    \raisebox{-0.5\height}{\includegraphics[width=0.12\linewidth]{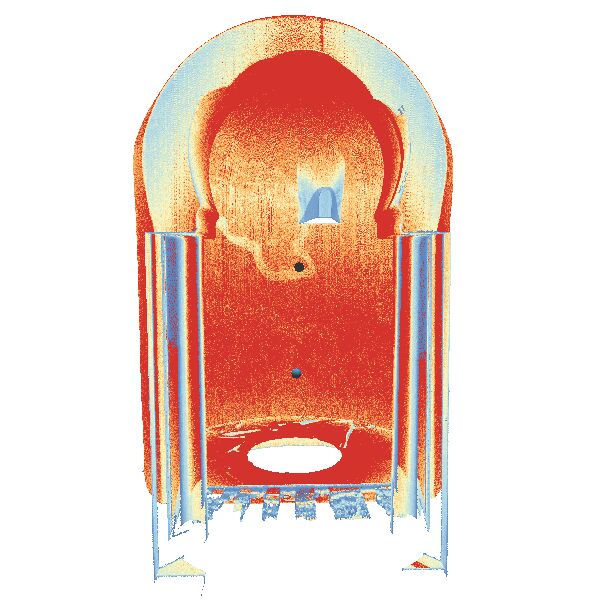}} &
    \raisebox{-0.5\height}{\includegraphics[width=0.12\linewidth]{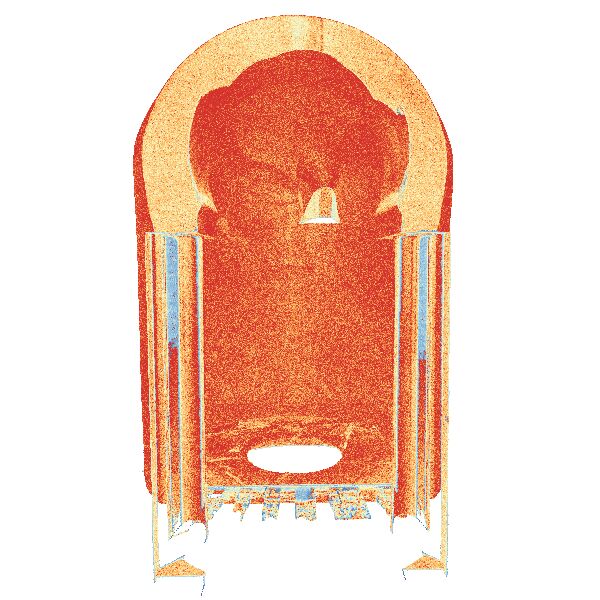}} &
    \raisebox{-0.5\height}{\includegraphics[width=0.12\linewidth]{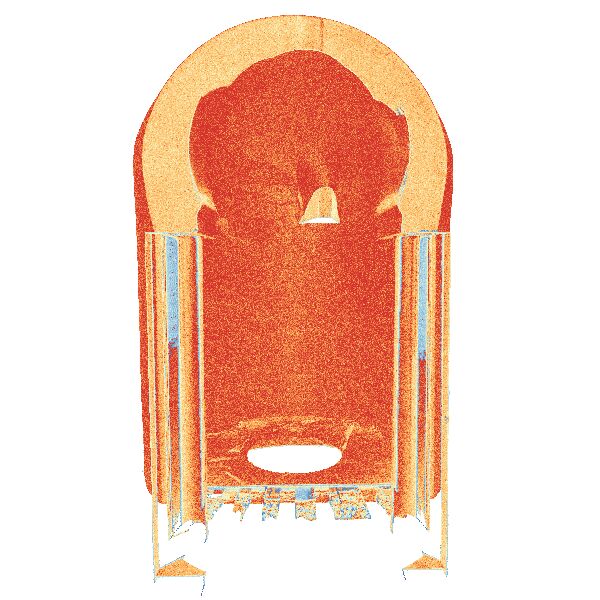}} &
    \raisebox{-0.5\height}{\includegraphics[width=0.12\linewidth]{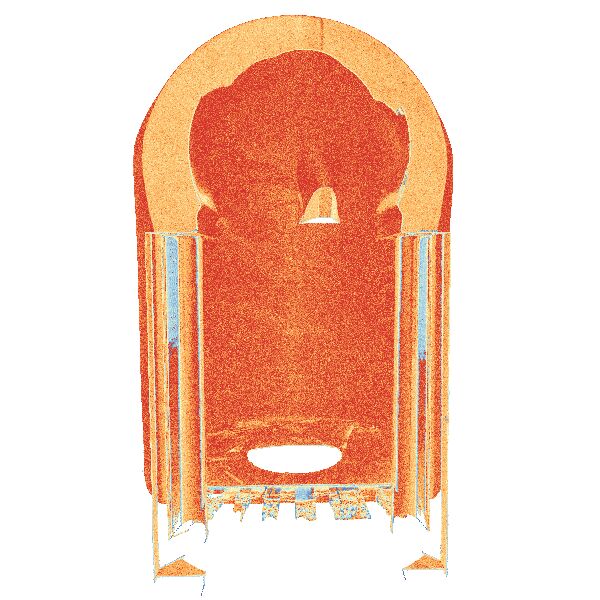}} &
    \raisebox{-0.5\height}{\includegraphics[width=0.12\linewidth]{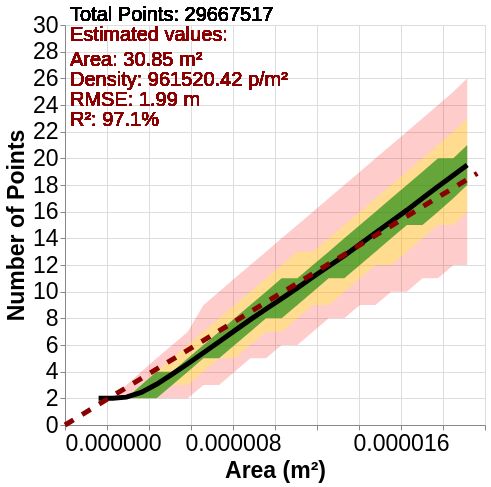}} 
    \\
    & &
    \figfull{synthetic_ss/legend}{$0\,p/m^2$}{$1e7\,p/m^2$}{white}{2.5} &
    \multicolumn{3}{c}{\figfull{synthetic_ss/legend}{$0\,p/m^2$}{$1e6\,p/m^2$}{white}{7.5}} &
    \\
    \rotatebox[origin=c]{90}{Market} \rotatebox[origin=c]{90}{372M Points} &
    \raisebox{-0.5\height}{\includegraphics[width=0.12\linewidth]{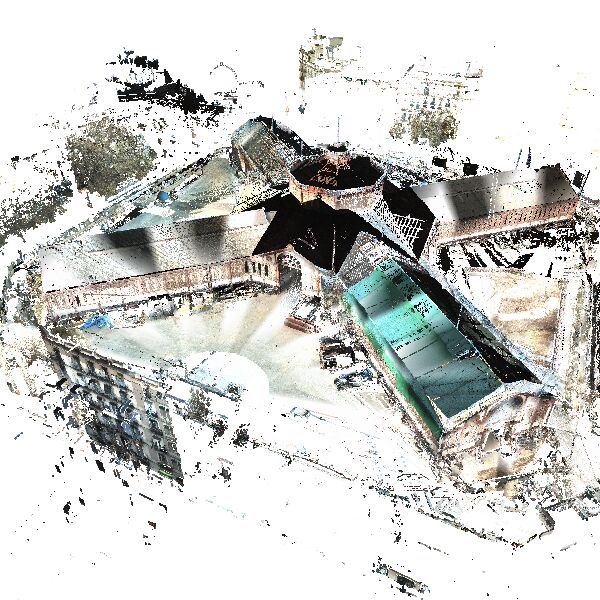}} &
    \raisebox{-0.5\height}{\includegraphics[width=0.12\linewidth]{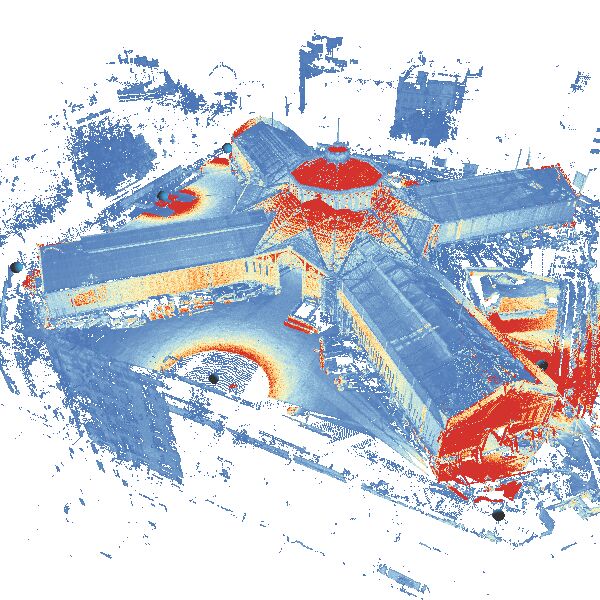}} &
    \raisebox{-0.5\height}{\includegraphics[width=0.12\linewidth]{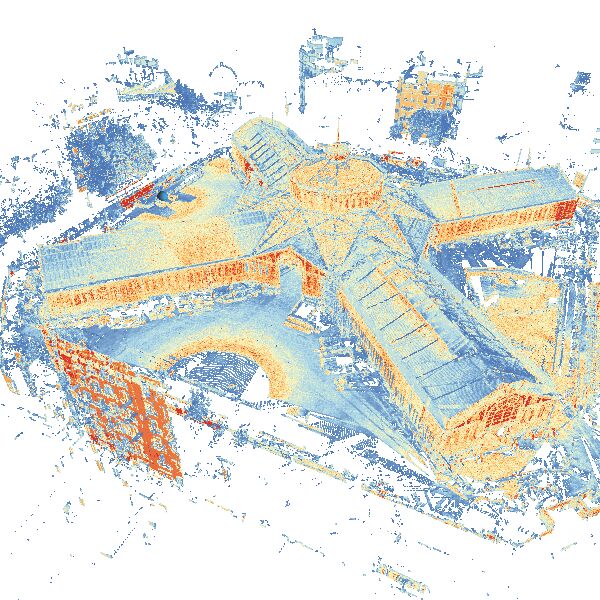}} &
    \raisebox{-0.5\height}{\includegraphics[width=0.12\linewidth]{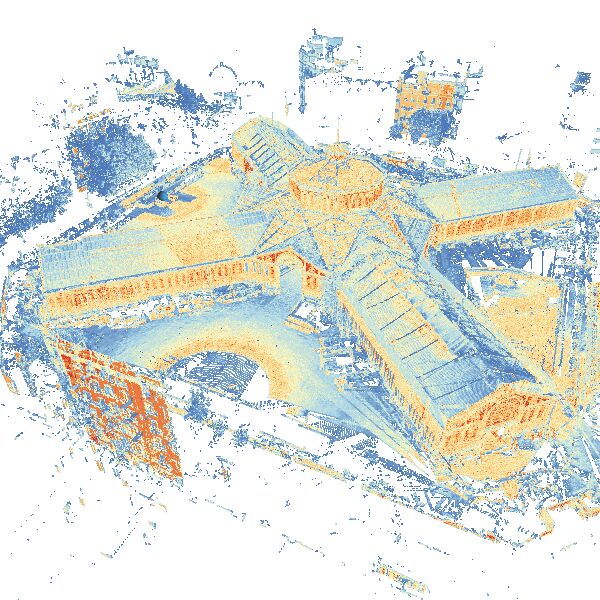}} &
    \raisebox{-0.5\height}{\includegraphics[width=0.12\linewidth]{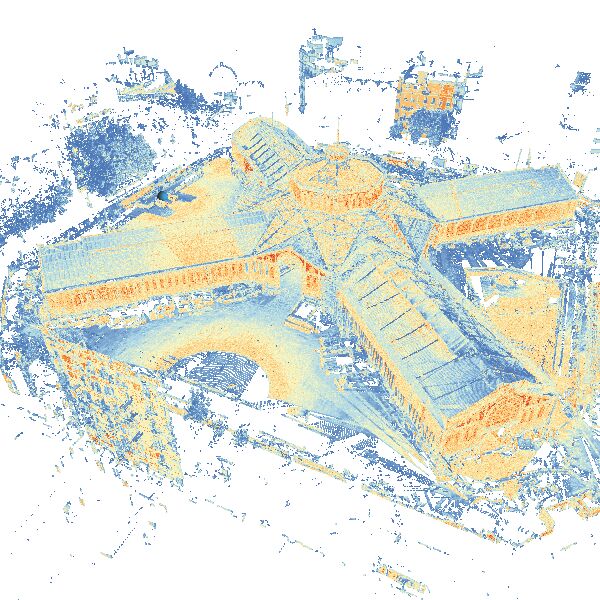}} &
    \raisebox{-0.5\height}{\includegraphics[width=0.12\linewidth]{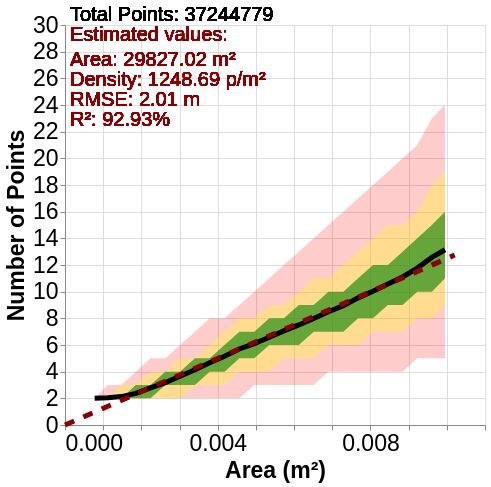}} 
    \\
    & &
    \figfull{synthetic_ss/legend}{$0\,p/m^2$}{$1e4\,p/m^2$}{white}{2.5} &
    \multicolumn{3}{c}{\figfull{synthetic_ss/legend}{$0\,p/m^2$}{$1e3\,p/m^2$}{white}{7.5}} &
    \\
    \vspace{2mm}
    \\
    \rotatebox[origin=c]{90}{Apse B} \rotatebox[origin=c]{90}{557M Points} &
    \raisebox{-0.5\height}{\includegraphics[width=0.12\linewidth]{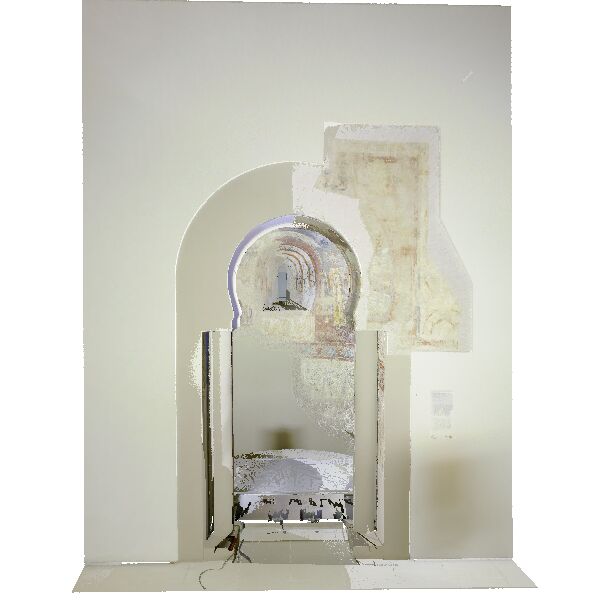}} &
    \raisebox{-0.5\height}{\includegraphics[width=0.12\linewidth]{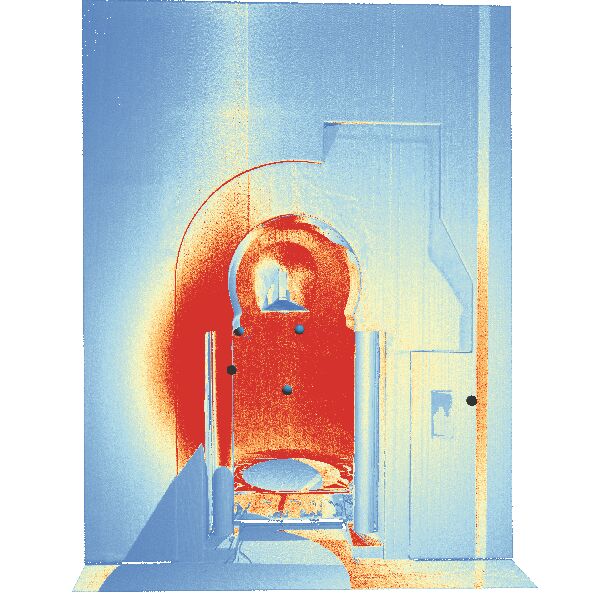}} &
    \raisebox{-0.5\height}{\includegraphics[width=0.12\linewidth]{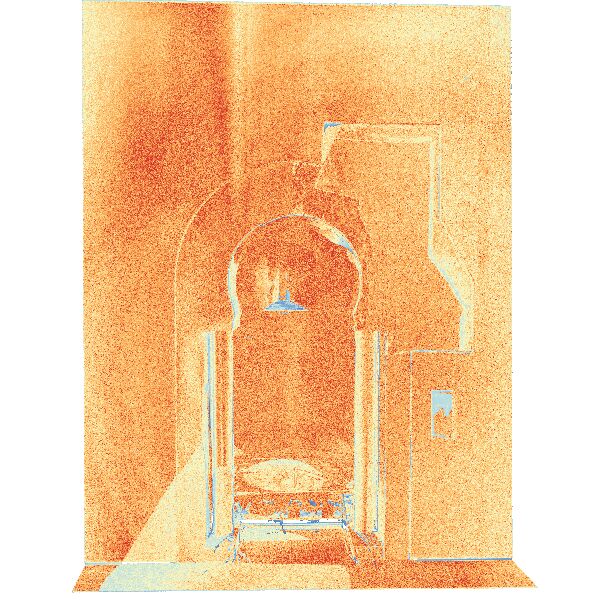}} &
    \raisebox{-0.5\height}{\includegraphics[width=0.12\linewidth]{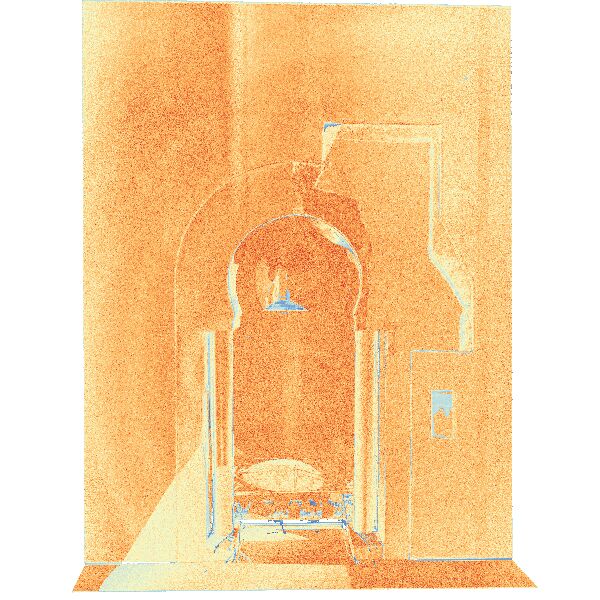}} &
    \raisebox{-0.5\height}{\includegraphics[width=0.12\linewidth]{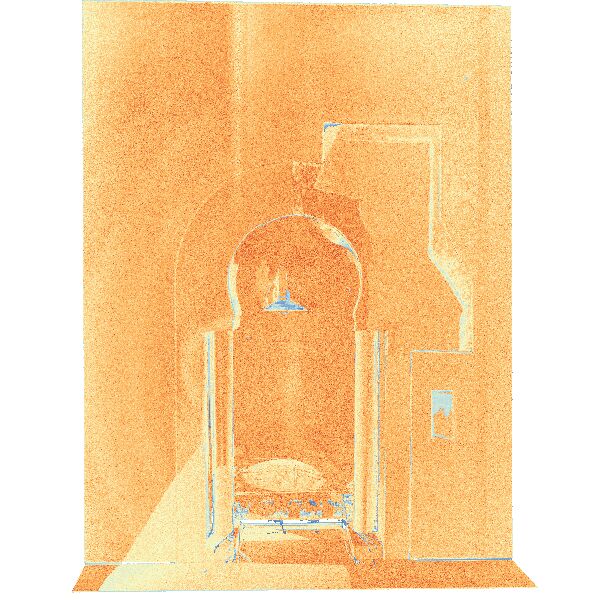}} &
    \raisebox{-0.5\height}{\includegraphics[width=0.12\linewidth]{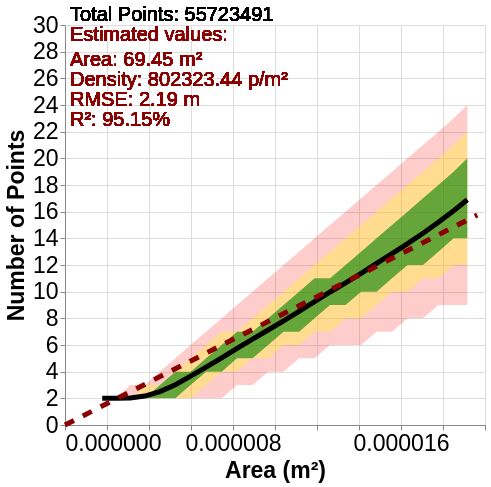}} 
    \\
    & &
    \figfull{synthetic_ss/legend}{$0\,p/m^2$}{$1e7\,p/m^2$}{white}{2.5} &
    \multicolumn{3}{c}{\figfull{synthetic_ss/legend}{$0\,p/m^2$}{$1e6\,p/m^2$}{white}{7.5}} &
    \\
    \rotatebox[origin=c]{90}{Church Interior} \rotatebox[origin=c]{90}{661M Points} &
    \raisebox{-0.5\height}{\includegraphics[width=0.12\linewidth]{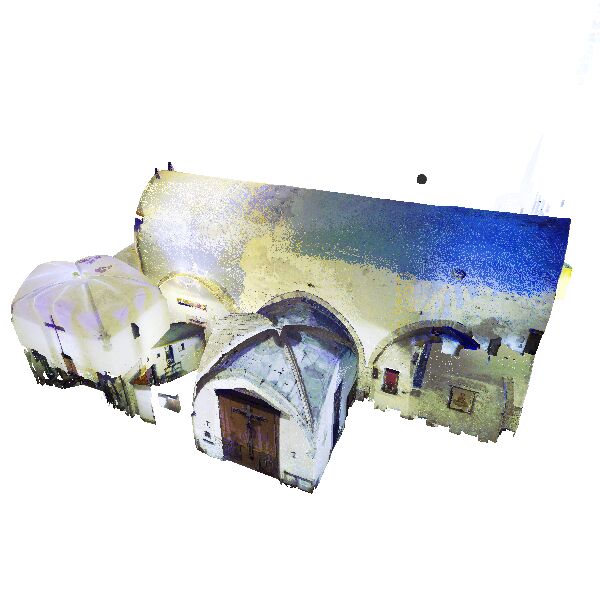}} &
    \raisebox{-0.5\height}{\includegraphics[width=0.12\linewidth]{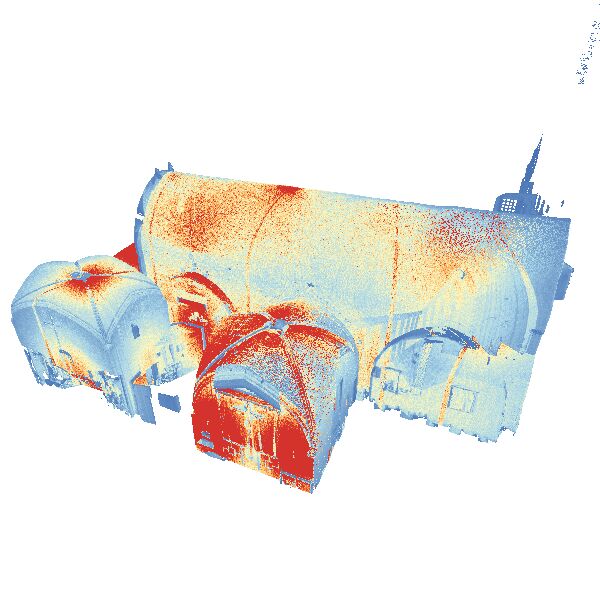}} &
    \raisebox{-0.5\height}{\includegraphics[width=0.12\linewidth]{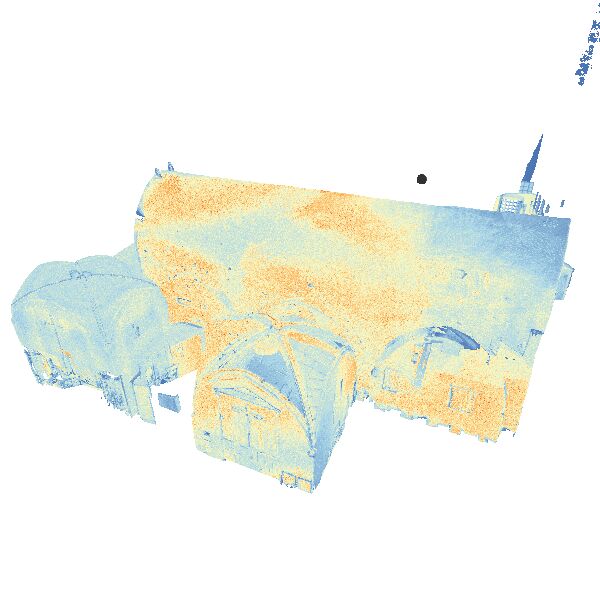}} &
    \raisebox{-0.5\height}{\includegraphics[width=0.12\linewidth]{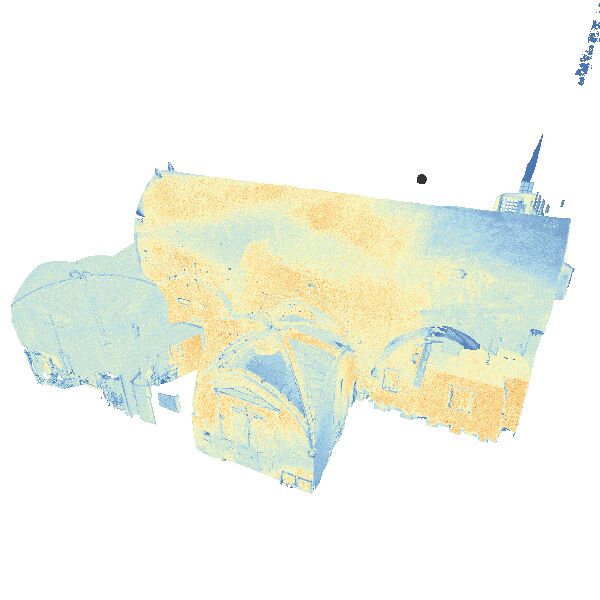}} &
    \raisebox{-0.5\height}{\includegraphics[width=0.12\linewidth]{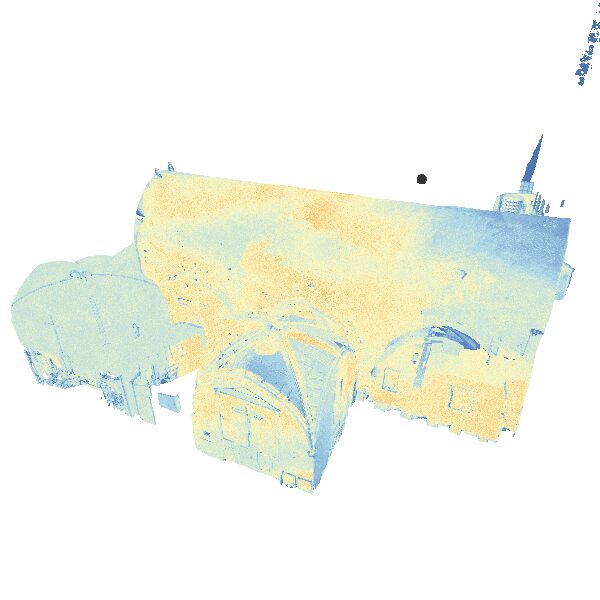}} &
    \raisebox{-0.5\height}{\includegraphics[width=0.12\linewidth]{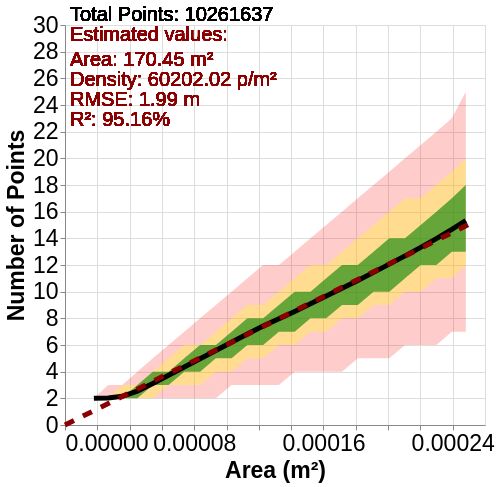}} 
    \\
    & &
    \figfull{synthetic_ss/legend}{$0\,p/m^2$}{$1e6\,p/m^2$}{white}{2.5} &
    \multicolumn{3}{c}{\figfull{synthetic_ss/legend}{$0\,p/m^2$}{$1e5\,p/m^2$}{white}{7.5}} &
    \\
    \rotatebox[origin=c]{90}{Church Exterior} \rotatebox[origin=c]{90}{1026M Points} &
    \raisebox{-0.5\height}{\includegraphics[width=0.12\linewidth]{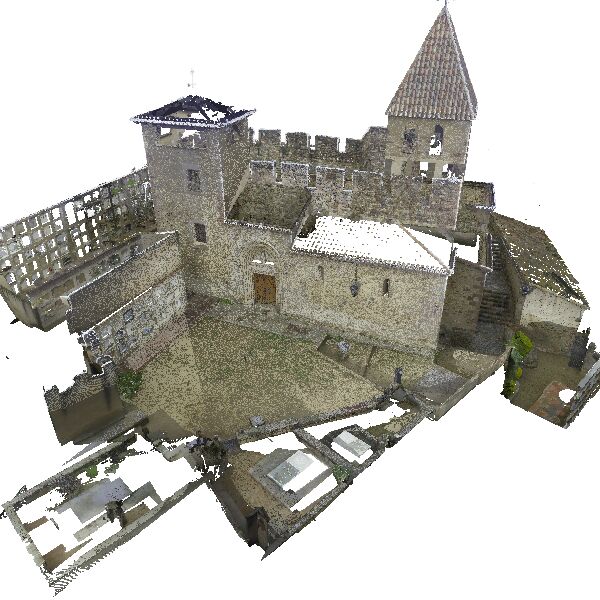}} &
    \raisebox{-0.5\height}{\includegraphics[width=0.12\linewidth]{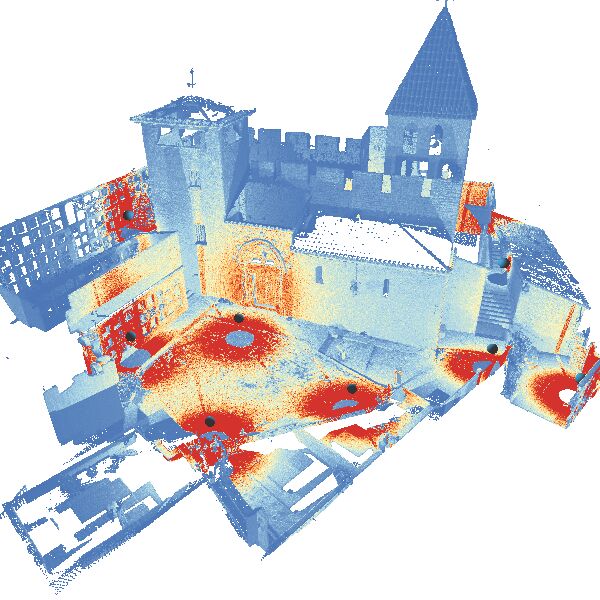}} &
    \raisebox{-0.5\height}{\includegraphics[width=0.12\linewidth]{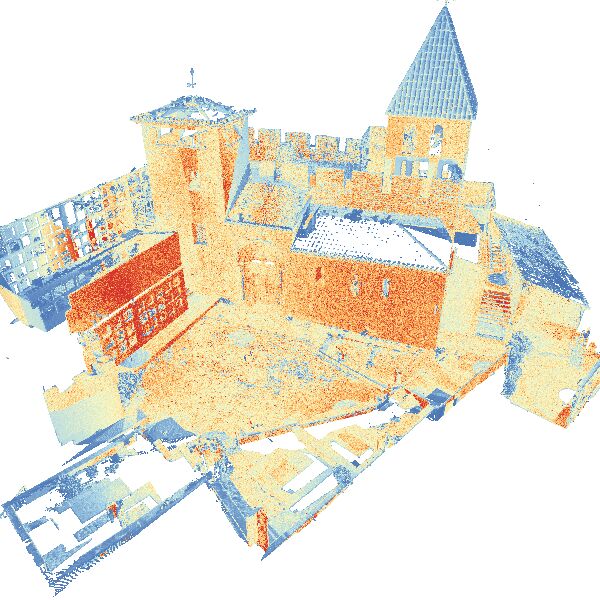}} &
    \raisebox{-0.5\height}{\includegraphics[width=0.12\linewidth]{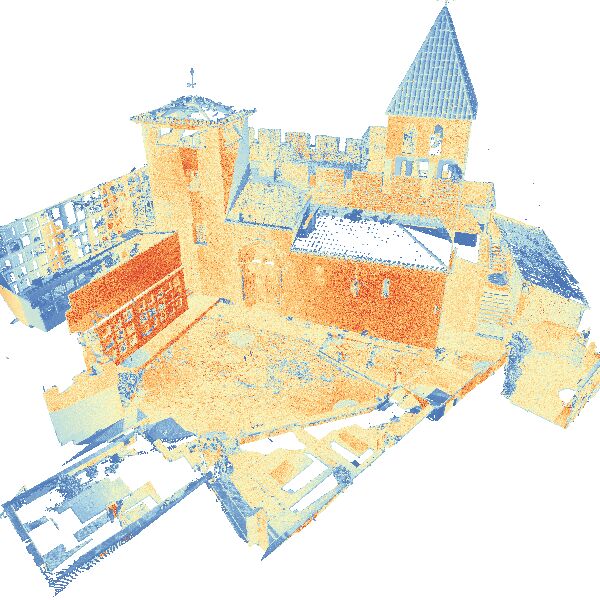}} &
    \raisebox{-0.5\height}{\includegraphics[width=0.12\linewidth]{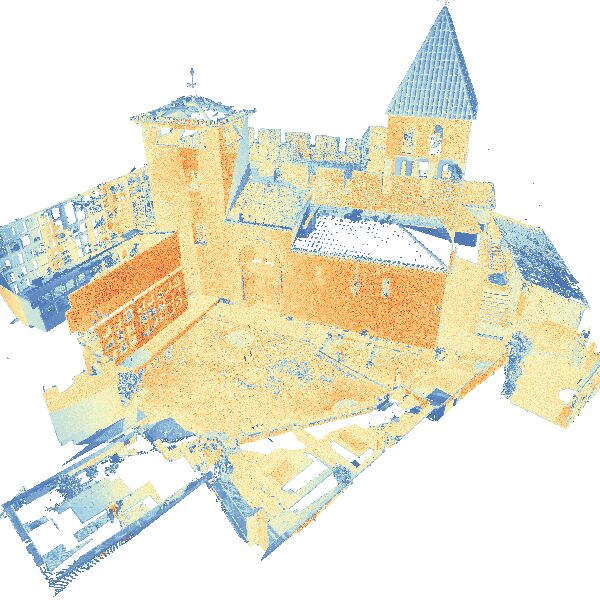}} &
    \raisebox{-0.5\height}{\includegraphics[width=0.12\linewidth]{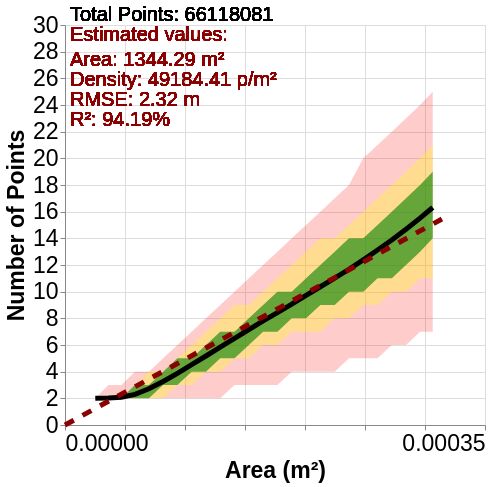}} 
    \\
    & &
    \figfull{synthetic_ss/legend}{$0\,p/m^2$}{$1e6\,p/m^2$}{white}{2.5} &
    \multicolumn{3}{c}{\figfull{synthetic_ss/legend}{$0\,p/m^2$}{$1e5\,p/m^2$}{white}{7.5}} &
    \\
    \rotatebox[origin=c]{90}{Chapel Exterior} \rotatebox[origin=c]{90}{1191M Points} &
    \raisebox{-0.5\height}{\includegraphics[width=0.12\linewidth]{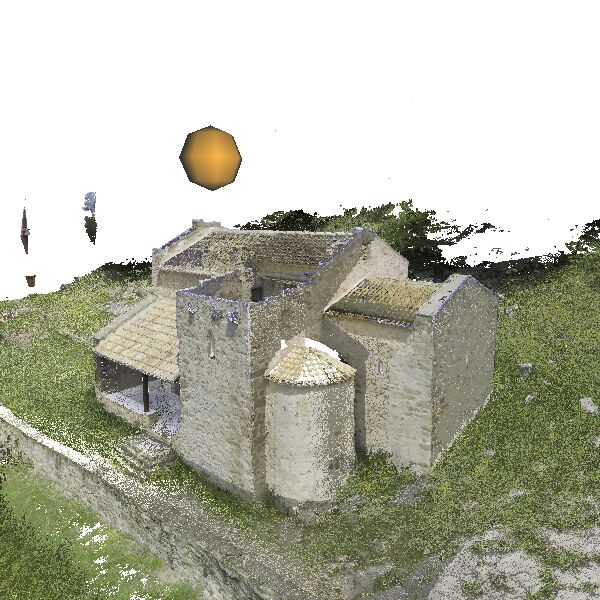}} &
    \raisebox{-0.5\height}{\includegraphics[width=0.12\linewidth]{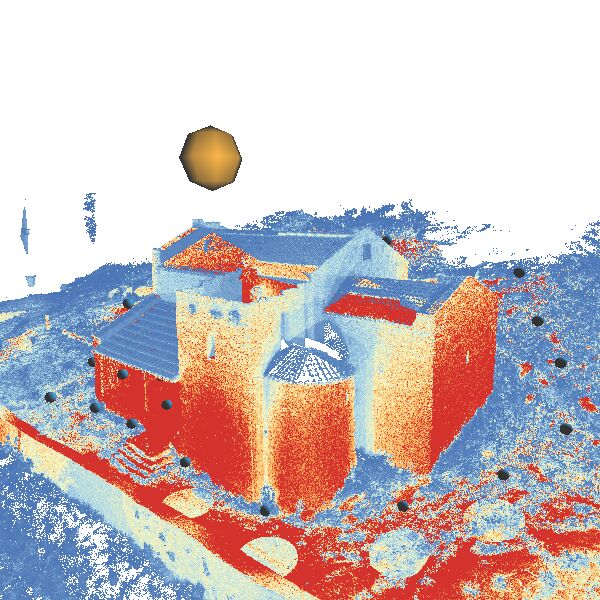}} &
    \raisebox{-0.5\height}{\includegraphics[width=0.12\linewidth]{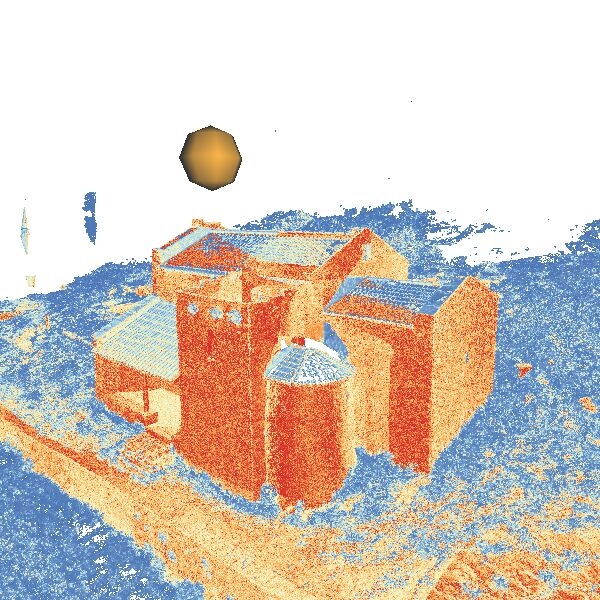}} &
    \raisebox{-0.5\height}{\includegraphics[width=0.12\linewidth]{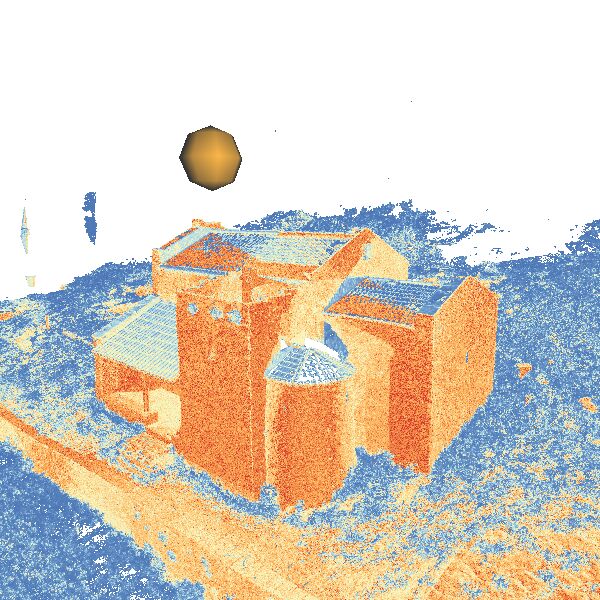}} &
    \raisebox{-0.5\height}{\includegraphics[width=0.12\linewidth]{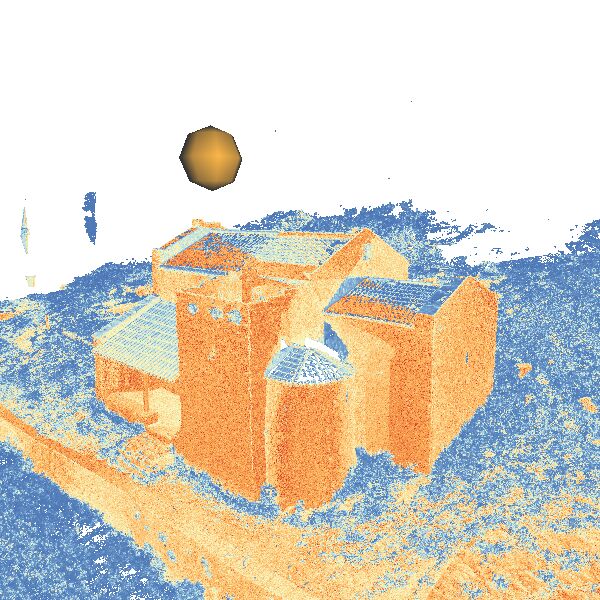}} &
    \raisebox{-0.5\height}{\includegraphics[width=0.12\linewidth]{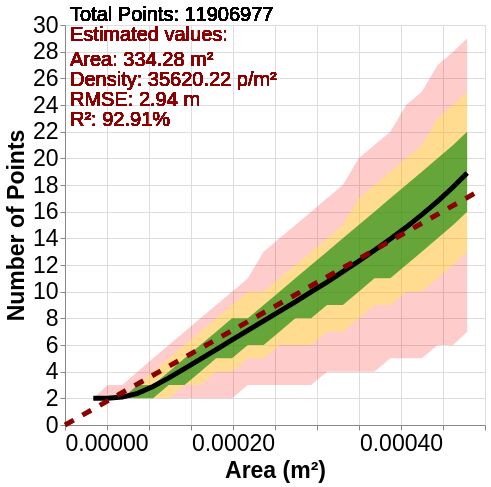}} 
    \\
    & &
    \figfull{synthetic_ss/legend}{$0\,p/m^2$}{$5e5\,p/m^2$}{white}{2.5} &
    \multicolumn{3}{c}{\figfull{synthetic_ss/legend}{$0\,p/m^2$}{$5e4\,p/m^2$}{white}{7.5}} &
    \\
    \rotatebox[origin=c]{90}{Chapel Interior} \rotatebox[origin=c]{90}{1213M Points} &
    \raisebox{-0.5\height}{\includegraphics[width=0.12\linewidth]{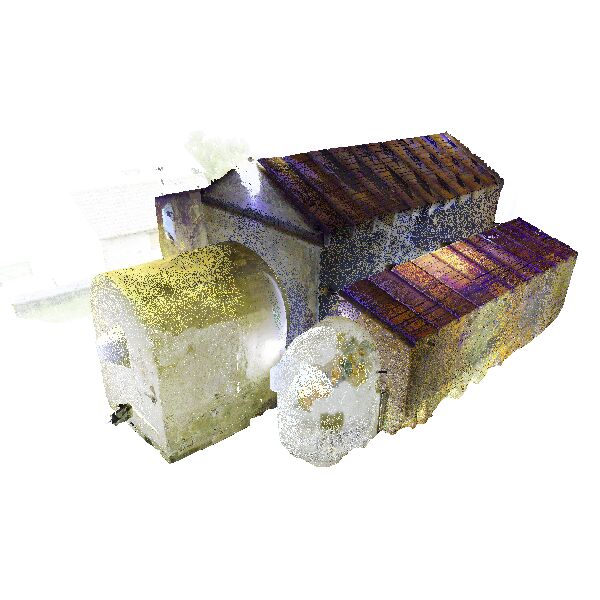}} &
    \raisebox{-0.5\height}{\includegraphics[width=0.12\linewidth]{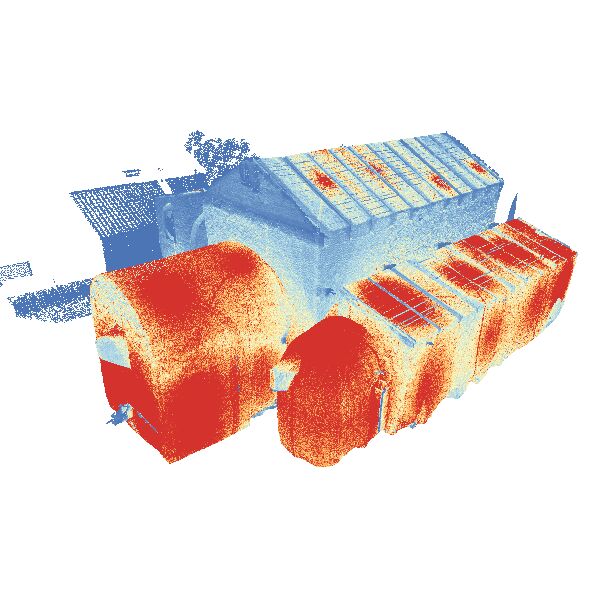}} &
    \raisebox{-0.5\height}{\includegraphics[width=0.12\linewidth]{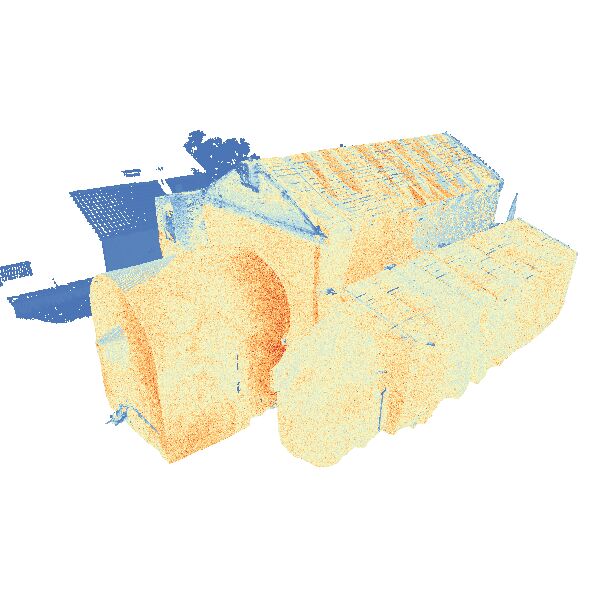}} &
    \raisebox{-0.5\height}{\includegraphics[width=0.12\linewidth]{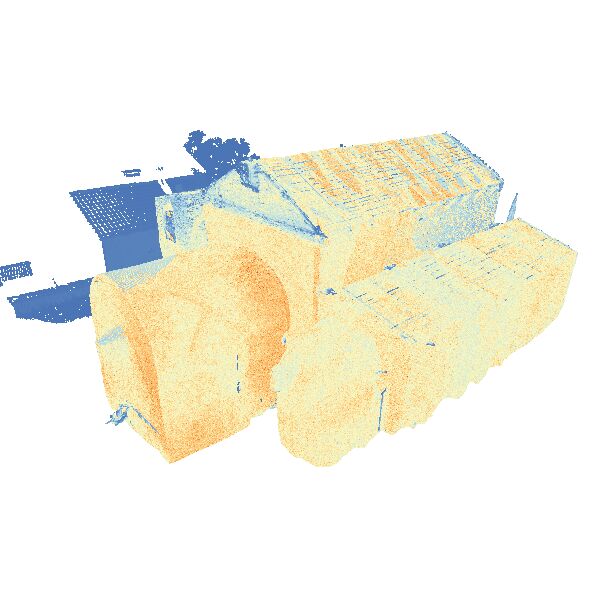}} &
    \raisebox{-0.5\height}{\includegraphics[width=0.12\linewidth]{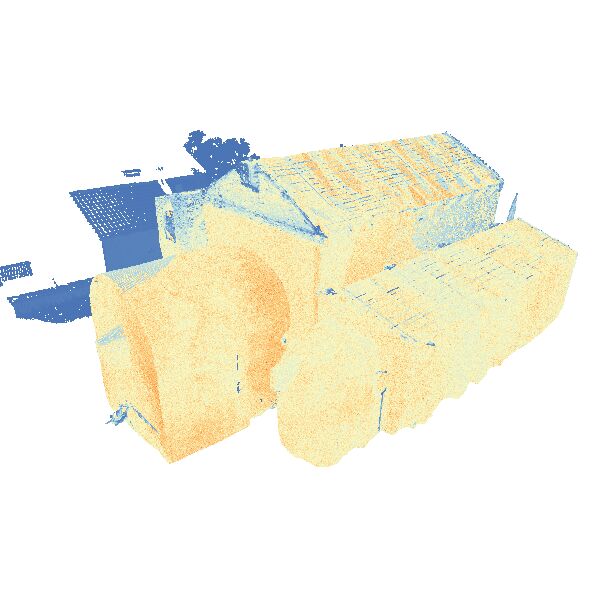}} &
    \raisebox{-0.5\height}{\includegraphics[width=0.12\linewidth]{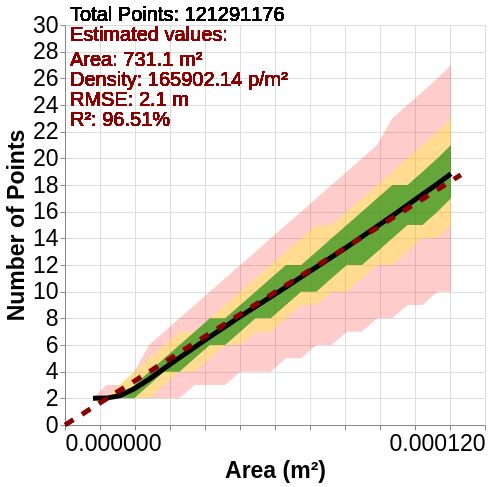}} 
    \\
    & &
    \figfull{synthetic_ss/legend}{$0\,p/m^2$}{$3e6\,p/m^2$}{white}{2.5} &
    \multicolumn{3}{c}{\figfull{synthetic_ss/legend}{$0\,p/m^2$}{$3e5\,p/m^2$}{white}{7.5}} &
    \\
    \rotatebox[origin=c]{90}{Museum} \rotatebox[origin=c]{90}{1250M Points} &
    \raisebox{-0.5\height}{\includegraphics[width=0.12\linewidth]{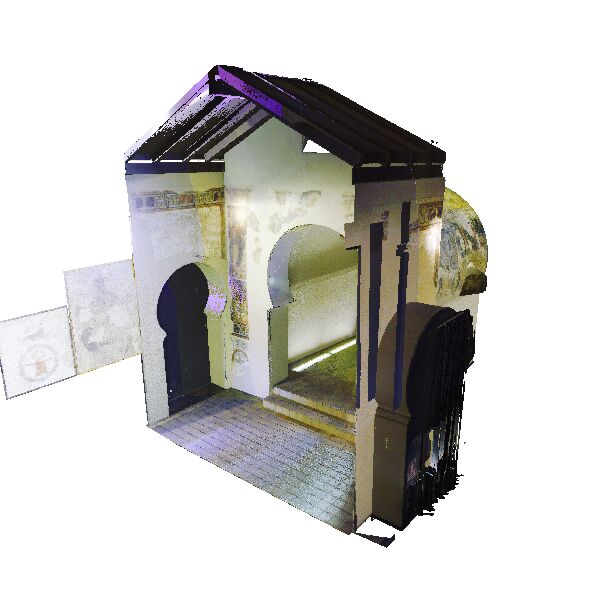}} &
    \raisebox{-0.5\height}{\includegraphics[width=0.12\linewidth]{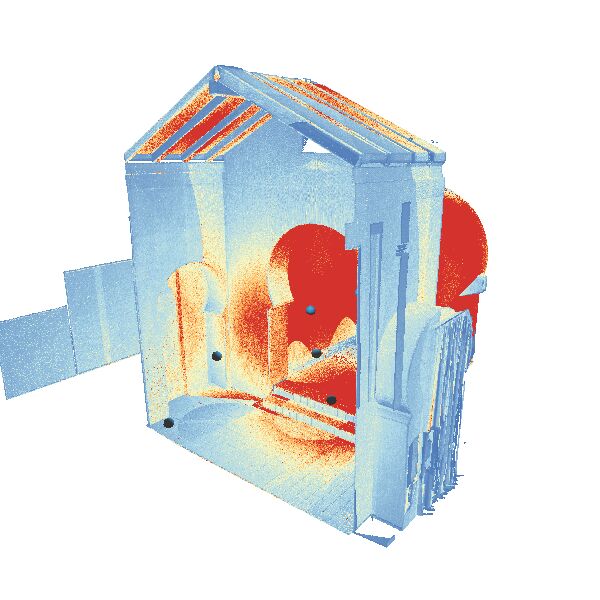}} &
    \raisebox{-0.5\height}{\includegraphics[width=0.12\linewidth]{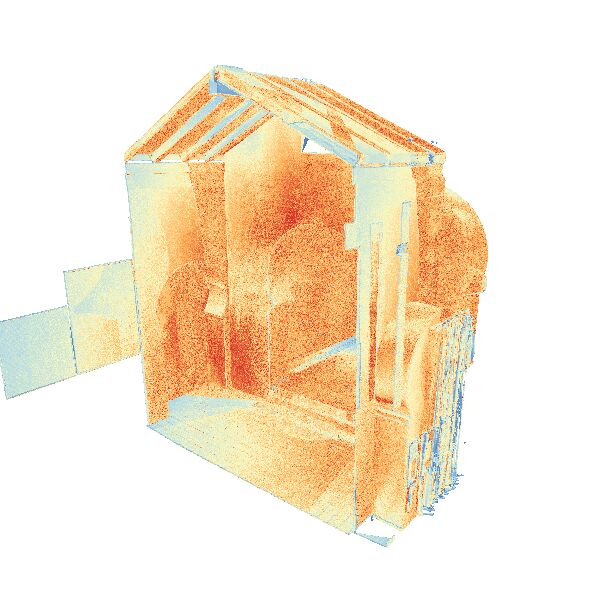}} &
    \raisebox{-0.5\height}{\includegraphics[width=0.12\linewidth]{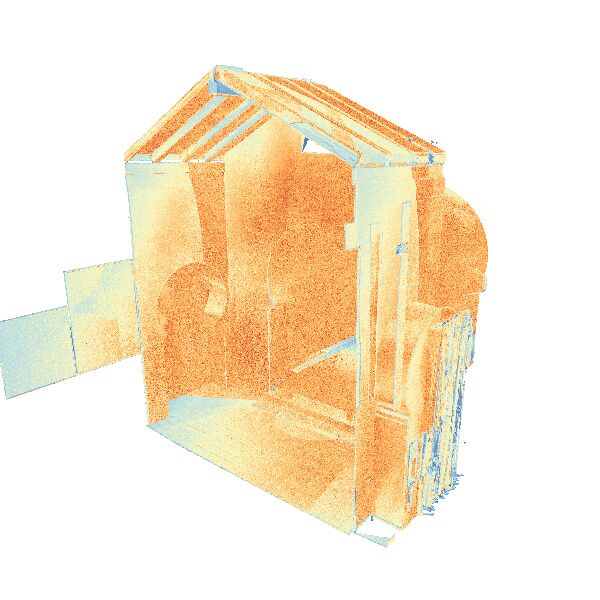}} &
    \raisebox{-0.5\height}{\includegraphics[width=0.12\linewidth]{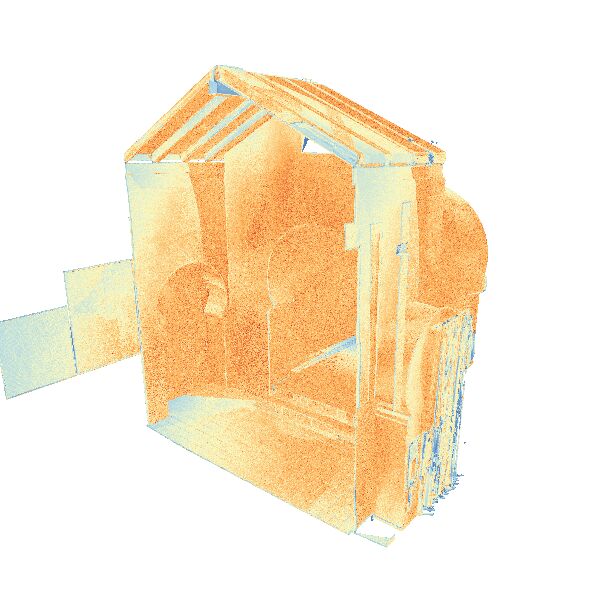}} &
    \raisebox{-0.5\height}{\includegraphics[width=0.12\linewidth]{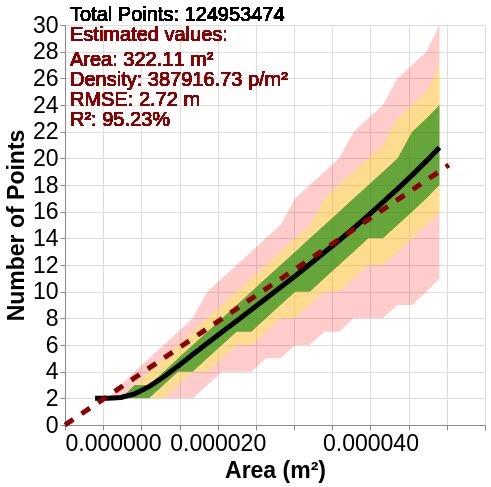}} 
    \\
    & &
    \figfull{synthetic_ss/legend}{$0\,p/m^2$}{$5e6\,p/m^2$}{white}{2.5} &
    \multicolumn{3}{c}{\figfull{synthetic_ss/legend}{$0\,p/m^2$}{$5e5\,p/m^2$}{white}{7.5}} &
\end{tabular*}
\caption{\label{fig:c4f8} Renders of the real datasets. From left to right: (1) scanned clouds with their original colors, (2) density distribution on the original clouds and (3-5) density distribution on the simplified clouds ($\lambda=0.1$). The color scale emphasizes the homogeneity/heterogeneity of the local point density across the cloud. Our method ($k=6$) shows the most homogeneous densities.}
\end{figure*}

\paragraph*{Real models} 
We now discuss the results with actual LiDAR data from high-end LiDAR scanners (see \Cref{fig:c4f8}). These models were too large to fit in core memory, and thus we could not compare our algorithm against Yuksel's and Corsini's methods. Since the scanner was mounted on a tripod, significant density variations on the floor are apparent in the original datasets, from highly dense regions near sensor locations to poorly sampled surfaces far away from these locations. Our decimation algorithm succeeded in equalizing densities and reaching exactly the target cloud size. In~\Cref{fig:c4f6} we show the running times for our algorithm with $k\in\{1, 3, 6\}$ on the different real datasets. Using $k=6$, running times varied from about 5 hours (300 million points) to 18 hours (1.2 billion points), always \carlos{taking less than}  12 iterations. In~\Cref{fig:c4f7} we further analyzed these times. We found an approximately linear empirical relation between the size of a cloud and the execution times for each iteration and the complete algorithm. This could be useful to predict the approximate processing time for a cloud of arbitrary size. 

\setlength{\tabcolsep}{8pt}
\renewcommand{\arraystretch}{0}

\setlength{\tabcolsep}{2.5pt}
\renewcommand{\arraystretch}{0}

\begin{figure}[!htb]
\centering
\begin{tabular*}{\linewidth}{cccc}
    & Ours ($k=6$) & \makecell{Ours ($k=6$) \\ + Normal Weights} & \makecell{Ours ($k=6$) \\+ Color Weights} \\
    \rotatebox[origin=c]{90}{Monastery} &
    \raisebox{-0.5\height}{\includegraphics[width=0.30\linewidth]{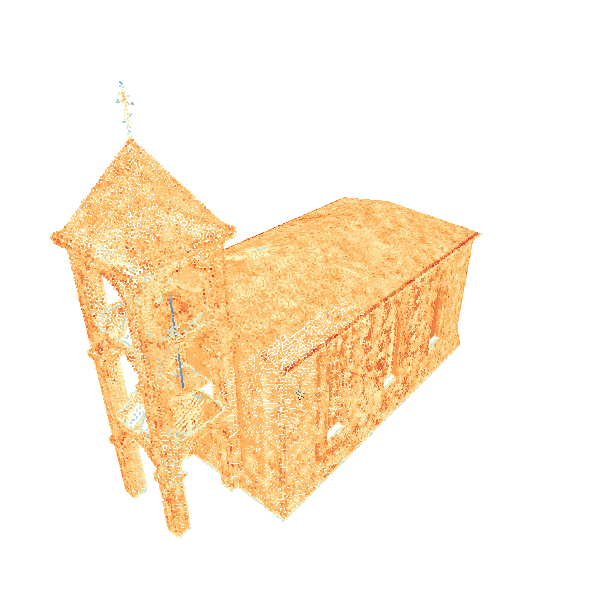}} &
    \raisebox{-0.5\height}{\includegraphics[width=0.30\linewidth]{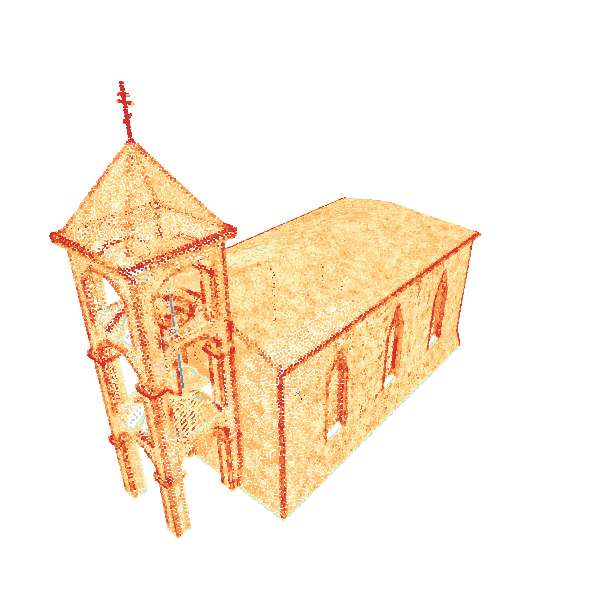}} &
    \raisebox{-0.5\height}{\includegraphics[width=0.30\linewidth]{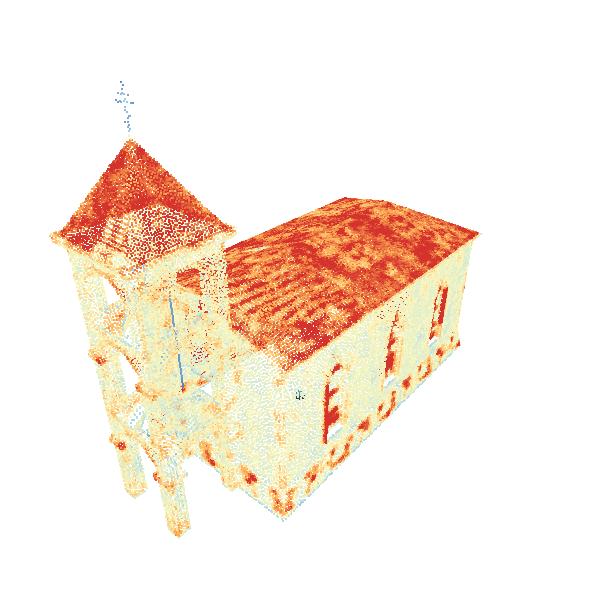}} 
    \\
    & 
    \multicolumn{3}{c}{\figfull{synthetic_ss/legend}{$0\,p/m^2$}{$3e2\,p/m^2$}{white}{7}}
    \\
    \vspace{2mm}
    \\
    \rotatebox[origin=c]{90}{San Francisco} &
    \raisebox{-0.5\height}{\includegraphics[width=0.30\linewidth]{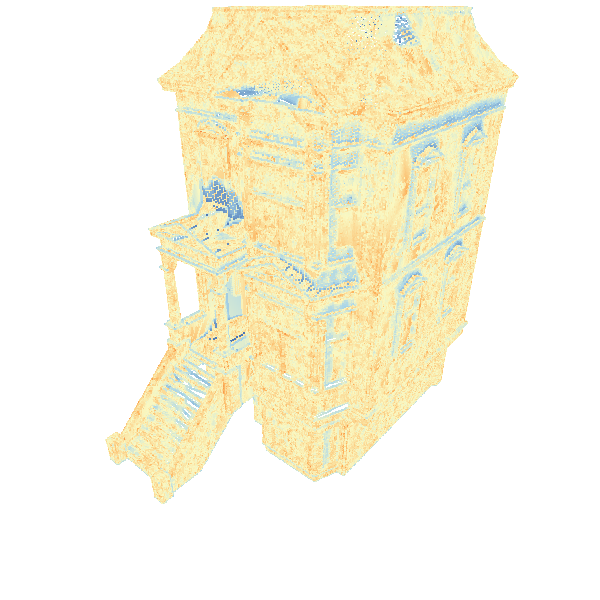}} &
    \raisebox{-0.5\height}{\includegraphics[width=0.30\linewidth]{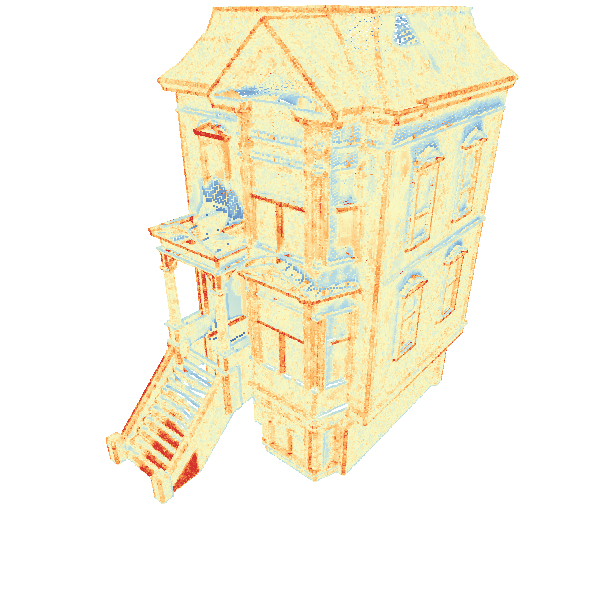}} &
    \raisebox{-0.5\height}{\includegraphics[width=0.30\linewidth]{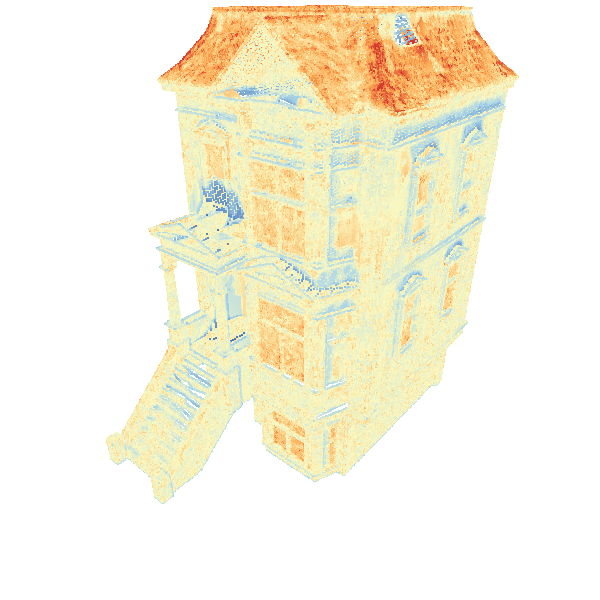}}
    \\
    & 
    \multicolumn{3}{c}{\figfull{synthetic_ss/legend}{$0\,p/m^2$}{$2e3\,p/m^2$}{white}{7}}
    \\
\end{tabular*}
\caption{\label{fig:c4f9} Renders of some synthetic datasets. From left to right: (1) simplified cloud using our method with $k=6$, (2) simplified cloud using our method with $k=6$ and normal weights, and (3) simplified cloud using our method with $k=6$ and color weights. The color scale emphasizes the homogeneity/heterogeneity of the local point density across the clouds. Using normal weights, we achieve preserving a larger number of samples near the features and edges of the model. In contrast, color weights help preserving samples on high-frequency texture areas (such as the windows and tiled roofs of our models).  }
\end{figure}

\paragraph*{Additional cost terms}

The proposed cost function can be modified by incorporating orientation-based (assuming the input point cloud is oriented) and color-based terms. Concerning orientation, the idea is to modify the cost function to penalize the removal of nearby points whose normal vectors exhibit a large orientation difference. The motivation is twofold. First, when computing pair-wise distances, we should ideally consider geodesic rather than Euclidean distances. We could assume the geodesic distance is close to the Euclidean distance for close points with similar normal directions. In contrast, points with opposite normal directions might belong to different surface sheets (e.g., different sides of a thin object). Thus their true geodesic distance might deviate arbitrarily from the Euclidean one. Similarly, our overall goal of equalizing point densities should be understood as referring to densities associated with samples belonging to the same local surface patch. 

\Cref{fig:c4f9} shows some results using our algorithm with extended point costs $w_{\bp,\bnp}^{k}$ that incorporate a term $\cos\theta$ = $\bnp\cdot \mathbf{n}_{\bp'}$, where $\theta$ is the angle between the normal $\bnp$ of point $\bp$ and that of its $k$-th closest point $\bp'$:

\begin{equation}
\label{eq:c4e13}
    w_{\bp,\bnp}^{k} = \sum_{\bp' \in \bn{k}{\bp}} \frac{\bnp\cdot \mathbf{n}_{\bp'}}{\bdist{\bp}{\bp'}^2} 
\end{equation}

\Cref{fig:c4f9} illustrates how this factor improves sampling density, e.g., on both sides of thin structures near the features and edges of the models. We also show some results where we add a color-deviation term to penalize the removal of points near color edges so that density is preserved in those regions with high-frequency color details. This could be desirable, e.g., when decimating artifacts (e.g., frescoes) for which color information might be more relevant than shape for certain studies.

\section{Conclusions and future work}

We have presented a method for simplifying a point cloud by selecting a suitable subset of points. This subsampling strategy is preferable over re-sampling whenever we prioritize preserving accuracy guarantees. Manufacturers of high-end LiDAR equipment put significant efforts to provide such error guarantees, which allow their devices to provide usable data in scenarios where such guarantees are critical, e.g., forensic applications~\cite{walsh2015}. In some countries, a dataset without a complete known error rate would not be considered evidence~\cite{walsh2015}. On the other hand, datasets with significant noise errors (e.g., from low-cost range scanners) might instead benefit from a re-sampling method.

From a formulation point-of-view, the major novelty is that we directly optimize by maximizing the minimum distances between points. Farthest-point optimization methods~\cite{yan2015} also maximize the minimum distance but in the context of a relaxation-based optimization. On these, each point is allowed to move to the farthest point (e.g., the Voronoi vertex farthest from its immediate neighbors), which gives much flexibility to get Blue-noise properties. We have shown the performance and output quality benefits of this approach in a sub-sampling setting. In terms of quality, this strategy produces better distributions than competing approaches~\cite{yuksel2015,corsini2012}. However, we could hardly compare to ~\cite{corsini2012} since their algorithm has not been designed to get close to a given decimation factor. A qualitative advantage of our approach with respect to~\cite{corsini2012} is the trivial addition of orientation-based and color-based terms to the cost function.

Moreover, the major advantage of our method compared to most competing approaches is the ability to process arbitrarily large point clouds. The use of the closest neighbors makes costs highly sensitive to updates and makes such updates more expensive. Fortunately, thanks to our voxelization-based out-of-core scheme and different optimizations (such as the neighbor buffer), the execution time of our algorithm scales roughly linearly with respect to the size of the input cloud.

We assume the scanned data has been acquired with a high-end scanner for which accurate error models are known. Therefore, as future work, we plan to incorporate such error models into our cost strategy, making our sub-sampling sensor-aware in the same spirit as some point processing methods~\cite{comino2018}. For example, a particular region might include accurate samples from a nearby-frontal scan and other less-accurate samples from distant or tangential scans. Adding this sensor-aware term to the cost definition would prioritize removing inaccurate samples in regions where more accurate points are already available but preserve them otherwise.  

Finally, as future work, we also plan to investigate further the impact of voxel size and neighbor buffer size on the execution time. Furthermore, we plan to implement a fully parallel version of our method (which should be straightforward). By carefully tuning these parameters and parallelizing the code, we expect significantly reducing the overall execution times. 

We plan to release an open C++ standalone implementation of our method upon publication of this work.

\paragraph*{Acknowledgments}

This work has been partially funded by Ministeri de Ciència i Innovació (MICIN), Agencia Estatal de Investigación (AEI) and the Fons Europeu de Desenvolupament Regional (FEDER) (project PID2021-122136OB-C21 funded by MCIN/AEI/10.13039/501100011033/FEDER, UE), the JPICH-0127 EU project Enhancement of Heritage Experiences: the Middle Ages. Digital
Layered Models of Architecture and Mural Paintings over Time (EHEM) and by the Universidad Rey Juan Carlos through the Distinguished Researcher position INVESDIST-04 under the call from 17/12/2020.

\bibliographystyle{acm}
\bibliography{bibliography}

\end{document}